\documentclass[prx,aps,twocolumn,floatfix,citeautoscript,showpacs,letter,portrait]{revtex4-2}
\usepackage{pdfpages}

\makeatletter
\AtBeginDocument{\let\LS@rot\@undefined}
\makeatother

\usepackage{graphicx,rotating,subfigure,amsmath,amsfonts,amssymb,delarray,color}
\usepackage{hyperref}
\usepackage{xcolor}
\usepackage{soul}
\usepackage{dsfont}
\usepackage[T1]{fontenc}
\usepackage{physics}
\usepackage{multirow}
\usepackage{bbm}

\newcommand{\e}{\text{e}}

\def\12{\frac{1}{2}}

\newcommand{\js}[1]{\textcolor{black}{#1}}

\newcommand{\ch}[1]{\textcolor{black}{#1}}
\usepackage{tikz} 
\usetikzlibrary{arrows.meta}
\usetikzlibrary{shapes,arrows}

\numberwithin{equation}{section}

\begin{document}
\title{Operator Growth in Disordered Spin Chains:\\ Indications for the Absence of Many-Body Localization}
\author{A. Weisse}
\affiliation{Max-Planck-Institut f\"ur Mathematik, Vivatsgasse 7, 53111 Bonn, Germany}
\author{R. Gerstner}
\author{J. Sirker}
\affiliation{Department of Physics and Astronomy and Manitoba
Quantum Institute, University of Manitoba, Winnipeg, Canada R3T 2N2}
\date{\today}
\begin{abstract}
We consider the spreading of a local operator $A$ in one-dimensional many-body systems with Hamiltonian $H$ by calculating the $k$-fold commutator $[H,[H,[...,[H,A]]]]$. We derive bounds for the operator norm of this commutator in free and interacting systems with and without disorder thus directly connecting the {\it operator growth hypothesis} with questions of localization. \ch{We analytically show, in particular, that an almost factorial growth of the operator norm---as recently proven for the random Ising model and strongly suggested to also hold for the Heisenberg model with random fields here---is inconsistent with an exponential localization of $A$. Assuming that a quasi-local unitary $U$ exists which maps $H$ onto an effective Hamiltonian $\tilde H=UHU^\dagger=\sum_n E_n \tau^z_n +\sum_{i,j} J_{ij} \tau^z_i\tau^z_j+\dots$, we show that $\tilde A=UAU^\dagger$ is a quasi-local operator which in the many-body case, in contrast to the Anderson case, indeed {\it does not remain exponentially localized} in general leading to an almost factorial growth of the commutator norm. Therefore the unitary $U$ in many-body systems with maximal norm growth either does not exist and such systems are always ergodic or unusual non-ergodic phases described by $\tilde H$ do exist which violate the operator growth hypothesis and in which local operators spread over the entire lattice implying that transport will eventually set in. To investigate this issue further, we concentrate on the XXZ chain with random magnetic fields. We analytically and symbolically verify our general results for the non-interacting Anderson and Aubry-Andr\'e models.} For the XXX case, the symbolic calculations are consistent with a maximal norm growth. Furthermore, we find no indication of a weakened exponential localization of $A$, expected for strong disorder and low commutator orders if the unitary $U$ does exist. Finally, we study the differences between the interacting and non-interacting cases when trying to perturbatively construct $U$ by consecutive Schrieffer-Wolff transformations. While it is straightforward to show that this construction converges in the Anderson case, we find no indications for a convergence in the interacting case, suggesting that $U$ does not exist and that many-body localization is absent.
\end{abstract}

\maketitle

\section{Introduction}
In a one-dimensional fermionic lattice model with short-range hoppings and without interactions, any amount of disorder leads to a localization of the eigenfunctions. This is known as Anderson localization \cite{Anderson58,EdwardsThouless,AbrahamsAnderson,AndersonLocalization}. An open question for the last 65 years has been what happens if interactions are included. One of the difficulties in investigating this question is that localization can no longer be defined in terms of single-particle eigenfunctions; new indicators of localization are needed. 

Studying the possibility of localization in the many-body case has gained renewed interest in the last 17 years starting with the work by Basko, Aleiner, and Altshuler \cite{BaskoAleiner}. This was followed by a large number of numerical studies concentrating, in particular, on the Heisenberg (XXX) chain with random magnetic fields \cite{ZnidaricProsen,OganesyanHuse,PalHuse,BardarsonPollmann,Luitz1,Luitz2,SerbynPapic,VoskHusePRX,AltmanVoskReview,Doggen2018,SierantDelande,NandkishoreHuse,PotterVasseurPRX,HuseNandkishore}. By a Jordan-Wigner transform, this model is equivalent to a fermionic chain with nearest-neighbor hoppings, density-density interactions, and random on-site potentials. Based on an analysis of spectral properties such as the level spacing as well as of dynamical properties such as the growth of the entanglement entropy after a quantum quench from an unentangled initial state, an apparent consensus was reached that this model shows a phase transition in all eigenstates at a finite disorder strength, separating an ergodic from a many-body localized phase. This consensus was first called into question by studies showing that the transition point, defined for a finite system of size $L$, appears to slowly shift to infinite disorder strength for $L\to\infty$ \cite{SuntajsBonca,SuntajsBonca2}. These arguments were supported by studies of the number and Hartley entropies after a quantum quench \cite{KieferUnanyan1,KieferUnanyan2,KieferUnanyan3,KieferUnanyan4,KieferUnanyan5,KieferUnanyan6} and of the fidelity susceptibility \cite{SelsPolkovnikov,SelsPolkovnikov2}.

Here we want to propose and follow a new avenue by studying the properties of operators instead of wave functions \ch{which offer a unified perspective of the non-interacting and interacting cases}. Our motivation is the following observation: In an Anderson localized model, the projectors onto the eigenstates $P_n=|E_n\rangle\langle E_n|$ are conserved {\it quasi-local operators}, i.e., they are centered at a site in the lattice and decay exponentially away from this site. These operators are also often called local integrals of motion (LIOMs). Without loss of generality, we can order the operators such that $P_n$ is an operator localized around lattice site $n$. To be more precise, we can use the fermionic language in which the Anderson Hamiltonian and any unitary transformation of this Hamiltonian are bilinear. We can then write $P_n = \sum_{ij} a^n_{ij} c^\dagger_i c_j $ where $c_i^{(\dagger)}$ is a fermionic annihilation (creation) operator at site $i$ and $|a^n_{ij}|\sim \exp(|i-n|)\exp(|j-n|)$ are amplitudes. In particular, the contribution of bilinear operators to $P_n$ is exponentially suppressed with the distance they cover. 

There is a unitary transformation $U$ which diagonalizes the microscopic, non-interacting Anderson Hamiltonian $H$, bringing it into the form $\tilde H=UHU^\dagger=\sum_n E_n P_n$ where $P_n$ is quasi-local and conserved, $[H,P_n]=0$. Importantly, this unitary transformation itself is quasi-local and will always map local microscopic operators $A_i$ onto quasi-local operators, $\tilde A_i=U A_i U^\dagger$. As we will show, this has important implications for the spreading of a local operator in Euclidean time. We can write
\begin{eqnarray}
    \label{intro2}
    A_i(\tau)&=&\e^{\tau H}A_i\e^{-\tau H}\\
    &=&\sum_{k=0}^\infty \underbrace{[H,[H,[...,[H,A_i]]]]}_{\textrm{k-times}} \frac{\tau^k}{k!} . \nonumber
\end{eqnarray}
I.e., the spreading of the operator is encoded in the $k$-fold commutator $[H,A_i]^{(k)}$ with the Hamiltonian. It is important to note that the Frobenius norm of this commutator is not affected by the unitary transformation  $||[H,A_i]^{(k)}||_2=||[\tilde H,\tilde A_i]^{(k)}||_2$. \ch{We also note that the same $k$-fold commutator needs to be considered when calculating real-time two-point correlation functions as, for example, a current-current correlation function which determines the transport properties of a system.}

In a many-body localized phase, it has been argued that---similar to the Anderson case---a unitary transformation of the microscopic Hamiltonian $H$ to a Hamiltonian $\tilde H$ of conserved quasi-local charges exists. The main difference is assumed to be that these quasi-local charges $\tau^z_n$ are interacting with each other \cite{HuseNandkishore,NandkishoreHuse}
\begin{equation}
    \label{intro3}
    \tilde H = \sum_n E_n \tau^z_n +\sum_{i,j} J_{ij} \tau^z_i\tau^z_j+\dots \, .
\end{equation}
Here, $J_{ij}$ are exponentially decaying coupling constants which describe a slow dephasing between the conserved charges. For the Anderson case, $J_{ij}=0$ and the $\tau^z_n$ can be chosen to be the projection operators $P_n$. We note that the choice of the quasi-local conserved operators $\tau^z_n$ is not unique. \ch{If we define a many-body localized phase by the property that a local operator in the microscopic model remains local---up to exponential tails---under Euclidean time evolution as in the non-interacting case, then this implies, as we will show, that the total operator norm {\it in both cases} can at most grow exponentially with the commutator order $k$. We will also show that if local operators spread over the entire system then this strongly suggests that transport will eventually set in.}

This has to be contrasted with the general, ergodic case. In one dimension, there is no phase transition in a clean system at finite temperatures. Thus, the time evolution is always analytic and Eq.~\eqref{intro2} implies that the norm of the $k$-fold commutator $[H,A_i]^{(k)}$ has to grow slower than factorially. In Ref.~\cite{AvdoshkinDymarsky}, it has been shown that this is indeed the case; however, the derived bound for the total norm grows much faster than exponentially and almost factorially: $||[H,A_i]^{(k)}||_2\sim(k/\ln k)^k$. The {\it operator growth hypothesis} states that any generic quantum chaotic system will asymptotically saturate this bound \cite{AvdoshkinDymarsky,ParkerCao}. Thus, we have a sharp contrast between the growth of the operator norm of the $k$-fold commutator of a local operator with the Hamiltonian between a \ch{strictly exponentially localized}, non-ergodic system, where it can grow at most exponentially, and a non-localized system, where an almost-factorial growth is expected in general. This qualitative difference in the operator norm growth is what we will investigate here. To avoid misunderstandings, we want to stress already here that what we mean by a local operator and what we will investigate in the following is an operator $A_i$ which resides on lattice site $i$ in the microscopic model. In contrast, operators of the form $A=\sum_i A_i$ are also often called local conserved charges if $[H,A]=0$ and are of particular importance in integrable models. Operators of the type $A$, which are sums of local densities but which extend over the entire lattice, are also the type of operators whose growth has been investigated in Refs.~\cite{ParkerCao,HevelingGemmer}. The latter type of operators are not of interest for a study of localization and we will not consider them here. \ch{From hereon we will therefore drop the lattice index with the understanding that $A$ is always a local operator.}

We note that in contrast to real-time evolution, the Euclidean time evolution of operators can be directly connected to questions of quantum chaos and the eigenstate thermalization hypothesis (ETH) \cite{AvdoshkinDymarsky,ParkerCao}. First, the Frobenius norm of a local operator $A$ under Euclidean time evolution in the energy eigenbasis for a finite system of size $L$ is given by 
\begin{equation}
    \label{Frobenius}
    ||A(\tau)||_2^2 =\frac{\tr(A^\dagger(\tau)A(\tau))}{\tr\mathbbm{1}}=\sum_{ij}\e^{\tau(E_i-E_j)}\frac{|A_{ij}|^2}{\tr\mathbbm{1}}.
\end{equation}
For a system satisfying ETH, most matrix elements $A_{ij}$ are expected to be non-zero even for extensive energy differences $|E_i-E_j|$. This implies that $||A(\tau)||_2^2\sim\exp(J\tau L)$ where $J$ is the local energy scale. If a system is non-interacting or has local conservation laws on the other hand, and thus does not satisfy ETH, matrix elements will be zero for extensive energy differences and we thus expect $||A(\tau)||_2^2\sim\exp(J\tau)$. By deriving strict bounds directly for the Frobenius norm $||A(\tau)||_2$, one can use Eq.~\eqref{Frobenius} to make precise statements about the scaling of off-diagonal matrix elements $A_{ij}$ as a function of $|E_i-E_j|$. For the generic, chaotic case this scaling has been considered in Ref.~\cite{AvdoshkinDymarsky} and is consistent with numerical investigations \cite{DalessioKafri,LeBlondMalayya}. 

Another quantity which can be obtained from considering the Euclidean time evolution of local operators are Lieb-Robinson bounds. A clean, one-dimensional system with short-range interactions shows a linear spreading of information under real-time evolution independent of whether the system is chaotic or not \cite{LiebRobinson,BravyiHastings}. In contrast, under Euclidean time evolution information spreads exponentially fast, $\ell\sim\e^\tau$, over a distance $\ell$ in a 1D chaotic system \cite{AvdoshkinDymarsky} while it only spreads ballistically in a non-interacting system or a system with local conservation laws. It has also been speculated that the time for the spreading of information in Euclidean time in a 1D chaotic system, $\tau\sim\ln(\ell)$, is indicative of the Thouless time in a system without conservation laws (for example a Floquet system), i.e., the time after which the system can be described by a random matrix. Furthermore, it is also possible to connect the characteristic time of Euclidean operator growth with the delocalization of this operator in Krylov space \cite{ParkerCao}. Overall, there is strong evidence that in a quantum chaotic system, an operator grows asymptotically at its maximally allowed rate and that a parametrically slower growth is related to ergodicity breaking. 

Here we want to expand the study of operator growth to disordered systems. There is, in particular, a direct conflict between the operator growth hypothesis which suggests that an almost factorial growth is indicative of ergodicity and systems such as the random transverse Ising model, which has been suggested to have a many-body localized phase but always shows almost factorial operator growth \ch{independent of the disorder strength} \cite{Cao}. While we will consider Lieb-Robinson bounds and the decay of off-diagonal matrix elements in the Suppl.~Mat.~\cite{SupplMat_LIOMs}, it is important to note that these quantities are obtained by taking sums over $k$-fold commutators $[H,A]^{(k)}$ and thus some of the information, in particular about the spatial support of such commutators which is relevant for the question of localization, is lost. Here we follow a novel path by directly obtaining analytical results for the asymptotic scaling of $k$-fold commutators resolved by the extent of their spatial support. Our results \ch{are a step towards resolving} the above mentioned conflict and suggest that the operator growth hypothesis is valid and that many-body systems, even for strong disorder, do not localize but rather stay ergodic.

In addition to these analytical results, we will also present exact results from symbolic calculations of commutators and norms. To be able to consider high orders $k$ in the commutator, we have written code in Julia with Nemo~\cite{FiekerEtalNemo} which is able to handle symbolic computations with up to $\sim 200$ million terms, limited only by the available memory. Nemo also provides arbitrary precision complex ball arithmetic which tracks error bounds rigorously. We typically use $200$ or $400$ digit precision for the amplitudes of the operators. While the absolute errors for large commutator orders can become large, we keep the relative errors always below $10^{-50}$ such that error bars are not visible in many of the plots presented in the following. We note that by using symbolic calculations we are able to consider system sizes for nearest-neighbor models which are about twice as large as the ones typically considered in exact diagonalizations.

\color{black}
\section{Outline and main results}
Here we want to give a summary of our main results and provide a guide to allow the reader to find these results and to point out parts of the paper which can be skipped in a first reading.

It has recently come to light that in the Ising model with random transverse and longitudinal fields \cite{ImbrieMBL,Imbrie2016} 
\begin{equation}
\label{ergodic_scaling}
||[H,\sigma^z_0]^{(k)} ||_2\sim \left(\frac{k}{\ln k}\right)^k
\end{equation}
is {\it both} an upper and a lower bound \cite{Cao}. This asymptotically maximal growth suggests, according to the operator growth hypothesis, that this model is ergodic for all disorder strengths. In Sec.~\ref{Sec_XXZ} we show data from symbolic calculations which indicate that the same maximal growth also holds for the Heisenberg model with random fields. This means that either both models do not show an MBL phase or that the operator growth hypothesis has to be modified such that a maximal operator growth is also possible in certain non-ergodic models.

\begin{figure}
    \centering   
    \includegraphics*[width=0.99\columnwidth]{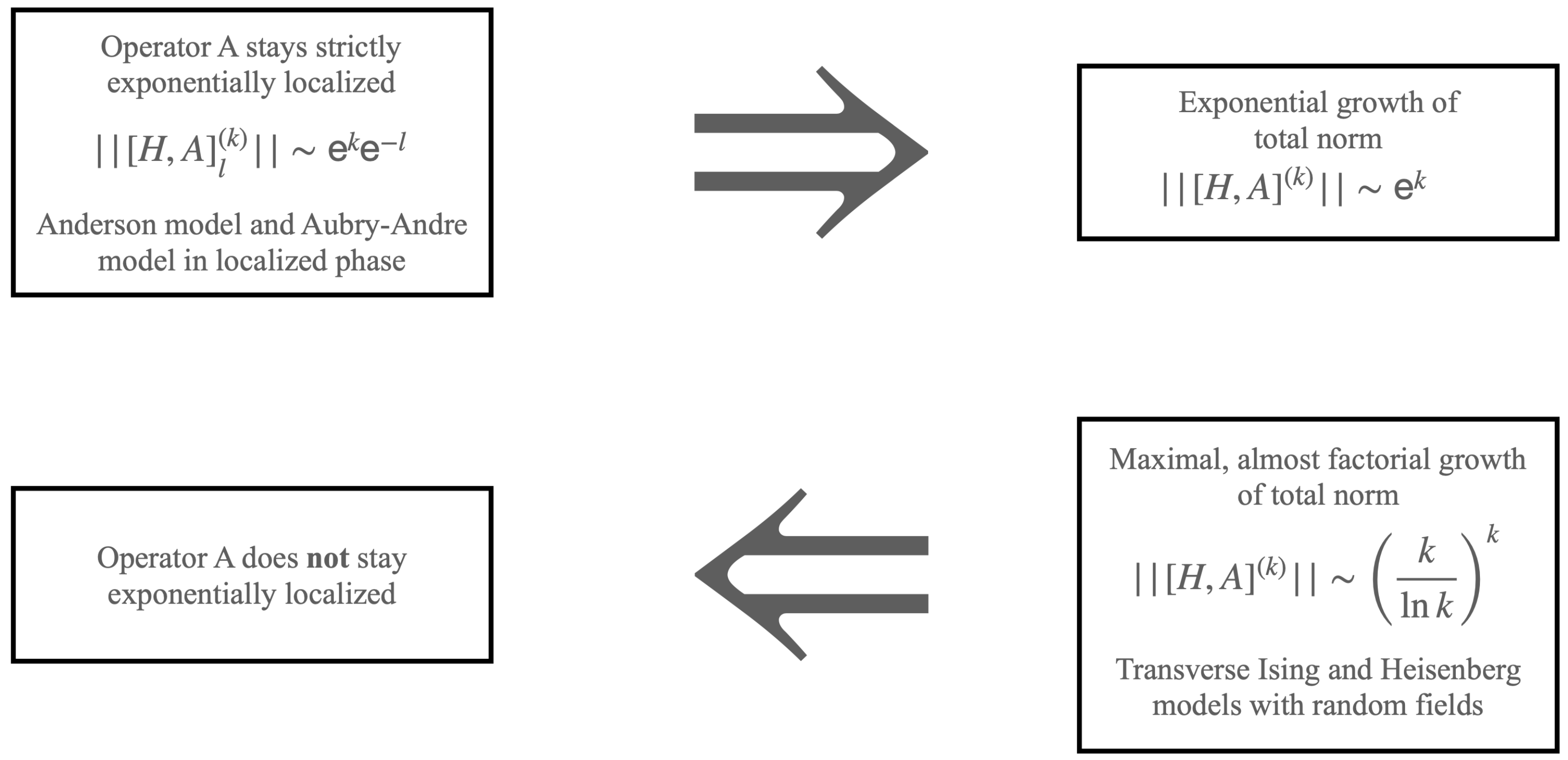}
    \caption{If local operators remain strictly exponentially localized then the total norm cannot grow faster than exponential. Conversely, in many-body systems such as the Ising and Heisenberg models with random fields the almost factorial growth of the total norm implies that local operators will not remain exponentially localized.}
    \label{Fig_Summary1}
\end{figure}

To investigate this issue further, we resolve the commutator by the number of sites $l$ it covers. At commutator order $k$, the commutator $[H,A]^{(k)}$ of a local operator $A$ acting on a single site with a nearest-neighbor Hamiltonian contains operators which span at most $l=k+1$ sites. We show in Sec.~\ref{Op_growth_general} that the contribution of operators with length $l$ to the total norm is always bounded by 
\begin{equation}
    \label{Gen_local_bound}
||[H,A]_l^{(k)} ||_2 \leq (2CJ)^k 2^l S(k,l)   
\end{equation}
where $CJ$ is a constant and $S(k,l)$ the Stirling number of the second kind. This bound, which is expected to be asymptotically tight for an ergodic system, has a maximum at some $l(k)$ which shifts to infinity for $k\to\infty$. I.e., the operator spreads through the entire lattice. Summing over all operator lengths $l$ leads to the almost factorial bound \eqref{ergodic_scaling}. On the other hand, we might expect that in a strictly exponentially localized system, the operator $A$ remains quasi-local, $||[H,A]_l^{(k)} ||_2\sim \e^{-l}$. In this case the total norm growth is dominated by terms with support on a few sites only. This can be understood in more detail by representing the commutator as graphs which we present in Sec.~\ref{Sec_non-int} for the non-interacting case and in Sec.~\ref{Sec_int} for the interacting case. In the strictly exponentially localized case we then find that 
\begin{equation}
    \label{Localized_local_bound}
||[H,A]_l^{(k)} ||_2 \sim \e^k \e^{-l} \,.  
\end{equation}
This implies, in particular, that in the strictly exponentially localized case the total norm cannot grow faster than exponentially with the commutator order $k$. Conversely, a faster, almost factorial growth of the total norm as in the many-body random-field Ising and Heisenberg models is {\it incompatible} with an exponential localization of operators. This important result is summarized in Fig.~\ref{Fig_Summary1}.

In the non-interacting case, we prove that in a localized phase, local operators indeed remain strictly exponentially localized. In Sec.~\ref{Sec_non-int} we demonstrate, in particular, that in an Anderson localized phase Eq.~\eqref{Localized_local_bound} holds while in a non-localized phase $||[H,A]_l^{(k)} ||_2 \sim \e^k$ independent of $l$ for $k\gg l$. I.e., in the latter case the operator spreads through the entire lattice. We show, furthermore, that this difference in scaling allows one to unambiguously identify the phase transition in the Aubry-Andr\'e model. We do find it important to rigorously establish this novel operator perspective on localization by studying the non-interacting case in detail first. However, the reader who does accept the non-interacting results and is mostly interested in what this approach has to say about MBL might skip Sec.~\ref{Sec_non-int} in its entirety. 

If one takes strict exponential localization of local operators under commutation with the Hamiltonian as the very definition of a localized phase, then one has to conclude that the Ising model with random transverse and longitudinal fields and also---based on the analysis presented in this paper---the Heisenberg model with random fields do not possess a localized phase because in these two models the total norm growth is faster than exponential and described by Eq.~\eqref{ergodic_scaling}, independent of the disorder strength. 

On the other hand, if one takes the existence of a unitary transformation $U$ which maps these microscopic models onto an effective model \eqref{intro3} as the definition then there appears to be potentially room for a faster than exponential growth of the total norm in the interacting case as shown by the strict upper bounds derived in Sec.~\ref{localization}. Here we summarize the main results. First, it is important to note that the Frobenius norm is invariant under unitary transformations $||[H,A]^{(k)} ||_2=||[\tilde H,\tilde A]^{(k)} ||_2$ with $\tilde H=UHU^\dagger$ and $\tilde A=UAU^\dagger$. It is easy to show that for a local operator such as $\sigma^x_0$ we have $||[\tilde H,\sigma^x_0]^{(k)} ||_2\sim \e^k\e^{-l}$ with $\tilde H$ as given in Eq.~\eqref{intro3}. I.e., the operator growth hypothesis does hold for $\tilde H$ clearly indicating that the model is not ergodic. However, it is crucial to note that a local operator $A$ in the microscopic model gets mapped onto a {\it quasi-local} operator $\tilde A$ under the unitary transformation $U$. In the non-interacting case, this is still a one-body operator and we can show that this operator remains strictly exponentially localized. I.e., in the non-interacting case, the definitions of a localized phase by requiring that $A$ stays strictly exponentially localized and the alternative definition that a quasi-local unitary $U$ exists which maps $H$ to $\tilde H$ are equivalent.

\begin{figure}
    \centering   
    \includegraphics*[width=0.99\columnwidth]{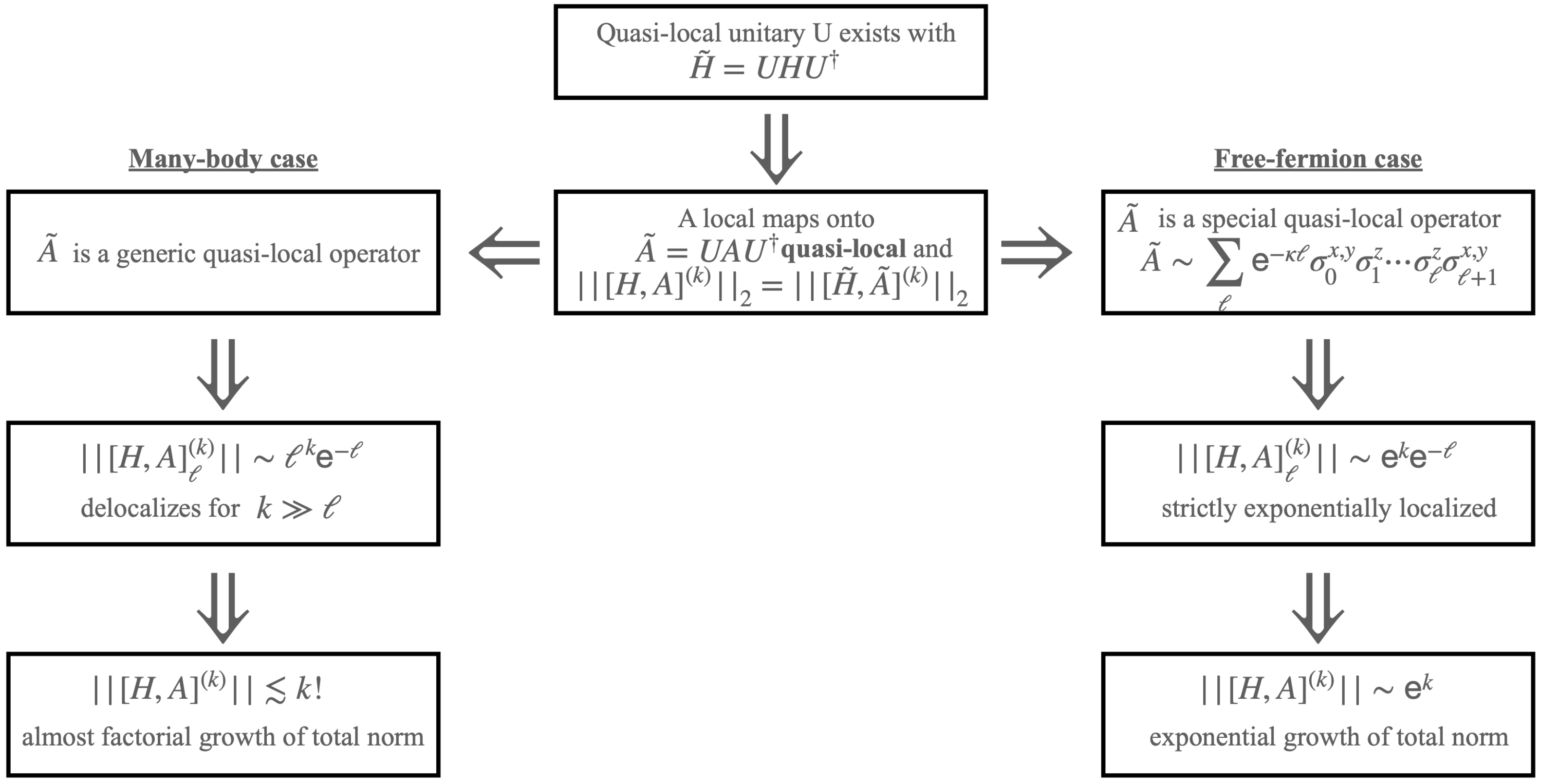}
    \caption{Assuming that a quasi-local unitary $U$ with $\tilde H=UHU^\dagger$ exists, local operators $A$ are mapped onto quasi-local operators $\tilde A=UAU^\dagger$. In the free-fermion case, $\tilde A$ remains a one-body operator, implying that $A$ is strictly exponentially localized and the total norm growth is exponential. In contrast, $\tilde A$ is a generic quasi-local operator in the many-body case in general, implying that $A$ eventually delocalizes and that the total norm growth is almost factorial.} 
    \label{Fig_Summary2}
\end{figure}

In the interacting case, however, $\tilde A$ is in general a generic quasi-local operator and, if no additional microscopic restrictions exist, then $||[\tilde H,\tilde A]^{(k)} ||_2$ will show an almost factorial growth. Furthermore, we will demonstrate in Sec.~\ref{localization} that we can obtain the following upper bound for the spreading of a quasi-local operator
\begin{equation}
    \label{local_bound}
    ||[\tilde H,\tilde A]_l^{(k)} ||_2\lesssim l^k \e^{-\kappa l}
\end{equation}
where $\kappa=\kappa(D)$ increases with disorder strength $D$. We argue in Sec.~\ref{localization} that $||[H,A]_l^{(k)}||_2$ should show the same functional dependence as Eq.~\eqref{local_bound} for strong disorder. We note that this function has a maximum at $l=k/\kappa$, i.e., for large commutator order $k$ the local operator $A$ spreads over the entire lattice. Furthermore, for very large $k$ the bound \eqref{Gen_local_bound} will eventually be tighter, implying that the operator will spread through the lattice in the exact same way as in an ergodic system. This then leads to two possible scenarios: (1) The quasi-local unitary $U$ does exist and the many-body Hamiltonian $H$ has an unusual non-ergodic phase where it is equivalent to $\tilde H$ but shows maximal norm growth and local operators $A$ delocalize. We show in Sec.~\ref{realtime} that this strongly suggests that transport eventually sets in at long times. I.e., even if such a phase exists it should not be called a many-body localized phase. (2) The unitary $U$ does not exist. These results are summarized in Fig.~\ref{Fig_Summary2}.

One way to investigate this question further is by symbolic calculations of $||[H,A]_l^{(k)}||_2$ which, according to Eq.~\ref{local_bound}, should be dominated by an exponential decay for strong disorder and finite commutator order $k$. Our results presented in Sec.~\ref{Sec_int} are not inconsistent with Eq.~\eqref{local_bound}, however, even for disorder strengths much larger than the hopping amplitude, $D\gg J$, the exponential decay is never fully dominant. We therefore also study by symbolic calculations in Sec.~\ref{Sec_SW} whether a mapping $H\to\tilde H$ by successive Schrieffer-Wolff transformations is convergent. This has been suggested to be possible by Imbrie \cite{Imbrie2016} for the Ising model with random fields and is supposed to work similarly to the Anderson case. However, there are important differences: (i) In the Anderson case, one-body operators remain one-body operators under Schrieffer-Wolff transformations. This means that two sites which are a distance $\ell$ apart do only get coupled by $c_j^\dagger c_{j+\ell}+h.c.$ and that this coupling is suppressed by $(J/D)^{2\ell}$ if we ignore resonances. In the interacting case, on the other hand, there is no such restriction and we can place any of the operators $\{\mathbbm{I},\sigma^x,\sigma^y,\sigma^z\}$ on the sites $j+1,\cdots,j+\ell-1$ and a Pauli operator on sites $j,\ell$ leading to $3\cdot4^{\ell-1}$ different operators coupling site $j$ and site $\ell$. Even if each individual term is suppressed as $D^{-2\ell}$ the parameter for the total contribution $\sim 4^\ell D^{-2\ell}$ is no longer always small. This is, of course, not necessarily inconsistent with having a transition at some finite $D$. (ii) The smallest pairwise difference  of random fields on a chain segment of length $\ell$ scales as $\ell^{-2}$. This implies that the effective coupling scales as $J_{j,j+\ell}\sim D^{-2\ell}\ell^2$ in the non-interacting case and resonances therefore do not proliferate. However, in the interacting case we have many-body resonances caused by some linear combination of {\it all} the random fields on the segment. We show in Fig.~\ref{SW_fig} results of symbolic Schrieffer-Wolff transformations which do not show any evidence that the amplitudes of the generated long-range terms do converge. (iii) Finally, we note that for a finite system we want to stop the Schrieffer-Wolff transformations once the norm of the off-diagonal parts of the effective Hamiltonian falls below some threshold $\varepsilon$. For the non-interacting case this leads to the condition that $k\sim\ln(L/\varepsilon)$ steps are required for a chain of length $L$ which is fully consistent with our symbolic calculations. In the interacting case, on the other hand, the number of generated off-diagonal terms scales exponentially with system size. If we assume that each individual contribution is still suppressed as $D^{-2k}$, i.e. resonances do not proliferate, then this leads to the condition $k\sim (L-\ln\varepsilon)/2\ln D$. In particular, we now expect a linear scaling with system size, $k\sim L$, as compared to the logarithmic scaling in the non-interacting case. However, we find symbolically that the Schrieffer-Wolff transformations for typical disorder distributions do not converge in $k\sim L$ steps. Instead, our results appear to be consistent with $k\sim \e^L$, i.e., all the terms generated will need to be eliminated. This suggests that many-body resonances prohibit the perturbative construction of an effective Hamiltonian via Schrieffer-Wolff transformations in the thermodynamic limit.

Our paper is organized as follows: In Sec.~\ref{Sec_general}, we discuss general bounds for the growth of the norm $||[H,A]^{(k)}||$ for a local operator $A$ and a nearest-neighbor Hamiltonian $H$. We review the result in the general interacting case \cite{AvdoshkinDymarsky} before discussing the free-fermion case. In Sec.~\ref{localization} we then derive bounds for systems where local operators remain strictly exponentially localized and for systems which are unitarily equivalent to an effective Hamiltonian $\tilde H$ of local conserved charges. We also discuss the implications of our results for real-time dynamics and transport. In Sec.~\ref{Sec_XXZ}, we then consider a specific model, the $XXZ$ spin chain. We start with the free-fermion (XX) model without disorder before discussing the Anderson case of quenched disorder. We also consider the Aubry-Andr\'e model which has a periodic potential incommensurate with the lattice and shows a phase transition at a finite disorder strength which, as we will demonstrate, is clearly visible in the norm growth. We then turn to the interacting Heisenberg (XXX) model and show that results obtained by symbolic computations are inconsistemt with a strict exponential localization for all studied disorder strengths. In Sec.~\ref{Sec_SW}, we consider Schrieffer-Wolff transformations which can be used to perturbatively transform the Hamiltonian into a basis of quasi-local conserved operators. We show that such a perturbative construction succeeds in the non-interacting case while it appears to fail once interactions are included. The final section contains a summary of our main results and a discussion of the remaining open issues.
\color{black}

\section{General bounds on operator norm growth}
\label{Sec_general}
We want to consider how the norm of the $k$-th order commutator
\begin{equation}
    \label{Commutator}
    [H,A]^{(k)}\equiv \underbrace{[H,[H,[...,[H,A]]]]}_{\textrm{k-times}}
\end{equation}
between a one-dimensional lattice Hamiltonian 
\begin{equation}
    \label{Ham_gen}
    H = \sum_I h_{I,I+1}\equiv \sum_I h_I 
\end{equation}
which acts only locally at nearest-neighbor sites $I$ and $I+1$ and a one-site operator $A$ grows with the order of the commutator $k$. We consider the Hamiltonian as acting on $L$ lattice sites which therefore can be described by an $M\times M$ matrix
\begin{equation}
    \label{Ham2_gen}
    H = \sum_I \mathbbm{1}_1\otimes\cdots\otimes\mathbbm{1}_{I-1}\otimes h_{I,I+1}\otimes\mathbbm{1}_{I+2}\otimes\cdots\otimes\mathbbm{1}_L 
\end{equation}
with $M=n^L$ where $n$ is the number of degrees of freedom per site. We place the operator $A$ in the middle of the chain and choose chain lengths sufficiently large so that the support of $[H,A]^{(k)}$ is always less than $L$.

We consider two entry-wise matrix norms for a matrix $A$ with elements $A_{ij}$:
\begin{eqnarray}
    \label{Norms}
    ||A||_1&=&\sum_{i,j} |A_{ij}|/\sqrt{\tr\mathbbm{1}_{M\times M}} \nonumber \\
    ||A||_2 &=& \sqrt{\frac{\tr(A^\dagger A)}{\tr\mathbbm{1}_{M\times M}}} =\sqrt{\frac{\sum_{i,j} |A_{ij}|^2}{\tr\mathbbm{1}_{M\times M}}} \, .
\end{eqnarray}
We note that the $2$-norm is also called the Frobenius norm. For finite-dimensional spaces it is also equivalent to the Hilbert-Schmidt norm which is consistent with the standard scalar product in quantum mechanics at infinite temperatures. We divide here by the constant factor $\tr\mathbbm{1}_{M\times M}$ such that the identity operator acting on the entire system is normalized to one. \js{We note that the $2$-norm is invariant under unitary transformations. I.e., the operator growth of a local operator $A$ in the microscopic model with Hamiltonian $H$ under Euclidean time evolution is exactly the same as that of the transformed operator $\tilde A=UAU^\dagger$ in the effective model $\tilde H=UHU^\dagger$ if $U$ describes a unitary transformation.} The $1$-norm, on the other hand, is not invariant under basis changes. Let us therefore explain how we define the matrix $A$ corresponding to an operator. {\color{black} If $\{\sigma_i^\alpha\}$ is a $f$-dimensional local operator basis, then we can always write 
\begin{equation}
    \label{Op_basis}
    A = \sum_{\{\alpha\}} a_{\alpha_1,\dots,\alpha_L} \sigma^{\alpha_1}_1\otimes \cdots\otimes\sigma^{\alpha_L}_L \, .
\end{equation}
There are $f^L$ many coefficients $a_{\alpha_1,\dots,\alpha_L}$, and the $2$-norm is given by $||A||_2=\sqrt{\sum_{\{\alpha\}} |a_{\alpha_1,\dots,\alpha_L}|^2}$. We can therefore consider as entries of our matrix $A$ the $f^L$ many coefficients $\sqrt{\tr\mathbbm{1}_{M\times M}}a_{\alpha_1,\dots,\alpha_L}$. For the norms the order of these coefficients does not matter.  The $1$-norm is then given by  $||A||_1=\sum_{\{\alpha\}} |a_{\alpha_1,\dots,\alpha_L}|$. In symbolic calculations, where operators are obtained in the form \eqref{Op_basis}, this means that we simply have to sum up the absolute values, or the absolute values squared, of the operator $A$ written in the local basis $\{\sigma^\alpha_i\}$.

The two matrix norms which we have defined in Eq.~\eqref{Norms} are sub-additive and the $1$-norm is an upper bound for the $2$-norm
\begin{eqnarray}
    \label{Norm_prop}
    ||A+B||_p &\leq& ||A||_p + ||B||_p \nonumber \\
    ||A||_2 &\leq& ||A||_1 \, .
\end{eqnarray}
However, due to the rescaling by the factor $(\tr\mathbbm{1}_{M\times M})^{-1/2}$ the two norms are no longer sub-multiplicative. We note, however, that if  $A = J_A \sigma^{\alpha_1}_1\otimes \cdots\otimes\sigma^{\alpha_L}_L$ and $B = J_B \sigma^{\beta_1}_1\otimes \cdots\otimes\sigma^{\beta_L}_L$, i.e., both operators have only one coefficient in Eq.~\eqref{Op_basis} which is non-zero, then
\begin{equation}
    \label{multiplicativity}
    ||AB||_p = ||A||_p||B||_p = |J_A| |J_B|\, .
\end{equation}
While the $2$-norm (Hilbert-Schmidt norm) is physically more relevant because it is induced by the standard scalar product in quantum mechanics, the $1$-norm is often easier to calculate and provides an upper bound for the $2$-norm.} 

\subsection{General Case}
\label{Op_growth_general}
The case of the Euclidean norm growth for a one-dimensional nearest-neighbor model without any further restrictions on the form of the nearest-neighbor Hamiltonian has been considered in Ref.~\cite{AvdoshkinDymarsky}. Here we briefly recapitulate the derivation of the general bound to contrast it with the tighter bounds which we derive below in the special cases of free-fermion and localized models.

Let us define $H=\sum_I\sum_{a=1}^C h_I^a$ where $h_I^a$ are local Hamiltonian densities of the form $h_I^a=J^a_I \sigma_I^{a_1}\otimes\sigma_{I+1}^{a_2}$. 
We note that the commutator \eqref{Commutator} is zero if the local operator $A$ and the local Hamiltonian densities $h^a_I$ do not form a connected cluster. Let $\{I_1,\cdots,I_k\}$ be a set of points which fulfills the adjacency condition, then
\begin{eqnarray}
    \label{gen1part1}
    ||[H,A]^{(k)}|| &=& ||\!\!\!\sum_{\{I_1,\cdots,I_k\}}\sum_{\{a_I\}} [h^{a_{I_1}}_{I_1},[h^{a_{I_2}}_{I_2},\cdots,[h^{a_{I_k}}_{I_k},A]]]|| \nonumber \\
    &\leq& \!\!\!\sum_{\{I_1,\cdots,I_k\}}\sum_{\{a_I\}} ||[h^{a_{I_1}}_{I_1},[h^{a_{I_2}}_{I_2},\cdots,[h^{a_{I_k}}_{I_k},A]]]|| \nonumber \\
   \end{eqnarray}
where we have used the sub-additivity property \eqref{Norm_prop}. \js{The commutator now is determined by terms which are products of single operators in the $\{\sigma_i^\alpha\}$ basis and we can use Eq.~\eqref{multiplicativity} implying that
\begin{equation}
    \label{mult2}
    || h^{a_{I_1}}_{I_1} h^{a_{I_2}}_{I_2} \cdots h^{a_{I_k}}_{I_k} A|| = || h^{a_{I_1}}_{I_1}||\, ||h^{a_{I_2}}_{I_2}|| \cdots ||h^{a_{I_k}}_{I_k}||\, ||A|| \, .
\end{equation}
There are $2^k$ terms of this kind and we therefore find
\begin{eqnarray}
    \label{gen1}
    ||[H,A]^{(k)}|| &\leq& 2^k \sum_{\{I_1,\cdots,I_k\}}\sum_{\{a_I\}} \underbrace{||h||\cdots ||h||}_{\text{$k$-many}}\, ||A|| \nonumber \\
    &\leq & (2CJ)^k ||A|| \sum_{\{I_1,\cdots,I_k\}}
\end{eqnarray}
where we have used that $||h^a_I||=J^a_I\leq J=\text{max}_{I,a}\{J^a_I\}$ and $\sum_{\{a_I\}}=C^k$.} 

Eq.~\eqref{gen1} is now a purely combinatorial problem. We need to figure out how many different arrangements of the lattice points $\{I_1,\cdots,I_k\}$ exist such that a connected cluster is formed. Without repeated indices, $j$ local adjacent Hamiltonians form a cluster with $j$ bonds. At each step, one can either attach a local Hamiltonian to the right or to the left of the existing cluster. Thus, as for the random walk in one dimension, there are $2^j$ possibilities to form a cluster with $j$ bonds. Lastly, consider the case that we have a cluster of length $j$ and $k\geq j$ local Hamiltonians $h_I$. Then the question is how many ways there are to `distribute' those local Hamiltonians on the $j$ bonds of the cluster. This is the same question as asking how many ways there are to partition $k$ elements into $j$ non-empty sets. The answer is given by the Stirling numbers of the second kind, $S(k,j)$. We therefore obtain
\begin{equation}
    \label{Bell}
   \sum_{\{I_1,\cdots,I_k\}} = \sum_{j=1}^k 2^j S(k,j) =B_k(2) 
\end{equation}
where $B_k(x)$ is the Bell (Touchard) polynomial. The general bound is therefore given by 
\begin{equation}
    \label{generalB}
   s^{(p)}(k)\equiv ||[H,A]^{(k)}||_p\leq ||A||_p (2CJ)^k B_k(2) 
\end{equation}
where $p=1,2$ is the $1$-norm or the $2$-norm defined earlier. We note that this bound is also valid for any sub-multiplicative norm.

We are, in particular, interested in the asymptotic growth of the operator norm for large $k$. In this limit, the Bell polynomial scales as
\begin{equation}
    \label{Bell_asymp}
    B_k(2)\sim\frac{\exp[-k(1+\ln W(k/2)-W^{-1}(k/2))]}{\sqrt{W(k/2)+1}} \e^{-2}k^k
\end{equation}
where $W(x)$ is the Lambert W function \cite{Dominici}. Using the Stirling formula $k^k\sim k! \exp(k)/\sqrt{2\pi k}$ we can rewrite the asymptotics as
\begin{eqnarray}
    \label{Bell2}
    B_k(2)&\sim&\frac{\e^{-2}k!}{\sqrt{2\pi k(W(k/2)+1})}\\
    &\times&\exp\left[k\frac{1-W(k/2)\ln W(k/2)}{W(k/2)}\right]. \nonumber
\end{eqnarray}
This means that the bound \eqref{generalB} grows faster than exponentially and almost factorially. We note that the term in the exponential in the second line of \eqref{Bell2} only becomes negative for $k \gtrsim 20$ thus suppressing the initial almost factorial growth. \js{For large $k$, one can also use that $W(k)\sim\ln k-\ln\ln k$ to see from Eq.~\eqref{Bell_asymp} that $B_k(2)\sim(k/\ln k)^k$.} 

We can also ask the question how many {\it distinct} terms are possible in the commutator $[H,A]^{(k)}$. Note that this question is different from the question of how many connected clusters exist which we addressed above because connected clusters formed in different ways can result in the same operator. Let us assume that our system has a basis of $f-1$ local operators plus the local identity $\mathbb{I}_j$. At order $k$, clusters have at most length $k+1$ and, if the operator $A$ was placed on site $j=0$, extend from sites $-k,-k+1,\dots,0$ to the right. I.e., there are $k+1$ clusters of length $k+1$ with each shifted by one lattice site to the right compared to the previous one. For the cluster starting at $-k$, we can put any of the $f-1$ local operators or the identity on each of the $k+1$ sites leading to $f^{k+1}$ possible operators. For the cluster starting at site $-k+1$, we have to put one of the $f-1$ local operators at the right most site, $j=1$, to obtain operators distinct from the ones already constructed. Thus, this cluster will add $(f-1)f^k$ possible operators. The same construction also works for the remaining $k-1$ clusters. We can therefore bound the total number of possible distinct operators occuring in the commutator $[H,A]^{(k)}$ by
\begin{equation}
    \label{nk_general}
    n(k)\leq f^{k+1}+k(f-1)f^k = (kf-k+f)f^k \, .
\end{equation}

For a generic system, we thus expect that the number of distinct terms $n(k)$ in the commutator grows exponentially. As we will discuss in more detail below, the terms occurring in the commutator can be represented as nodes in a graph and $n(k)$ is directly related to the complexity of this graph. 

To summarize, if a system does not have {\it local} operators which are conserved, then the hypothesis is that the bound on the total norm \eqref{generalB} is asymptotically tight \cite{AvdoshkinDymarsky}. Importantly, a norm growth $s(k)\sim(k/\ln k)^k$ is not possible in regular graphs or Cayley trees where the norm cannot grow faster than exponential. Many-body systems and the question of localization in such systems are therefore qualitatively different from studies of localization on regular graphs \cite{KravtsovKhaymovich,KravtsovAltshuler}.
Next, we consider non-interacting systems which are one example of systems which do correspond to regular graphs and where stricter, exponential bounds on the norm can be derived.

\subsection{Free Fermions}
The main difference between a general, interacting system and a free-fermion system is that in the latter case the commutator of a one-body operator with the Hamiltonian always remains a one-body operator. Therefore many of the connected clusters leading to the the bound \eqref{Bell} do not contribute. Here, a free-fermion model is a system with a bilinear Hamiltonian of the form $H=\sum_{j,k}h_{jk} c^\dagger_j c_k$ where $c_j^{(\dagger)}$ is a fermionic annihilation (creation) operator at site $j$. 

Our starting point to prove a bound in the free-fermion case is the last line of Eq.~\eqref{gen1}. We now want to understand how many distinct connected clusters there are which give a non-zero contribution to the sum. The Hamiltonian in this case is, in general, a sum of $C$ one-body terms, $H=\sum_I\sum_{a=1}^C h_I^a$ such as nearest-neighbor hoppings and local potentials. Now consider a connected cluster $\{I\}=\{I_1,\dots,I_k\}$ which gives a non-zero contribution to the sum in Eq.~\eqref{gen1}. Since the commutator $[H,A]^{(k)}$ is always a one-body operator if $A$ is a one-body operator, there is one fermionic operator on a site $i$ and another fermionic operator on a site $j$ for each such cluster. Then, growing the cluster the only contributions which can potentially be non-zero will come from adding a one-body term $h^a$ on one of the bonds $I_{i-1},I_i,I_j,I_{j+1}$. I.e., there are at most $4$ possibilities to build a new non-zero cluster with $k+1$ elements. While this is, in general, overcounting the number of possibilities we can thus iteratively show that
\begin{equation}
    \label{fF1}
    \sum_{\{I_1,\dots,I_k\}} \leq 4^k 
\end{equation}
resulting in the bound
\begin{equation}
    \label{fFB}
    ||[H,A]^{(k)}||\leq ||A|| (8CJ)^k \, .
\end{equation}
For a free-fermion system, the operator norm growth is thus at most exponential \cite{Sels2022}.

We can also again consider the question how many distinct terms can at most occur in the commutator $[H,A]^{(k)}$. As in the general case, the clusters have length of at most $k+1$ and have $k+1$ different starting points (leftmost lattice site in the cluster). The difference is that we now have to distribute only two operators, a fermionic creation operator $c^\dagger_i$ and a fermionic annihilation operator $c_j$, over the cluster. For the first cluster, we have $(k+1)$ possibilities to place the creation and $(k+1)$ possibilities to place the annihilation operator for a total of $(k+1)^2$ possible distinct terms. For the other clusters, we have to place either $c^\dagger$ or $c$ on the rightmost site to get new distinct terms. The other operator can then be placed on any of the $k+1$ sites for a total of $2(k+1)-1 = 2k+1$ (placing both operators on the rightmost site results in only one new term) distinct new terms for each of the $k$ remaining clusters. A bound for the total number of distinct terms is thus given by
\begin{equation}
    \label{nk_fF}
    n(k)\leq (k+1)^2 +k(2k+1) = 3k(k+1)+1 \, .
\end{equation}
In a free-fermion system, the number of distinct operators in the commutator $[H,A]^{(k)}$ thus grows  quadratically with the order of the commutator $k$. As we will show below when considering the example of the XX chain, the quadratic growth of the number of distinct operators is directly related to a regular structure of the corresponding graph with a constant number of nodes and edges at each level.

\section{Localization}
\label{localization}
\color{black}
Studying the localization of operators when being commuted with the Hamiltonian has the advantage that it is applicable both to the non-interacting and the interacting case. This is in contrast to the localization of wave functions which is only a useful criterion in the non-interacting Anderson case. However, we then have to define precisely what we mean by a local operator staying localized. In the non-interacting Anderson case there is a simple answer: as we will show below, the usual criterion of the single-particle wave function becoming exponentially localized is equivalent to a local operator remaining strictly exponentially localized when commuted with the Hamiltonian,
\begin{equation}
    \label{loc_criterion}
    \frac{s_l(k)}{s_1(k)}\sim \exp(-l) \, ,
\end{equation}
with $s_l(k)=||[H,A]_l^{(k)}||$. As we will also show below, however, strict exponential localization implies that the total operator norm $s(k)\leq\sum_l s_l(k)$ cannot grow faster than exponential. This does not lead to any contradiction in the Anderson case because we have already shown in Eq.~\eqref{fFB} that an exponential norm bound always holds in the free-fermion case. It does, however, cause an issue in the many-body case because it is analytically known that the norm always grows faster than exponential in the Ising model with random longitudinal and transverse fields and we will provide strong evidence from symbolic calculations in Sec.~\ref{Symbolic_localization} that the same is also true in the Heisenberg chain with random fields. I.e., we will demonstrate that a strict exponential localization is not possible in the many-body localized case. If one takes exponential localization of local operators as the defining feature of a localized system, then one has to conclude that many-body localization does not exist.

Another possible definition of many-body localization is, as we have already eluded to earlier, to demand that a unitary transformation exists which maps the microscopic Hamiltonian onto the effective Hamiltonian \eqref{intro3}. If one could show that the existence of such a transformation implies that local operators in the microscopic model remain exponentially localized when commuted with the Hamiltonian, then one would have proven that many-body localization is also excluded if defined in this way. We are, however, not able to prove a strict exponential bound but rather only the weaker bound \eqref{local_bound} which does not exclude an almost factorial growth of the total norm. However, even the bound \eqref{local_bound} predicts an ultimately almost exponential decay of the norm as a function of the length of the operator support if $l\gg k/\kappa$ at fixed order $k$. We will show in Sec.~\ref{Symbolic_localization} by symbolic computations that there is no evidence for such a weaker type of localization even at very strong disorder.

\subsection{Strict exponential localization}
In a localized phase, one might expect that a local operator $A$ remains localized to a finite region of space of length $\xi_{\rm loc}$ (up to exponentially small tails) when commuted $k$-times with the Hamiltonian. We call this strict exponential localization and prove in Sec.~\ref{Sec_Anderson} that it does hold in the Anderson case. 

In any system which is strictly exponentially localized, we expect that connected clusters of length $j\gg\xi_{\rm loc}$ give almost no contribution to the norm. Starting from the general bound \eqref{gen1}, which is valid both for non-interacting and interacting systems,
\color{black}
and neglecting such clusters leads to
\begin{equation}
    \label{loc1}
   \sum_{\{I_1,\cdots,I_k\}}\!\!\!\!\!\!\!{\vphantom{\sum}}' \; = \sum_{j=1}^{\min(k,\xi_{\rm loc})} 2^j S(k,j) \, .
\end{equation}
Asymptotically, for large $k$, we thus obtain
\begin{equation}
    \label{loc2}
    \sum_{j=1}^{\xi_{\rm loc}} 2^j S(k,j)\sim \sum_{j=1}^{\xi_{\rm loc}} \frac{2^j j^k}{k!} < \frac{2^{\xi_{\rm loc}}}{(\lceil\xi_{\rm loc}\rceil-1)!}\xi_{\rm loc}^k 
\end{equation}
where we have used the asymptotic scaling $S(k,j)\sim j^k/j!$ of the Stirling numbers of the second kind for $k\gg j$. The operator growth in a strictly localized phase is therefore at most exponential
\begin{equation}
    \label{loc3}
    ||[H,A]^{(k)}||\leq ||A|| \frac{2^{\xi_{\rm loc}}}{(\lceil\xi_{\rm loc}\rceil-1)!}(2\xi_{\rm loc}CJ)^k  \, .
\end{equation}
In deriving this bound, we have made the simplifying assumption that the contributions of all clusters with lengths $j\gg\xi_{\rm loc}$ can be neglected. Given that with increasing length also more possibilities exist to distribute the local Hamiltonians $h^a_I$, one might worry that this assumption is too naive. We will show in the Suppl. Mat.~\cite{SupplMat_LIOMs} that the argument remains valid if clusters with $j>\xi_{\rm loc}$ are included but are assumed to be exponentially suppressed with their length $l$.

\color{black}
Another way to think about strict exponential localization is to take the norm growth of terms in the commutator of order $k$ which have support on $l$ lattice sites 
\begin{equation}
    \label{loc4}
   s^{(p)}_l(k)\equiv ||[H,A]^{(k)}_l||_p\sim \e^k\e^{-l} 
\end{equation}
directly as the definition of such a phase. Here the exponential decay with $l$ is the strict exponential localization of the operator while the exponential growth with the order $k$ is a consequence of having a finite number of possible clusters with lengths less than $l$ which can be visited before returning to a cluster of length $l$. Along the commutator path the weights are multiplied leading to an $e^k$ growth. This is best understood in a graph theoretical language which is discussed in Sec.~\ref{Sec_XXZ}. From Eq.~\eqref{loc4} it follows immediately that the total norm $s^{(p)}(k)\leq\sum_l s^{(p)}_l(k) \sim \e^k$ because the sum over $l$ converges.

We note, furthermore, that in the free-fermion case both the non-localized and the localized case show an exponential growth of the operator norm. Thus, the total operator norm is insufficient to distinguish between them. What does distinguish them is that the local norm $s^{(p)}_l(k)$ will follow Eq.~\eqref{loc4} in the localized case while in a non-localized phase the initially local operator $A$ will spread over the entire lattice, implying that terms in the commutator with $l\ll k$ will contribute approximately equally to the total norm.

\subsection{Localization defined by local conservation laws}
Another way localization might possibly be defined is to demand that a unitary transformation $U$ does exist which maps the microscopic model onto an effective model $\tilde H$ as given in Eq.~\eqref{intro3} which has an infinite set of conserved charges $\tau^z_i$. These conserved charges are quasi-local when expressed in the basis of the original Pauli operators $\sigma^\alpha_i$ of the microscopic model. The effective model $\tilde H$ is strictly exponentially localized when considering operators which are fully local in the $\tau$-basis. This is a consequence of the exponentially decaying couplings $J_{ij}$ in $\tilde H$ and the fact that one of the operators in the Hamiltonian has to overlap with the local operator. If we consider, for example, the commutator of the nearest-neighbor interaction with the local operator $\tau^x_n$ then we find that $[\sum_iJ_{i,i+1}\tau^z_i\tau^z_{i+1},\tau^x_n]^{(k)}$ only produces terms of the form $\tau^x_n,\tau^y_n\tau^z_{n+1},\tau^z_n\tau^y_{n+1}$, and $\tau^z_{n-1}\tau^x_n\tau^z_{n+1}$. I.e., the length of the operators produced by the commutator cannot grow beyond three sites. Considering also longer-range interactions, terms with additional $\tau^z$'s which are further apart are produced as well but the corresponding couplings $J_{ij}$ are exponentially suppressed.

However, it is crucial to recognize that local operators in the microscopic model are generally mapped by the unitary transformation $U$ onto quasi-local operators. The question then becomes what we can say about the operator norm growth of a quasi-local operator $\tilde A =UAU^\dagger$ when commuted with the effective Hamiltonian $\tilde H$. We can assume w.l.o.g.~that for infinite disorder the conserved charges are given by $\tau^z_i\equiv\sigma^z_i$ and are fully local. For finite disorder the conserved charges then become quasi-local, $\tau^z_i=\sigma^z_i+\textrm{exp. decaying terms}$. For the derivation of a norm bound it suffices to consider just the leading term and to write $\tilde H=\sum_n\varepsilon_n\sigma^z_n$ where we have also dropped the interaction terms between the conserved charges. We consider a local operator $A$ which is situated at site $i=0$. This operator then gets mapped to a quasi-local operator which we can express as
\begin{equation}
    \label{Amapped}
    \tilde A = UAU^\dagger =\sum_{j,\ell}\sum_{\{\alpha\}} \e^{-\kappa_1 |j|} \e^{-\kappa_2\ell} \sigma_j^{\alpha_0}\otimes\cdots\otimes \sigma_{j+\ell}^{\alpha_\ell}
\end{equation}
with $\alpha_0,\alpha_\ell\neq 0$. Here $j$ is the leftmost site of the operator and $\ell$ its length. The factors $\kappa_1,\kappa_2$ encode the strength of the exponential decay of the quasi-local operator. Here we have used the fact that we can expand every operator in the local basis $\sigma^\alpha_i=\{\mathbbm{1}_i,\sigma^x_i,\sigma^y_i,\sigma^z_i \}$. We remind the reader that the Frobenius norm is invariant under unitary transformations, i.e., $||[H,A]^{(k)}||_2=||[\tilde H,\tilde A]^{(k)}||_2$. A bound for the latter thus will also be a bound for the former. Using the properties of the norm we find 
\begin{widetext}
\begin{eqnarray}
    \label{Local1}
    ||[\tilde H,\tilde A]^{(k)}||_2 
    &=& ||\sum_{j,\ell} \sum_{i_1=j}^{j+\ell}\cdots\sum_{i_k=j}^{j+\ell}\sum_{\{\alpha\}} \varepsilon_{i_1}\cdots\varepsilon_{i_k}\e^{-\kappa_1 |j|} \e^{-\kappa_2\ell} [\sigma^z_{i_1},[\sigma^z_{i_2},\cdots,[\sigma^z_{i_k},\sigma_j^{\alpha_0}\otimes\cdots\otimes \sigma_{j+\ell}^{\alpha_\ell}]]]||_2 \nonumber \\
    &\leq & \sum_{j,\ell} \sum_{i_1=j}^{j+\ell}\cdots\sum_{i_k=j}^{j+\ell}\sum_{\{\alpha\}} |\varepsilon_{i_1}|\cdots|\varepsilon_{i_k}|\e^{-\kappa_1 |j|} \e^{-\kappa_2\ell}||[\sigma^z_{i_1},[\sigma^z_{i_2},\cdots,[\sigma^z_{i_k},\sigma_j^{\alpha_0}\otimes\cdots\otimes \sigma_{j+\ell}^{\alpha_\ell}]]]||_2 \nonumber \\
    &\leq & 2^k\sum_{j,\ell} \sum_{i=j}^{j+\ell}\cdots\sum_{i_k=j}^{j+\ell}\sum_{\{\alpha\}} |\varepsilon_{i_1}|\cdots|\varepsilon_{i_k}|\e^{-\kappa_1 |j|} \e^{-\kappa_2\ell} 
    \leq (2J)^k\sum_{j}\e^{-\kappa_1 |j|} \sum_\ell (\ell+1)^k 4^{\ell+1}\e^{-\kappa_2\ell} \nonumber \\
    &=& \frac{4(2J)^k}{1-\e^{-\kappa_1}}\sum_\ell (\ell+1)^k 4^{\ell}\e^{-\kappa_2\ell} = \frac{4(2J)^k}{1-\e^{-\kappa_1}}\sum_\ell (\ell+1)^k \e^{-\tilde\kappa_2\ell} \, .
\end{eqnarray}
\end{widetext}
Here we have used that there are $2^k$-terms in the commutator and we have bounded $|\varepsilon_i|\leq J$. Furthermore, in the chosen normalization of the norm we have $||\sigma^z_{i_1}\sigma^z_{i_2}\cdots\sigma^z_{i_k}\sigma_j^{\alpha_0}\otimes\cdots\otimes \sigma_{j+\ell}^{\alpha_\ell}||_2= ||\sigma^z_{i_1}||_2 \, ||\sigma^z_{i_2}||_2 \cdots ||\sigma^z_{i_k}||_2 \,||\sigma_j^{\alpha_0}\otimes\cdots\otimes \sigma_{j+\ell}^{\alpha_\ell}||_2 =1$. Finally, we have defined $\tilde\kappa_2=\kappa_2-\ln 4$ and we assume that the disorder $D$ is strong enough such that $\kappa_2=\kappa_2(D)>\ln 4$ so that we still have an exponential decay. We note that this is consistent with the fact that in the general, interacting case a finite disorder strength is required for a potentially localized phase while, in contrast, any amount of disorder leads to localization in the non-interacting case.

\begin{figure}
    \centering   
    \includegraphics*[width=0.99\columnwidth]{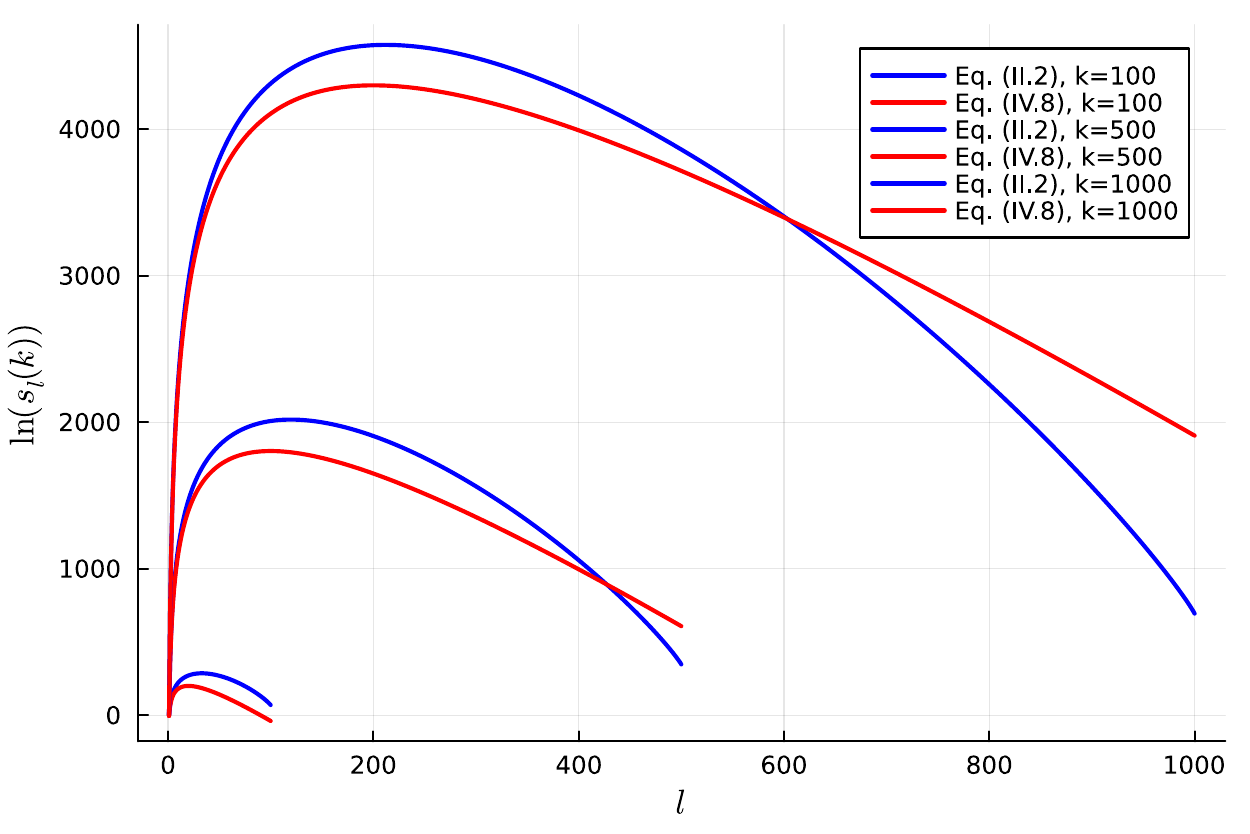}
    \caption{Comparison between the generic local bound for nearest-neighbor models, Eq.~\eqref{Gen_local_bound}, and the bound for quasi-local operators in a model described by $\tilde H$, see Eq.~\eqref{quasi-loc_bound} for $J=1$ and $\tilde\kappa_2=5$. Note that for very large $k$, the bound \eqref{quasi-loc_bound} will ultimately converge to the bound \eqref{Gen_local_bound} for $l\lesssim 
 k/\tilde\kappa_2$ and Eq.~\eqref{Gen_local_bound} will then be a tighter bound for $l\gtrsim 
 k/\tilde\kappa_2$.}
    \label{Fig_local_bounds}
\end{figure}

If this bound is tight, then it implies that quasi-local operators, in contrast to fully local operators, do not remain strictly exponentially localized but rather evolve as
\begin{equation}
    \label{quasi-loc_bound}
 s_l(k)=||[\tilde H,\tilde A]^{(k)}_\ell||_2  \sim  (J\ell)^k \e^{-\tilde\kappa_2\ell} \, .
\end{equation}
This function is first increasing up to a maximum at $\ell=k/\tilde\kappa_2$ before an almost exponential decrease with $\ell$ sets in. I.e., for any finite $k$ and sufficiently strong disorder the maximum will be at some small cluster length $\ell$ and we should observe a decay for larger length which is almost indistinguishable from being exponential. However, no matter how strong the disorder and, consequently, how large $\tilde\kappa_2$ is, {\it a generic quasi-local operator $\tilde A$ in the effective Hamiltonian $\tilde H$ will eventually fully delocalize for large $k$.} We also note that we have made no assumptions about the structure of the initial microscopic Hamiltonian. If this Hamiltonian only has nearest-neighbor terms, as we have assumed in Sec.~\ref{Sec_general}, then the bound \eqref{Gen_local_bound} will ultimately be the tighter bound for very large $k$. This is shown in Fig.~\ref{Fig_local_bounds} where the two bounds are compared. We conclude that generic quasi-local operators in a system described by $\tilde H$ will delocalize for very large $k$ in exactly the same way as they do in an ergodic system.  

Before we go any further, we have to understand if the bound \eqref{quasi-loc_bound} can be tight and how the Anderson case fits in which is unitarily equivalent to $\tilde H$ and where quasi-local operators $\tilde A=UAU^\dagger$ with $A$ fully local are expected to remain localized. First, we note that the factor $(\ell+1)^k$ comes from bounding the commutator in the second line of \eqref{Local1} by assuming the 'worst case scenario' that every $\sigma^{\alpha_n}_{j+n}$ in the expansion of the quasi-local operator in the microscopic Pauli basis is either a $\sigma^x_{j+n}$ or a $\sigma^y_{j+n}$ operator so that the commutator never vanishes no matter where the $\sigma^z_{i_m}$ are placed. Do operators which are local in the microscopic model indeed possibly unitarily map onto such type of quasi-local operators? We can first consider the non-interacting case. W.l.o.g.~we choose $A=\sigma^z=c^\dagger c-1/2$. Then, a unitary $U$ will always map one-body to one-body operators, i.e., the operator $\tilde \sigma^z=U\sigma U^\dagger$ will also only contain one fermionic creation and one annihilation operator which means in the spin language that the quasi-local operator $\tilde\sigma^z$ is a sum of terms of the form $\sigma^{x,y}\sigma^z\cdots\sigma^z\sigma^{x,y}$. I.e., each term has either a $\sigma^x$ or a $\sigma^y$ operators at the end, an arbitrary number of $\sigma^z$ operators in between, and an amplitude which decays exponentially with the length of the string. In this case the commutator in the second line of Eq.~\eqref{Local1} is only non-zero if $i_n=j$ or $i_n=j+\ell$. Therefore, the factor $(\ell+1)^k$ is replaced by $2^k$ and we find
\begin{eqnarray}
    \label{quasi-local_nonint}
&&||[H, A]^{(k)}||_2 = ||[\tilde H,\tilde A]^{(k)}||_2\leq \sum_\ell ||[\tilde H,\tilde A]^{(k)}_\ell||_2  \nonumber  \\
&\leq& \frac{4(4J)^k}{1-\e^{-\kappa_1}}\sum_\ell \e^{-\tilde\kappa_2\ell}=\frac{4(4J)^k}{(1-\e^{-\kappa_1})(1-\e^{-\tilde\kappa_2})}  \, .
\end{eqnarray}
This is consistent with our previous findings \eqref{fFB} that the total norm in the free fermion case can never grow faster than exponential and that local operators remain strictly exponentially localized, see Eq.~\eqref{loc4}.

In the interacting case, however, no such restrictions do exist and one-body operators can get mapped onto many-body operators. Consider, for example, again the case that $A=\sigma^z$ which has the property that $(\sigma^z)^2=\mathbbm{1}$. Because the transformation is unitary, this implies that $(\tilde\sigma^z)^2=U\sigma^zU^\dagger U\sigma^z U^\dagger =\mathbbm{1}$ as well. This poses some restrictions on the quasi-local operator $\tilde\sigma^z$ and there might be other, less trivial, conditions which can be imposed. We note that if such conditions can be found for a microscopic model and if they lead to a tighter bound then there might be a direct contradiction between an almost factorial norm growth and the existence of the unitary $U$. Without such additional constraints, we can explicitly construct an operator which is quasi-local, does fulfill the condition $(\tilde\sigma^z)^2=\mathbbm{1}$, and for which the bound \eqref{Local1} is asymptotically tight. Such an operator is, for example, given by  
\begin{equation}
\label{counterexample}
\tilde \sigma^z = \sqrt{1 - \sum_l e^{-2\kappa } } \sigma_0^z + \sum_l e^{-\kappa l} \sigma_0^y \otimes \cdots \otimes \sigma_{l-1}^y \otimes \sigma_l^x
\end{equation}
where we define to have just a single $\sigma^x$ in the sum for $l=0$. This example thus demonstrates that in the many-body case local operators might get unitarily mapped onto quasi-local operators which do not remain exponentially localized when commuted with the effective Hamiltonian.

Given that we therefore have to take the bound \eqref{Local1} seriously, there are two important remaining questions to address: (1) How does the total norm scale? (2) What does this bound imply for the spreading of the fully local operator $A$ in the microscopic model described by the Hamiltonian $H$? To answer the first question, we can perform the sum over the length $\ell$ of the terms of the quasi-local operator in Eq.~\eqref{Local1} to find
\begin{eqnarray}
    \label{quasi-local_total}
&&||[H, A]^{(k)}||_2 = ||[H_\textrm{eff},\tilde A]^{(k)}||_2 \\
&\leq & \frac{4(2J)^k}{1-\e^{-\kappa_1}}\sum_{\ell=0}^\infty (\ell+1)^k \e^{-\tilde\kappa_2\ell}=\frac{4(2J)^k\e^{\tilde\kappa_2}}{1-\e^{-\kappa_1}}\sum_{\ell=1}^\infty \ell^k \e^{-\tilde\kappa_2\ell}  \nonumber \\
&=& \frac{4(2J)^k\e^{\tilde\kappa_2}}{1-\e^{-\kappa_1}}\textrm{Li}_{-k}(\e^{-\tilde\kappa_2})\stackrel{k\gg 1}{\sim}\frac{4\e^{\tilde\kappa_2}}{\tilde\kappa_2(1-\e^{-\kappa_1)}}\left(\frac{2J}{\tilde\kappa_2}\right)^k k! \nonumber
\end{eqnarray}
where in the last step we have used the asymptotics of the polylogarithm $\textrm{Li}_{-k}(\e^{-\tilde\kappa_2})$ for large commutator order $k$. For a quasi-local operator $\tilde A$ in Eq.~\eqref{Amapped} where the contribution of terms decays quickly with their length $\tilde\kappa_2=\kappa_2-\ln 4\gg 1$, i.e. in the limit of strong disorder, we thus find again a bound which grows slightly slower than factorial, similar to the bound which we derived in Sec.~\ref{Op_growth_general}, see Eq.~\eqref{generalB}. Note that in deriving \eqref{quasi-local_total} we have not made any assumptions about the structure of the Hamiltonian $H$ and thus this bound is not quite as tight as \eqref{generalB} which was derived for a Hamiltonian with only nearest-neighbor terms. To answer the second question, we can use the fact that a unitary transformation leaves the Frobenius norm invariant and that the inverse transformation makes operators more local which leads to
\begin{eqnarray}
    \label{local_growth}
    &&||[H,A]^{(k)}_m|| = ||(U^\dagger [\tilde H,\tilde A]^{(k)}U)_m|| \\
    &\leq & \sum_{\ell=m}^{m+\xi} || \tilde H,\tilde A]^{(k)}_\ell || \leq \frac{4(2J)^k}{1-\e^{-\kappa_1}}\sum_{\ell=m}^{m+\xi}(\ell+1)^k\e^{-\tilde\kappa_2\ell}\nonumber \\
    &=&\frac{4(2J)^k}{1-\e^{-\kappa_1}}(m+1)^k\e^{-\tilde\kappa_2m}\left[1+\left(\frac{m+2}{m+1}\right)^k\e^{-\tilde\kappa_2}+\cdots\right] \nonumber \\
    &\approx & \frac{1+\e^{-\tilde\kappa_2}}{1-\e^{-\kappa_1}}4(2J)^k(m+1)^k\e^{-\tilde\kappa_2m} \, . \nonumber
\end{eqnarray}
I.e., mostly terms in the commutator $[\tilde H,\tilde A]^{(k)}$ with length $\ell\in[m,m+\xi]$ will unitarily map to terms in $[H,A]^{(k)}$ with length $m$. We, furthermore, assume that we are in the strong disorder regime where $\tilde\kappa_2\gg 1$ so that the function in the square brackets in the third line only has a weak $m,k$ dependence and can be approximated by $1+\e^{-\tilde\kappa_2}$. In other words, the norm of terms with length $m$ in the commutator $[H,A]^{(k)}$ is given by a 'running average' of the norms of terms with lengths $\ell\in[m,m+\xi]$ in the unitarily transformed commutator $[\tilde H,\tilde A]^{(k)}$. If the quasi-local operator $\tilde A$ decays fast, as is expected for strong disorder, then this will only renormalize the prefactor but will not change the functional form of the norm of such terms. We thus conclude that in a model where the unitary transformation $U$ maps the local operator $A$ onto a generic quasi-local operator $\tilde A$, the local operator will {\it not remain exponentially localized} but rather spread through the entire lattice, $||[H,A]^{(k)}_l||\sim(2Jl)^k\e^{-\tilde\kappa_2 l}$ which, for large commutator order $k$, means that the operator spreads exactly as in an ergodic system.

\subsection{Consequences for transport}
\label{realtime}
One might then ask what the consequences of the spreading of local operators for the transport properties of a system are. Let us consider a globally conserved charge $Q=\sum_i q_i$ where the local densities are, for example, $q_i=\sigma^z_i$ for a spin model or $q_i=c_i^\dagger c_i$ for a fermionic chain. Let us denote the corresponding current by $\mathcal{J}=\sum_i j_i$ where $j_i$ is the current density. We are interested in the transport properties in the infinite temperature limit. Assuming that the model has no Drude weight in this limit---which will always be the case once a possible integrability is broken by disorder---we can obtain the diffusion constant $\mathcal{D}$ by
\begin{equation}
    \label{diffusion}
    T\mathcal{D} = \lim_{\omega\to 0}\lim_{L\to\infty}\frac{1}{2\chi L}\int_{-\infty}^\infty \!\!\!\! dt\, \e^{i\omega t}\langle \mathcal{J}(t)\mathcal{J}(0)\rangle 
\end{equation}
where $\chi$ is the corresponding susceptibility \cite{SirkerLectureNotes}. We note that in the high-temperature limit, $T\mathcal{D}$ is the proper non-trivial quantity to consider. We can write the correlator as
\begin{eqnarray}
    \label{diff2}
 \langle \mathcal{J}(t)\mathcal{J}(0)\rangle &=& \sum_{n,m} \langle j_n(t) j_m(0)\rangle  \\
 &=& \sum_{n,m,k}\langle [H,j_n]^{(k)} j_m\rangle \frac{(it)^k}{k!} \, . \nonumber
\end{eqnarray}
I.e., the behavior of the correlator is determined by the same $k$-fold correlator we have been considering earlier.

For a translationally invariant system, we can write $\sum_{n,m} \langle j_n(t) j_m(0)\rangle =L\sum_m \langle j_0(t) j_m(0)\rangle\equiv L C(t)$ implying that $T\mathcal{D}=\frac{1}{2\chi}C(\omega\to 0)$ and is therefore in general non-zero in the thermodynamic limit. For an exponentially localized system, on the other hand, the amplitudes of terms in $[H,j_n]^{(k)}$ will decay exponentially with the length of their support on some length scale $\xi$, implying that $\sum_{n,m} \langle j_n(t) j_m(0)\rangle \approx \sum_n\sum_{n=m-\xi}^{n=m+\xi} \langle j_n(t) j_m(0)\rangle \equiv \sum_n C_n(t)$. For $L\gg\xi$ and rounding $\xi$ to the nearest integer we then obtain
\begin{eqnarray}
    \label{diff3}
\frac{1}{L}\sum_n C_n(t) &=& \frac{1}{L}\left[\left(C_0(t)+C_\xi(t)+\cdots\right) \right. \\
&+& \left.\left(C_1(t)+C_{1+\xi}(t)+\cdots\right)+\cdots\right] \nonumber \\
&\stackrel{L\to\infty}{\to}& 0 \nonumber
\end{eqnarray}
because correlators which are further apart than $\xi$ are essentially uncorrelated so that for each time $t$, the terms $\frac{1}{L}\left(C_i(t)+C_{i+\xi}(t)+\cdots\right)$ are averages over random numbers and thus tend to zero. We thus find that, as expected, the diffusion constant $T\mathcal{D}$ vanishes in a phase where local operators remain exponentially localized.

For the case where the current operator at large commutator orders eventually spreads through the entire lattice, $[H,j_n]^{(k)}_l\sim l^k \e^{-l}$, no similar argument for the vanshing of the diffusion constant can be made because all local current densities will become correlated with each other. We have also seen that for very large $k$ the asymptotics of $[H,j_n]^{(k)}$ will be exactly the same as for an ergodic system. In this case one would thus generically expect that $\frac{1}{L}\sum_n C_n(t)$ is non-zero in the thermodynamic limit indicating transport of the globally conserved charge. This finding is consistent with the results in Refs.~\cite{KieferUnanyan2,KieferUnanyan3,KieferUnanyan4} where the transport of charges between subsystems of the disordered Heisenberg chain was studied numerically. Of course, in contrast to Euclidean time evolution, the real-time evolution \eqref{diff2} cannot be related to a norm because the $i^k$-factor allows for interference effects. While one can imagine some fine-tuning where such interference effects lead to a vanishing of the diffusion constant, it is difficult to see how this can be a generic scenario which is stable against small changes in the microscopic parameters. 

We also want to point out that it is a common misconception that the operator growth considered here is related to short-time dynamics. As one can immediately see from Eq.~\eqref{diff2}, the short-time dynamics is determined by $[H,j_n]^{(2)}$ only while we have been focusing on the asymptotic behavior of the $k$-fold commutator for large $k$. It is correct though that the asymptotic behavior of the commutator is insufficient to fully determine the long-time behavior of real-time correlation functions. To do so, we would need in addition information about the behavior at small $k$ and possibly also information about sub-leading terms to the large $k$ asymptotics. This point is discussed in more detail in Ref.~\cite{ParkerCao}. For the discussion here this means that while the spreading of local current densities over the entire lattice implies transport of the globally conserved charge and diffusion seems to be the most natural scenario, other more exotic scenarios of anomalous diffusion are not ruled out.

\color{black}

\section{The XXZ model with (quasi-)random fields}
\label{Sec_XXZ}
As a specific example, we want to consider the XXZ chain with random or quasi-random magnetic fields
\begin{equation}
    \label{Ham}
H = \sum_j \left(\sigma^x_j\sigma^x_{j+1} + \sigma^y_j\sigma^y_{j+1} + \Delta \sigma^z_j\sigma^z_{j+1} + 2h_j \sigma^z_j\right) \,.
\end{equation}
Here, $\sigma^{x,y,z}$ are Pauli spin-1/2 matrices, $\Delta$ characterizes the spin exchange anisotropy, and $h_j$ are magnetic fields. As the operator which we will commute with the Hamiltonian $H$ we consider a single local $\sigma^z_j$. The model \eqref{Ham} is equivalent to a spinless fermion model using the Jordan-Wigner transformation
\begin{eqnarray}
    \label{JW}
    \sigma^+_j &=& S_j c_j^\dagger, \quad \sigma^-_j = S^\dagger_j c_j,\quad \sigma^z_j = 2c^\dagger_j c_j -1, \nonumber \\
    S_j &=& \exp\left[-i\pi\sum_{k=1}^{j-1}c_k^\dagger c_k\right] =\prod_{k=1}^{j-1}\left[1-2c_k^\dagger c_k\right] 
\end{eqnarray}
with $\sigma_j^\pm=(\sigma_j^x\pm i\sigma_j^y)/2$ leading to 
\begin{eqnarray}
\label{Ham2}
 \frac{H}{4} &=& \frac{1}{2}\sum_j \left\{c^\dagger_j c_{j+1} + c_{j+1}^\dagger c_j \right\} \\
 &+& \sum_j \left\{ \Delta(n_j-1/2)(n_{j+1}-1/2) + h_j (n_j-1/2)\right\}. \nonumber 
\end{eqnarray}
Here $c_j^{(\dagger)}$ are fermionic annihilation (creation) operators and $n_j=c_j^\dagger c_j$. In the following, we will investigate the non-interacting case, $\Delta=0$, and the interacting isotropic Heisenberg case, $\Delta=1$.

\subsection{Non-interacting case}
\label{Sec_non-int}
We will first concentrate on the non-interacting free-fermion case, $\Delta=0$, where localization is well understood. Here we want to show that the general bounds for the total norm of the $k$-th order commutator derived earlier do apply and that we can distinguish localized from non-localized phases by resolving the contributions to the norm by the length of the spatial support of each term in the commutator.

\subsubsection{No disorder}
The first specific case we will consider is the free-fermion case, $\Delta=0$, without any magnetic fields, $h_j=0$. In this case, the entire structure of the commutator $[H,\sigma^z_0]^{(k)}$ can be worked out analytically. We find these results instructive and will discuss them in detail here to motivate a description of the commutator by graphs where the vertices are the terms generated and the edges denote their contribution to the 1-norm. 

\begin{figure}
    \includegraphics*[width=0.99\columnwidth]{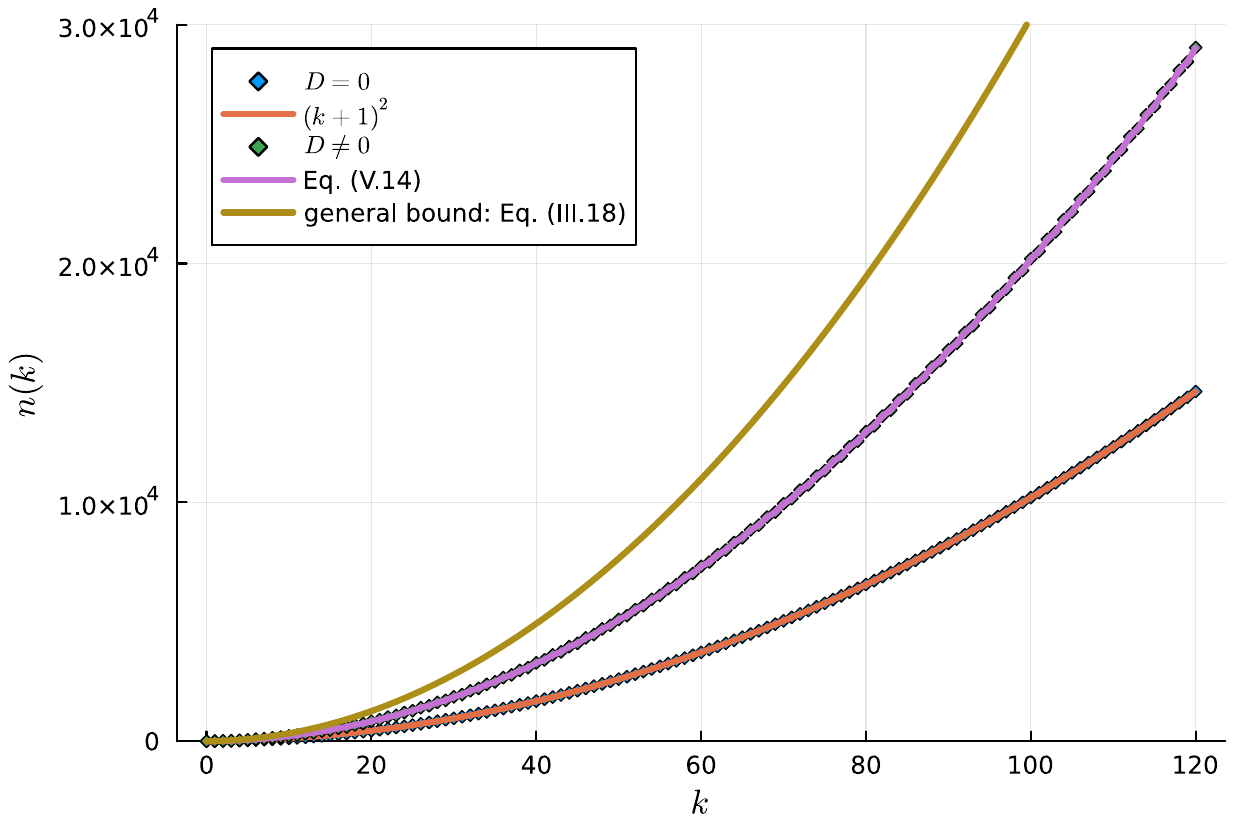}
\caption{XX model: Without disorder, the number of distinct terms $n(k)$ in $[H,\sigma^z_0]^{(k)}$ grows as $(k+1)^2$ while it grows according to Eq.~\eqref{Anderson_num_terms} with disorder. The general bound \eqref{nk_fF}, valid for any free-fermion model, is also shown.}
\label{XX_num_terms_2}
\end{figure}

The full analytical solution is best understood in the fermionic language: the commutator $[H,c_0^\dagger c_0]^{(k)}$ will only contain terms which are one-particle operators, i.e., terms which contain exactly one annihilation and one creation operator. More specifically, we find for the case that the order of the commutator $k$ is odd that
\begin{eqnarray}
\label{FF1}
[H, \sigma_0^z]^{(k)} &=& \sum_{l = 0}^{\frac{k - 1}{2}} \sum_{s = -\frac{k+1}{2} - l}^{\frac{k-1}{2} - l} a_{kls}\left\{ 
 \sigma_s^x \bigotimes_{j = s + 1}^{s + 2l} \sigma_j^z \, \sigma_{s + 2l + 1}^y\right. \nonumber \\
&-&\left. \sigma_s^y \bigotimes_{j = s + 1}^{s + 2l} \sigma_j^z \,  \sigma_{s + 2l + 1}^x
\right\} \\
&=& -\frac{2}{i}\sum_{l = 0}^{\frac{k - 1}{2}} \sum_{s = -\frac{k+1}{2} - l}^{\frac{k-1}{2} - l} a_{kls}\left\{ c_s^\dagger c_{s+2l+1} - h.c. \right\} \nonumber 
\end{eqnarray}
with coefficients $a_{kls}$ which are discussed below. For $k$ even, we find instead
\begin{eqnarray}
\label{FF2}
 [H, \sigma_0^z]^{(k)} &=& \sum_{s=-\frac{k}{2}}^{\frac{k}{2}} a_{k0s}\, \sigma_s^z \nonumber \\
&+& \sum_{l = 1}^{\frac{k}{2}} \sum_{s = -\frac{k}{2} - l}^{\frac{k}{2} - l} a_{kls}\left\{ \sigma_s^x \bigotimes_{j = s + 1}^{s + 2l - 1} \sigma_j^z \, \sigma_{s + 2l}^x\right. \nonumber \\
&+& \left. \sigma_s^y \bigotimes_{j = s + 1}^{s + 2l - 1} \sigma_j^z  \, \sigma_{s + 2l }^y
\right\} \\
&=& 2\sum_{s=-\frac{k}{2}}^{\frac{k}{2}} a_{k0s}\, (n_s-1/2) \nonumber \\
&-&2\sum_{l = 1}^{\frac{k}{2}} \sum_{s = -\frac{k}{2} - l}^{\frac{k}{2} - l} a_{kls} \left\{c_s^\dagger c_{s+2l} + h.c.  \right\} .\nonumber
\end{eqnarray}
Note that the commutator has the form of a current density for $k$ odd and that of a charge density for $k$ even. From the formulas \eqref{FF1} and \eqref{FF2} we can immediately read off that there are $n(k)=\frac{k+1}{2}(k+1)\cdot 2=k+1+\frac{k}{2}(k+1)\cdot 2=(k+1)^2$ different terms generated at order $k$, both in the odd and in the even case. We confirm this result by calculating the commutator using symbolic manipulations up to high order, see Fig.~\ref{XX_num_terms_2}. 
\begin{figure}
    \includegraphics*[width=0.99\columnwidth]{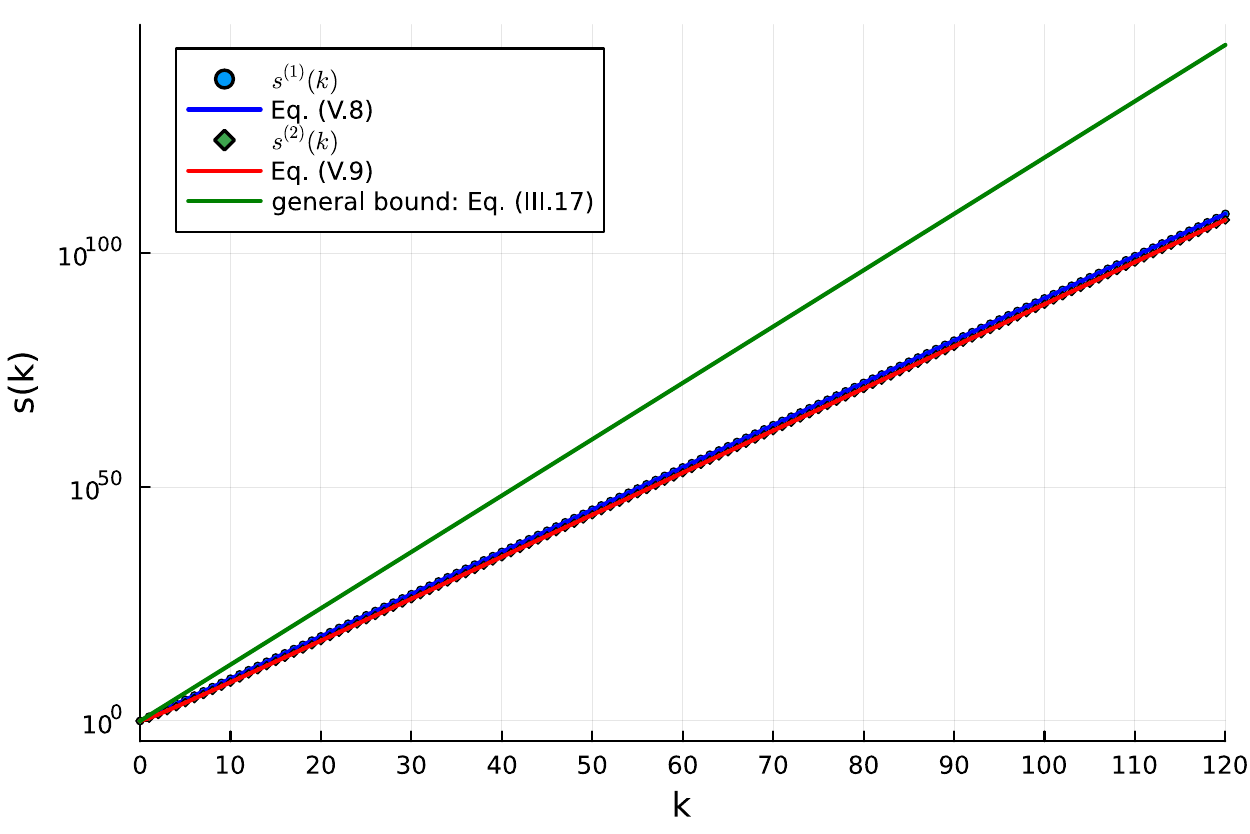}
\caption{In the XX model without disorder the norms grow as $s^{(1)}(k) = 8^k$ and $s^{(2)}(k)=2^{k}\binom{2k}{k}$. Note that on this scale both results are very close to each other. The $2$-norm is only slightly smaller than the $1$-norm. Also shown is the general bound, Eq.~\eqref{fFB}, with $C=2$.}
\label{Fig2}
\end{figure}
We want to stress that the quadratic growth of the number of terms is a special property of a free fermionic model where a one-body operator, when commuted with the Hamiltonian, always remains a one-body operator. What changes is only the separation between the creation and annihilation operator (support of the one-body operator) and the number of different sites where the first of the fermionic operators is located. Both the allowed separations as well as the possible sites for the first fermionic operator grow linearly in $k$, leading to the overall quadratic growth in the number of distinct terms in the commutator. We already want to note here that this behavior is expected to change dramatically in the interacting case.

In the simple, non-disordered XX case considered here, we can even take it one step further and calculate the coefficients $a_{kls}$ exactly. These coefficients are completely determined by the coefficient matrices $C_k$ which are defined by 
\begin{equation}
\label{Coeff_Matrix}
(C_k)_{ij} = 2^k {{k}\choose{i-1}} {{k}\choose{j-1}}
\end{equation}
with $1 \leq i,j \leq k+1$. In the following, we concentrate on the case of $k$ odd which has a slightly simpler structure, see Eq.~\eqref{FF1}. In this case we find, in particular, that 
\begin{equation}
\label{akls}
a_{kls} = \text{i}(-1)^s (C_k)_{s + \frac{k+3}{2} + l, \frac{k + 1}{2} - l}. 
\end{equation}

From the result \eqref{akls} we can now determine the total norms of the commutator and, furthermore, we can even determine which operators contribute how much to the total norm. From Eq.~\eqref{FF1} and Eq.~\eqref{akls} we find that the set 
 of the coefficients $\{|a_{kls}|\}$ is the same as $\{(C_k)_{ij}\}^2$, i.e., each entry in $C_k$ appears two times. We therefore find
 \begin{equation}
 \label{1norm_nondisordered}
s^{(1)}(k)\equiv || [H, \sigma^z_0]^{(k)} ||_1 = 2 \sum_{i = 1}^{k+1} \sum_{j = 1}^{\frac{k + 1}{2}} (C_k)_{ij} = 8^k 
\end{equation}
where we have used the explicit form of the coefficient matrix \eqref{Coeff_Matrix}. Similarly, we find
\begin{equation}
(s^{(2)}(k))^2\equiv || [H, \sigma^z_0]^{(k)} ||^2_2 = 2 \sum_{i = 1}^{k + 1} \sum_{j = 1}^{\frac{k + 1}{2}} (C_k)_{ij}^2 = 2^{2k} {{2k}\choose{k}}^2.
\end{equation}
Asymptotically, the $2$-norm therefore grows as
\begin{equation}
 s^{(2)}(k) =2^{k} {{2k}\choose{k}} \approx \frac{8^k}{\sqrt{\pi k}}
\end{equation}
where we have used Stirling's formula. This shows that the $2$-norm grows also exponentially in $k$ but with a correction which makes it always smaller than the $1$-norm as expected on general grounds, see Eq.~\eqref{Norm_prop}. We can again check these results for finite $k$ by symbolic manipulations, see Fig.~\ref{Fig2}.

\begin{figure}
    \includegraphics*[width=0.99\columnwidth]{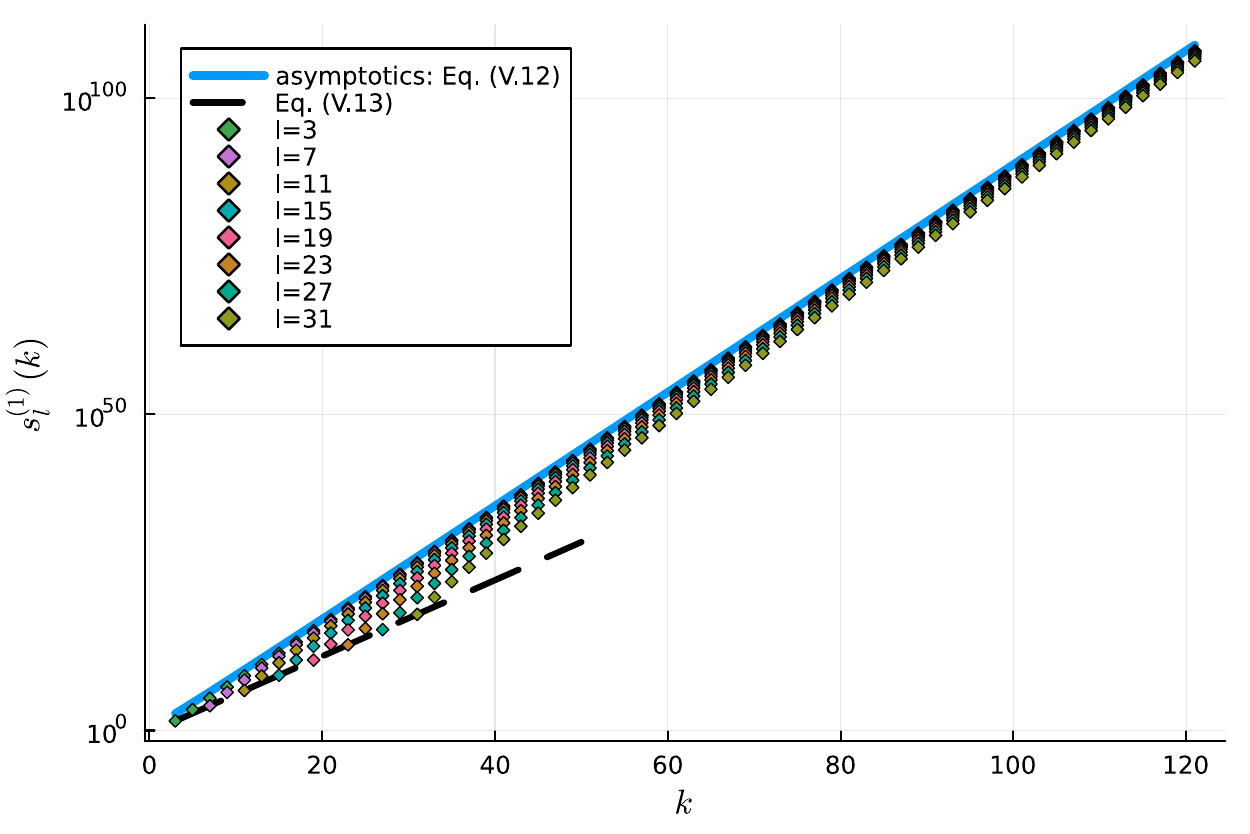}
\caption{For $k\gg l$ the norm of operators with support on $l$ sites is asymptotically described by Eq.~\eqref{spat_structure} and is independent of $l$. The exact value \eqref{largest_support} when $s_l^{(1)}(k)$ first becomes non-zero is also shown.}
\label{Fig3}
\end{figure}

Finally, we can study how much terms which extend over a distance $l$ contribute to the total norm. We consider the case $k$ odd where the amplitude $a_{kls}$ for terms which have support on $2l+2$ sites is given by Eq.~\eqref{akls}. We find again that the same entries appear multiple times and that all prefactors for a given $l$ are represented by the column $j=\frac{k+1}{2}-l$ of the coefficient matrix $(C_k)_{ij}$. Summing over all initial sites we thus obtain 
\begin{eqnarray}
\label{1norm_support_nondisordered}
s^{(1)}_{2l+2}(k)&=&||[H,\sigma^z_0]^{(k)}_{2l+2}||_1=\sum_i |a_{kli}|\\ 
&=& 2\sum_{i=1}^{k+1} (C_k)_{i, \frac{k + 1}{2} - l} 
= 2^{2k+1} {{k}\choose{\frac{k - 1}{2} - l}} \nonumber \\
&\rightarrow& s^{(1)}_{l}(k)=2^{2k+1} {{k}\choose{\frac{k+1-l}{2}}} \nonumber
\end{eqnarray}
where in the last line we transformed the equation to show the support on $l$ sites rather than on $2l+2$ sites. In the limit $k \gg l$ we can apply Stirling's formula and find
\begin{equation}
\label{spat_structure}
s^{(1)}_{l}(k) \approx \sqrt{\frac{8}{k\pi}}8^k \, .
\end{equation}
This means that all the terms {\it contribute equally independent of the length of their support $l$} which clearly indicates delocalization. We also note that the largest possible support at commutator order $k-1$ are $k$ sites with
\begin{equation}
    \label{largest_support}
s_k^{(1)}(k-1)=2^{2k-1} \, .    
\end{equation}
These analytical results can be checked by symbolic manipulations, see Fig.~\ref{Fig3}.

While we have been able to exactly calculate the $k$-th order commutator in the non-interacting, non-disordered case, see Eqs.~(\ref{FF1}, \ref{FF2}, \ref{akls}), this is clearly no longer possible, even in principle, if random fields are included. On a positive note, this is also more than what we need. To distinguish between free-fermionic localized and non-localized phases and interacting localized and non-localized phases it is sufficient to consider the total commutator norm and the norm of the terms in the commutator which are supported on $l$ sites. This can be achieved more efficiently by developing graphs where the vertices represent the possible terms at order $k$ and the edges represent the contribution to the $1$-norm. Fig.~\ref{Fig_nonint_nodes} shows such a graph for the non-interacting model. The case without random magnetic fields discussed so far corresponds to the case $D=0$ where only the terms in the left column are present. 
\begin{figure}
\centering
\begin{tikzpicture}
\begin{scope}[every node/.style={rectangle,draw}]
    \node (A) at (0,0) {$\sigma^z$};
    \node (B) at (0,-1.5) {$\sigma^x \sigma^y$};
    \node (C) at (0,-3.25) {$\sigma^x \sigma^z \sigma^x$};
    \node (D) at (0,-5) {$\sigma^x \sigma^z \sigma^z \sigma^y$};
    \node (E) at (0,-6.75) {$\sigma^x \sigma^z \sigma^z \sigma^z \sigma^x$};
    \node (F) at (4,-1.5) {$\sigma^x \sigma^x$};
    \node (G) at (4,-3.25) {$\sigma^x \sigma^z \sigma^y$};
    \node (H) at (4,-5) {$\sigma^x \sigma^z \sigma^z \sigma^x$};
    \node (I) at (4,-6.75) {$\sigma^x \sigma^z \sigma^z \sigma^z \sigma^y$};
\end{scope}

\begin{scope}[>={Stealth[black]},
              every node/.style={fill=white,circle},
              every edge/.style={draw=black,very thick}]
    \path [->] (A) edge node {$8$} (B);
    \path [<->] (B) edge node {$4$} (C);
    \path [<->] (C) edge node {$4$} (D);
    \path [<->] (D) edge node {$4$} (E);
    \path [->] (B) edge[bend right=60] node {$4$} (A); 
    \path [<->] (F) edge node {$4$} (G);
    \path [<->] (G) edge node {$4$} (H);
    \path [<->] (H) edge node {$4$} (I);
    \path [<->] (B) edge node {$8D$} (F);
    \path [<->] (C) edge node {$8D$} (G);
    \path [<->] (D) edge node {$8D$} (H);
    \path [<->] (E) edge node {$8D$} (I);
\end{scope}
\end{tikzpicture}
\newline
\hspace*{-3em} \large{\textbf{\vdots} \hspace{9em}\textbf{\vdots}}
\caption{Graph for the commutator $[H,\sigma^z_0]^{(k)}$ in the non-interacting case, $\Delta=0$. The vertices represent the possible terms with support on $l$ sites ($l=1,2,\dots$ from top to bottom) while the edges represent the contribution to the $1$-norm. The number of traversed edges equals the order of the commutator $k$. Terms obtained by $\sigma^x\leftrightarrow\sigma^y$ are grouped together into a single vertex. Without disorder, $D=0$, only the terms in the left column will be present.}
\label{Fig_nonint_nodes}
\end{figure}
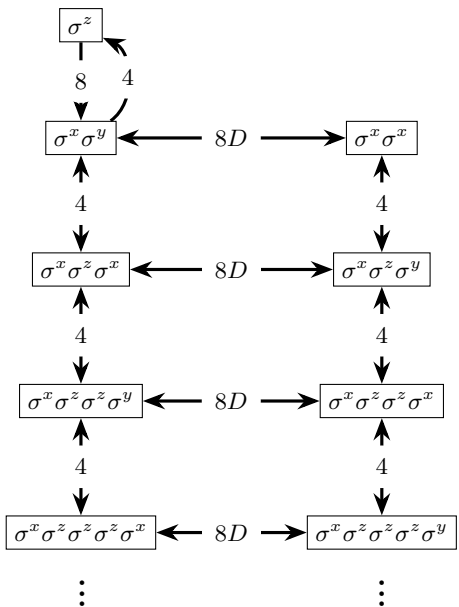
\begin{figure}
    \includegraphics*[width=0.99\columnwidth]{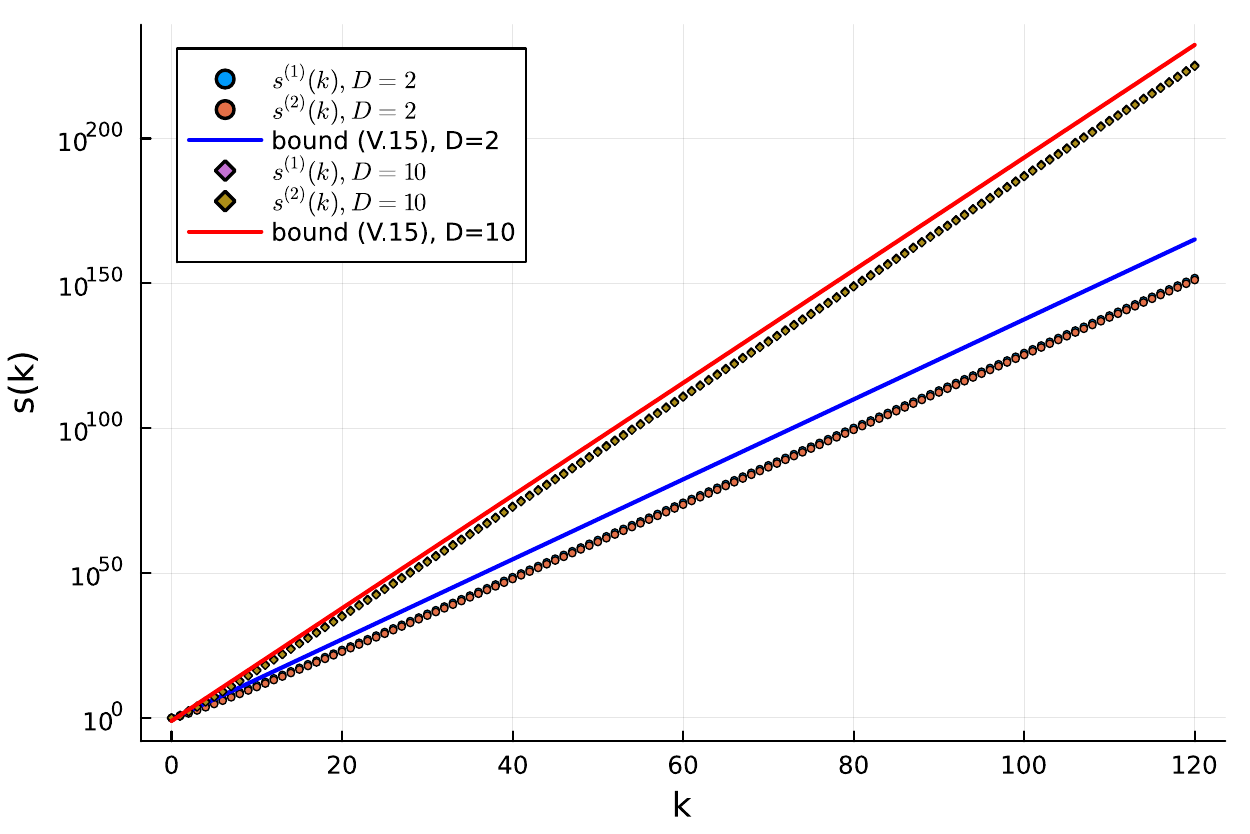}
\caption{$s^{(1)}(k)$ and $s^{(2)}(k)$ for the Anderson model with $D=2$ and $D=10$. The results are averaged over $100$ samples. Note that the $2$-norm is only slightly smaller than the $1$-norm and the two norms are almost indistinguishable on this scale. The bound \eqref{1norm_bound_disordered} is shown as well.}
\label{Fig_XX_disorder}
\end{figure} 
In this case, the graph is a $2$-regular graph with one node per level and two edges per node. The constant number of nodes per level is directly reflected in the quadratic growth of the number of distinct terms $n(k)$.

To obtain $s^{(1)}(k)$, we have to sum over all weighted walks through the graph which start at the top and have length $k$ by multiplying the values of all traversed edges, irrespective of the vertex the walk ends on. If, on the other hand, we want to calculate only the contributions of terms which have support on $l$ sites, $s^{(1)}_l(k)$, then we have to sum only over those weighted walks with $k$ steps which end at a vertex with terms on $l$ sites. We discuss this approach to calculating or bounding the norm of the commutator in more detail in the Suppl.~Mat. Fig.~\ref{Fig_nonint_nodes} also already shows how the graph has to be modified in the Anderson case which we discuss next.

\subsubsection{Anderson Localization}
\label{Sec_Anderson}
Next, we consider the case of quenched disorder where the magnetic fields $h_j$ in the Hamiltonian \eqref{Ham} are drawn randomly from a box distribution, $h_j\in [-D,D]$. We note that in the one-dimensional case considered here any amount of disorder is a relevant perturbation. The model is localized for all $D>0$. Since $h_j$ is random, we cannot expect exact results for the norm. Instead, we will derive strict upper bounds.

What we can do exactly, however, is to count the number of distinct terms $n(k)$ in the commutator which occur at order $k$ and which is independent of the specific disorder realization. This is easiest done by making use of the graph, see Fig.~\ref{Fig_nonint_nodes}, and by taking also into account the different possibilities where in the lattice these terms can be located. Doing so we find 
\begin{equation}
    \label{Anderson_num_terms}
 n(k)=2k(k+1)+\left\{
 \begin{array}{cc}1 &\mbox{even}\\
 1-k &\mbox{odd.}\end{array}\right.  
\end{equation}
This result is compared to symbolic calculations in Fig.~\ref{XX_num_terms_2}. We stress once more that Eq.~\eqref{Anderson_num_terms} is independent of the type of disorder or the specific disorder configuration and is thus also valid in the Aubry-Andr\'e case discussed in Sec.~\ref{Aubry-Andre}. Note that while the precise number of terms has changed as compared to the non-disordered case, importantly the number of terms still only increases quadratically with $k$.

\begin{figure*}[!htp]
    \includegraphics*[width=0.99\columnwidth]{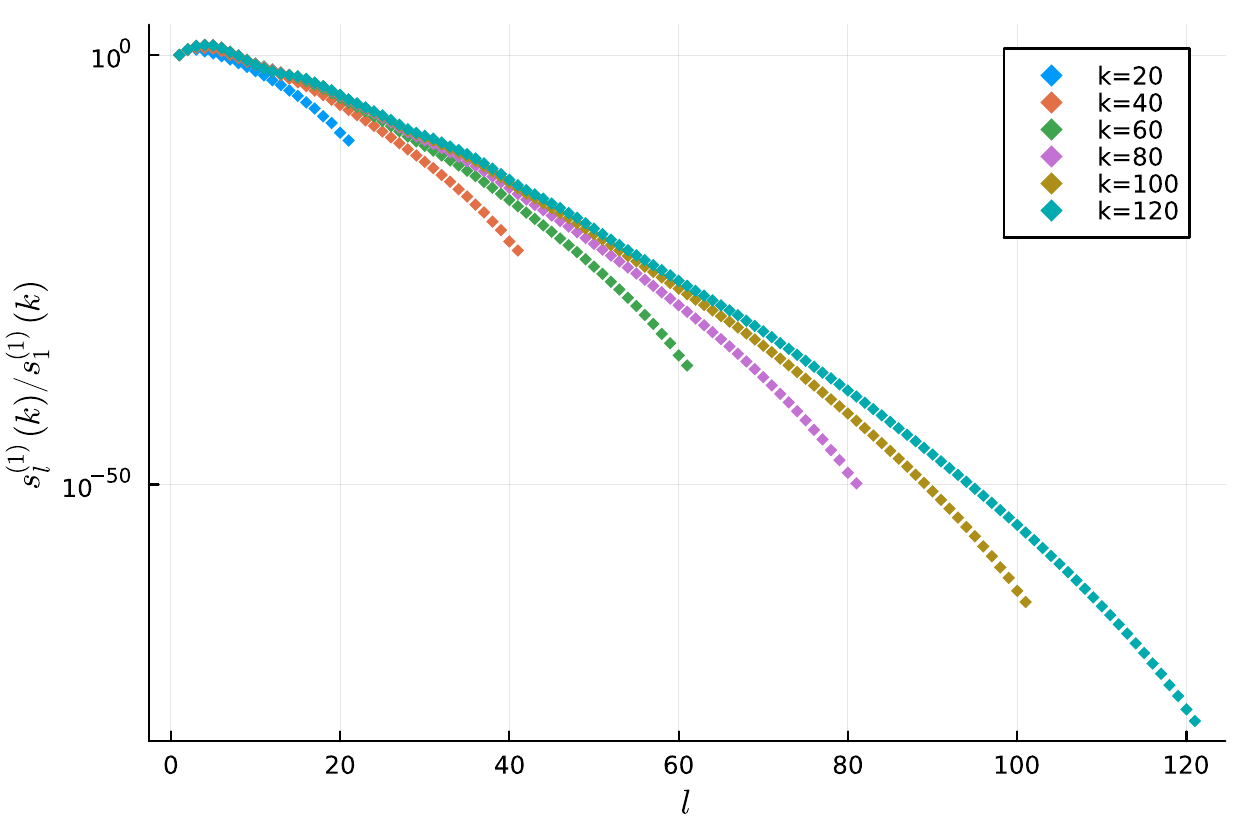}
    \includegraphics*[width=0.99\columnwidth]{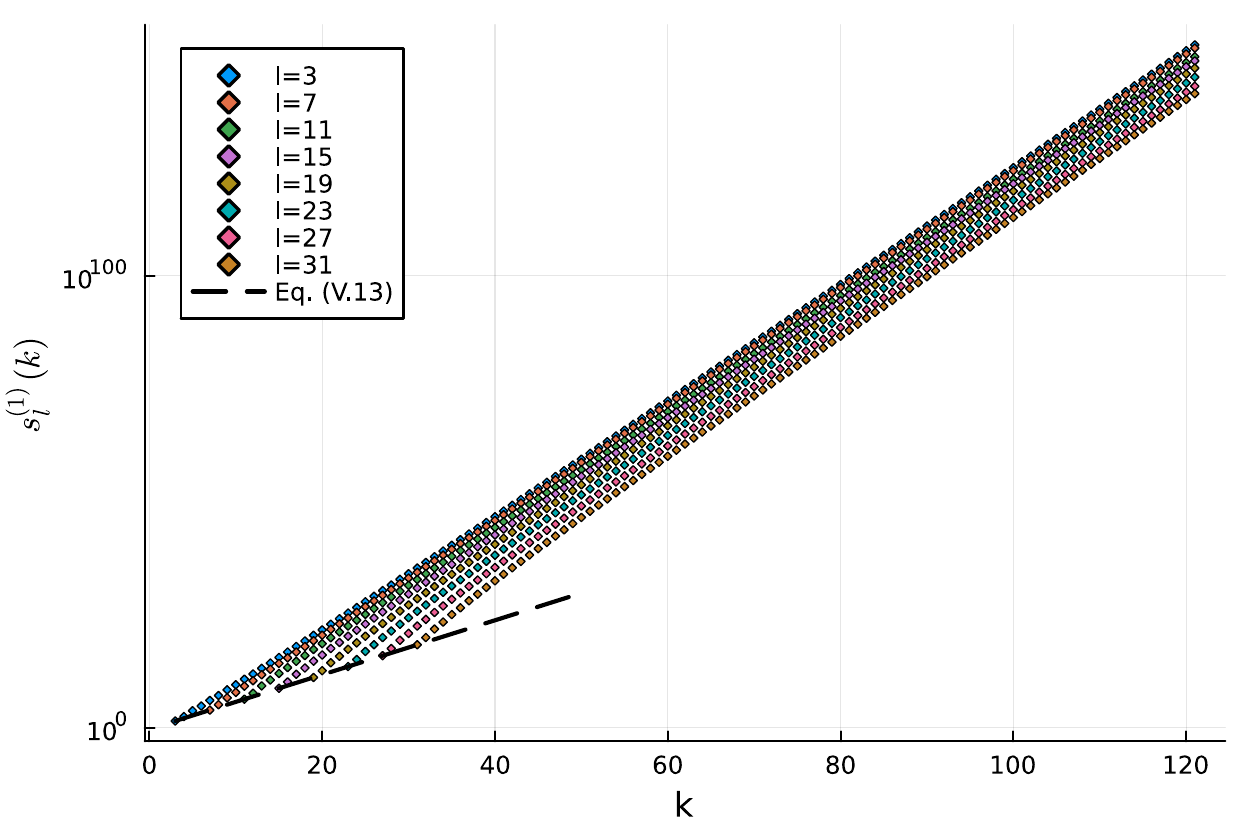}
    \includegraphics*[width=0.99\columnwidth]{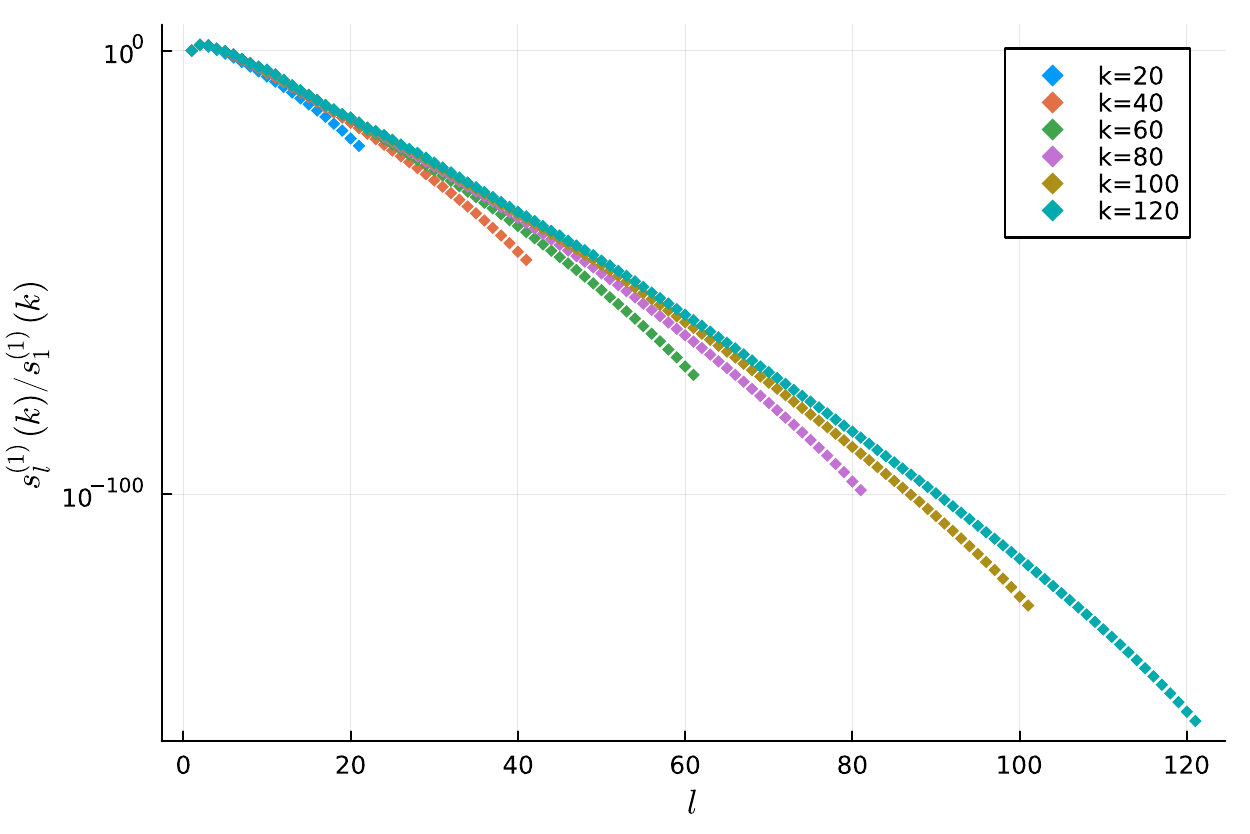}
    \includegraphics*[width=0.99\columnwidth]{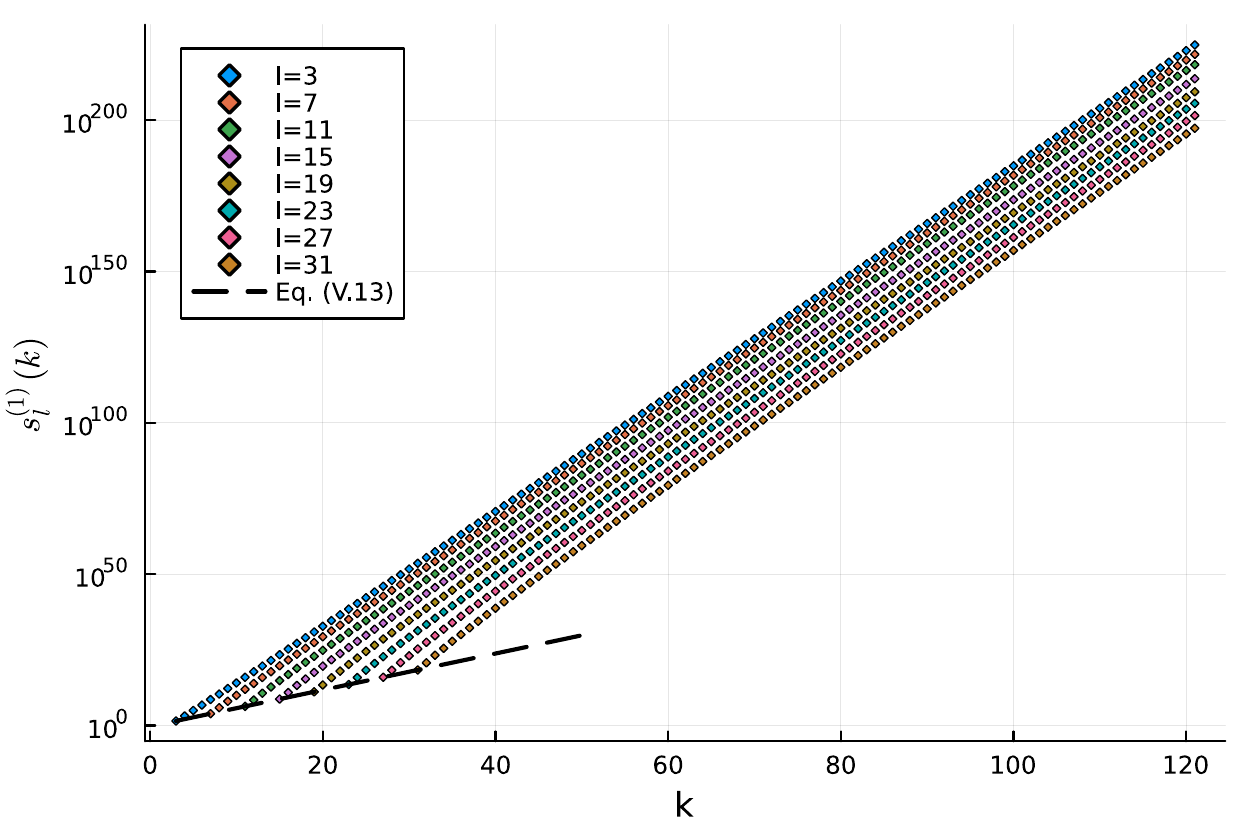}
\caption{Left column: Norm ratio $s^{(1)}_l(k)/s^{(1)}_1(k)$ for a fixed $k$ as a function of $l$ for the Anderson model with $D=2$ (top row) and $D=10$ (bottom row). Right column: $s^{(1)}_l(k)$ for a fixed $l$ as a
function of $k$. The data are averaged over $100$ samples. Note that the value of $s_l^{(1)}(k)$ when it first becomes non-zero at $k=l-1$ is not affected by disorder, see Eq.~\eqref{largest_support}.}
\label{Fig_XX_disorder_support}
\end{figure*} 
To bound the $1$-norm, we use the graph shown in Fig.~\ref{Fig_nonint_nodes}. With random fields, the additional terms in the right column and the corresponding additional paths are present. In the first step, we collect a factor of $8$ in the $1$-norm. In every other step, we can collect at most a factor $8+8D$ where we have bounded the disorder by $|h_j|\leq D$. We therefore obtain the following bound
\begin{equation}
s^{(1)}(k) \leq 8(8+8D)^{k-1}=8^k (D + 1)^{k-1}.
\label{1norm_bound_disordered}
\end{equation}
As expected, we find that the norm has an exponential bound. Note that this bound increases with increasing disorder $D$, see Fig.~\ref{Fig_XX_disorder}. Also note that for $D=0$ we obtain the exact result for the norm in the non-disordered case, Eq.~\eqref{1norm_nondisordered}. Apart from the initial node and the node on the top right, the graph with disorder is a $3$-regular graph. I.e., each node is connected to three edges. We note that in every $n$-regular graph the norm $s^{(1)}(k)$ can not grow faster than exponential. This remains true even if the number of nodes increases with $k$ as, for example, in a Cayley tree as long as the number of edges per node is constant. This follows immediately from the fact that $s^{(1)}(k)$ is the sum of the product of edge values along all possible paths of length $k$. That $n(k)$ grows only quadratically means that the graph in the free-fermion case is particularly simple and has a constant number of nodes at each level.

To see the difference between a free-fermionic model, where an initially localized operator spreads over the entire chain, and a system where the operator remains localized it is thus not sufficient to study the total operator norm. Instead, we have to investigate the spatial structure of the terms in the commutator at order $k$. Based on the graph in Fig.~\ref{Fig_nonint_nodes}, we can derive a bound for the contribution of terms with support on $l$ sites to the norm at order $k$. We use again that $|h_j|\leq D$. The derivation of this bound is discussed in detail in the Suppl.~Mat. Here, we will consider the simplest possible approach where we only take into account those terms with support on $l$ sites which have the largest possible number of factors $h_j$. This approach is justified for $D\gg 1$ and, as we will argue, gives the same asymptotics in the limit $k\gg l\gg 1$ as the more sophisticated approach discussed in the Suppl.~Mat. Let us first consider terms which have support on a single site, i.e., which are of the form $\sigma^z_j$. As is obvious from the graph in Fig.~\ref{Fig_nonint_nodes}, such terms are only generated at even orders of $k$. We collect the largest number of factors $h_j$ if we go down one level, then travel $k-2$ times along the vertical direction, and then go back up to the top of the graph. The $1$-norm of such terms is thus bounded by
\begin{equation}
    \label{1-site}
s^{(1)}_1(k)\leq 8\times (8D)^{k-2}\times 4 = 2^{2k+1}(2D)^{k-2} \, .     
\end{equation}
Next, we consider the norm of terms with support on $l\geq 2$ sites with the largest number of possible $h_j$ factors. In the first step, we have to go one level down in the graph in Fig.~\ref{Fig_nonint_nodes} which contributes a factor of $8$. To get to terms with support on $l\geq 2$ sites we have to take another $l-2$ steps down, contributing a factor $4^{l-2}$. We can distribute these steps among a total of $k-1$ remaining steps, i.e., there are $\binom{k-1}{l-2}$ possibilities. Finally, we get a factor $(8D)^{k-1-(l-2)}$ from the remaining steps along the vertical lines in the graph. Thus, the one norm of such terms is bounded by
\begin{equation}
    \label{l-sites}
s^{(1)}_l(k)\leq \binom{k-1}{l-2}2^{2k+1}(2D)^{k-l+1} \, .    
\end{equation}
\begin{figure*}[!htp]
    \includegraphics*[width=0.99\columnwidth]{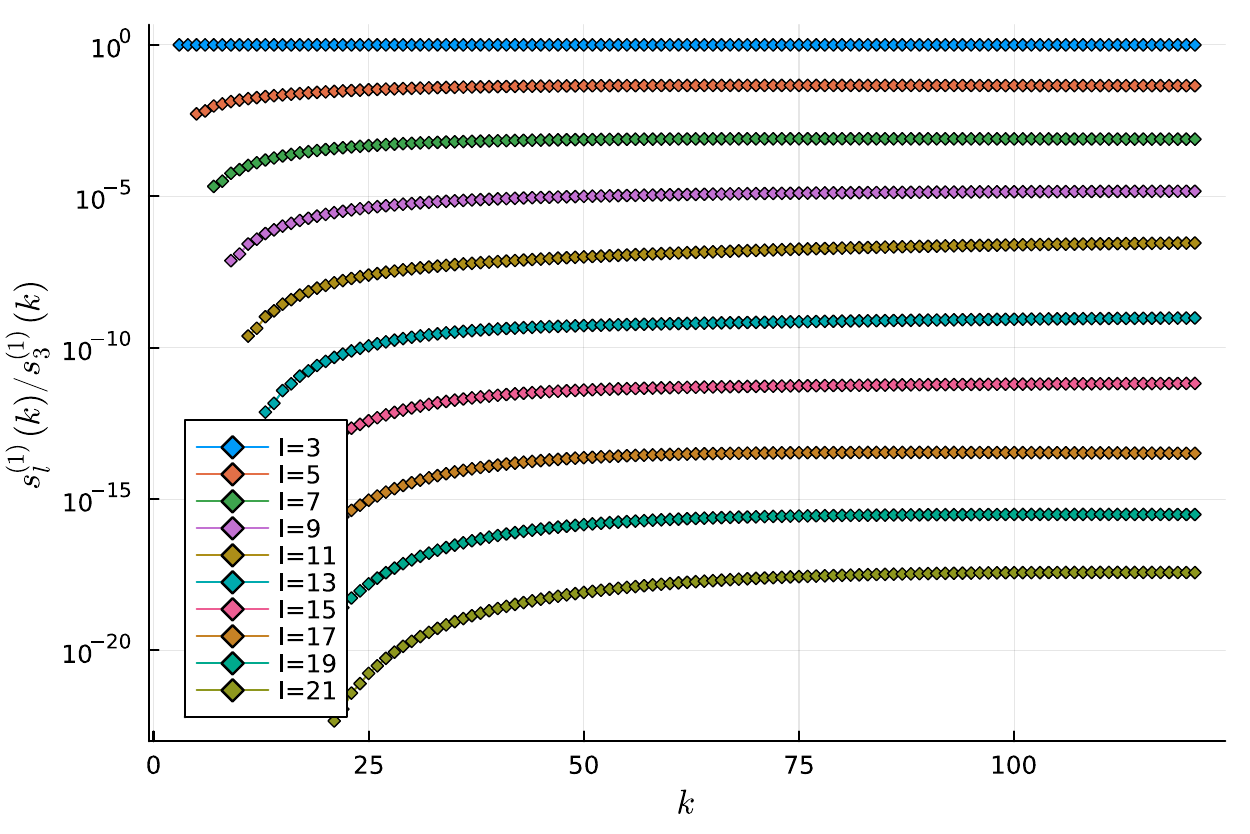}
    \includegraphics*[width=0.99\columnwidth]{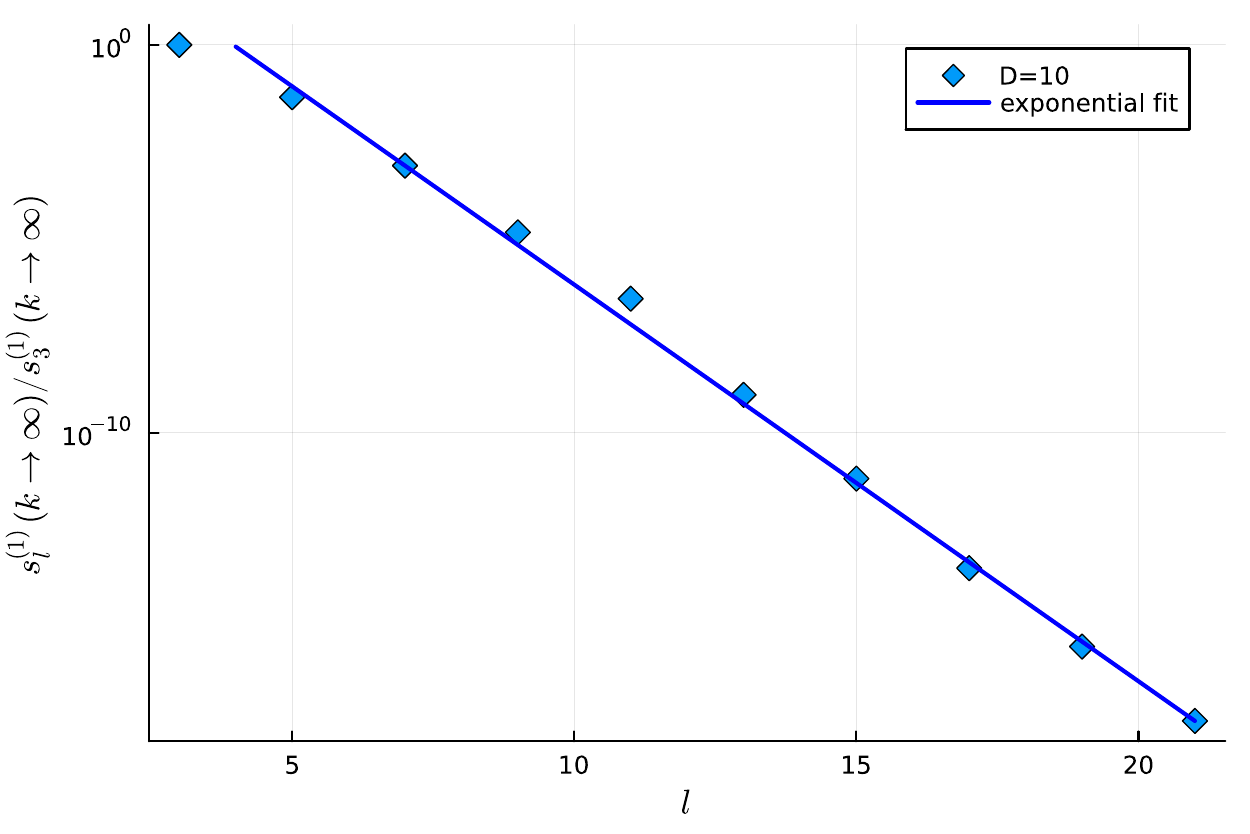}
    \caption{Anderson model with $D=10$. Left: Norm ratio $s^{(1)}_l(k)/s^{(1)}_3(k)$ for a fixed $l$ as a function of $k$. Right: The norm ratio $s^{(1)}_l(k\to\infty)/s^{(1)}_3(k\to\infty)$ decays exponentially with the support $l$. The data are averaged over $100$ samples.}
    \label{Fig_XX_disorder_analysis}
\end{figure*}
The relative contribution to the norm of terms with support on $l$ sites when normalized to the contribution coming from single-site operators is thus given by
\begin{equation}
    \label{ratio}
\frac{s_l(k)}{s_1(k)}\lesssim \binom{k-1}{l-2}(2D)^{3-l} 
\approx \frac{1}{\sqrt{2\pi l}} \left(\frac{ke}{l}\right)^l e^{-\frac{l^2}{2k}} (2D)^{3-l}\, .
\end{equation}
We note that while for a fixed $k$ we indeed find that the ratio \eqref{ratio} goes exponentially to zero for large $l$, there is a maximum in the ratio which shifts to larger $l$ with increasing $k$. This is not quite the behavior we expect. Instead, we expect that with increasing $k$ the ratio \eqref{ratio} converges to a function which is exponentially decreasing in $l$. Here it is important to note that \eqref{ratio} is only an upper bound. In particular, we have only considered those paths which have a maximum number of $h_j$ factors. There are, however, many additional paths and we expect that the prefactor in \eqref{ratio} will eventually converge with $k$ for a given $l$. This is consistent with what we observe from symbolic manipulations to finite orders of $k$, see Fig.~\ref{Fig_XX_disorder_support}. We note that in contrast to the non-disordered case where $s_l(k)$ grows exponentially and independent of $l$ for $k\gg l$, see Eq.~\eqref{spat_structure} and Fig.~\ref{Fig3}, we find that with quenched disorder
\begin{equation}
    \label{spat_structure_Anderson}
s_l(k) \sim \e^k D^{-l} \, ,
\end{equation}
see Fig.~\ref{Fig_XX_disorder_analysis}. The spatial structure of the norm contributions $s_l(k)$ thus provide a clear distinction between the non-localized and localized phases. 

\subsubsection{Aubry-Andr\'e model}
\label{Aubry-Andre}
\begin{figure*}
    \includegraphics*[width=0.99\columnwidth]{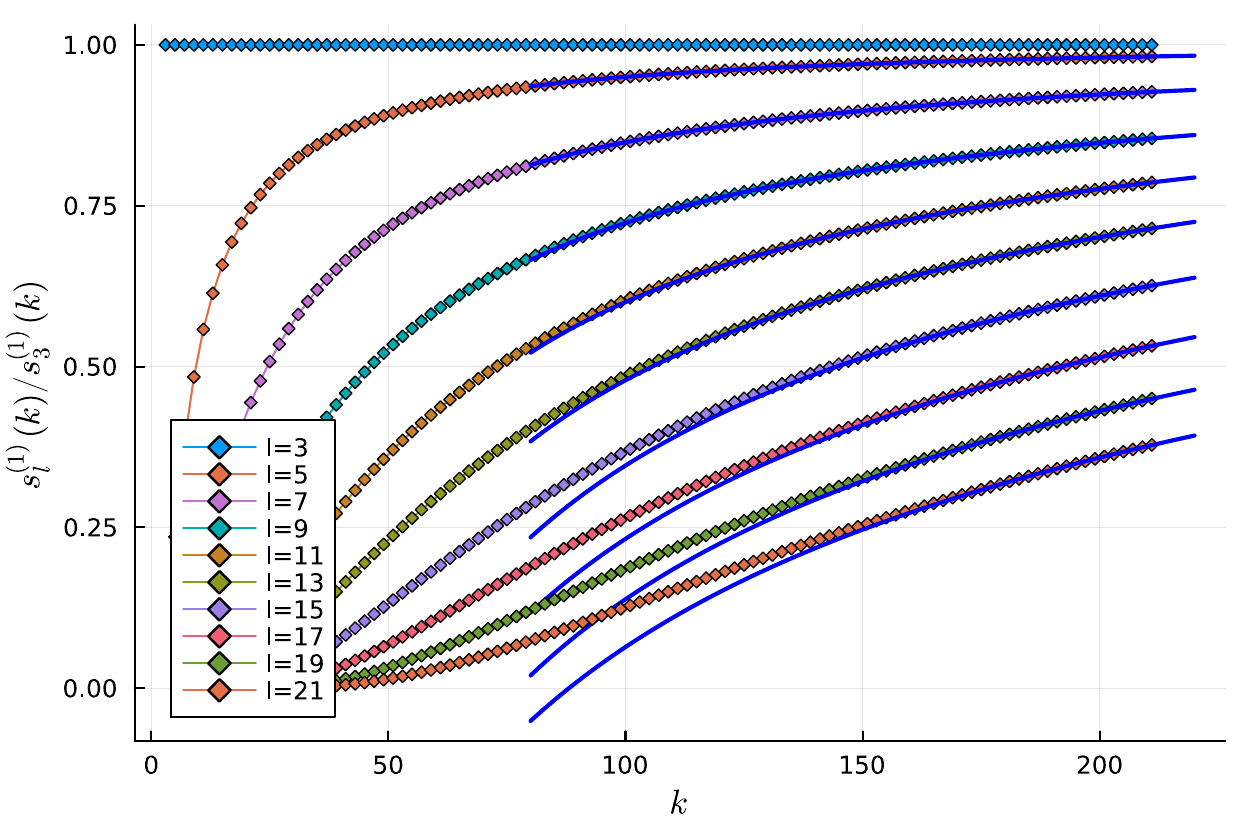}
    \includegraphics*[width=0.99\columnwidth]{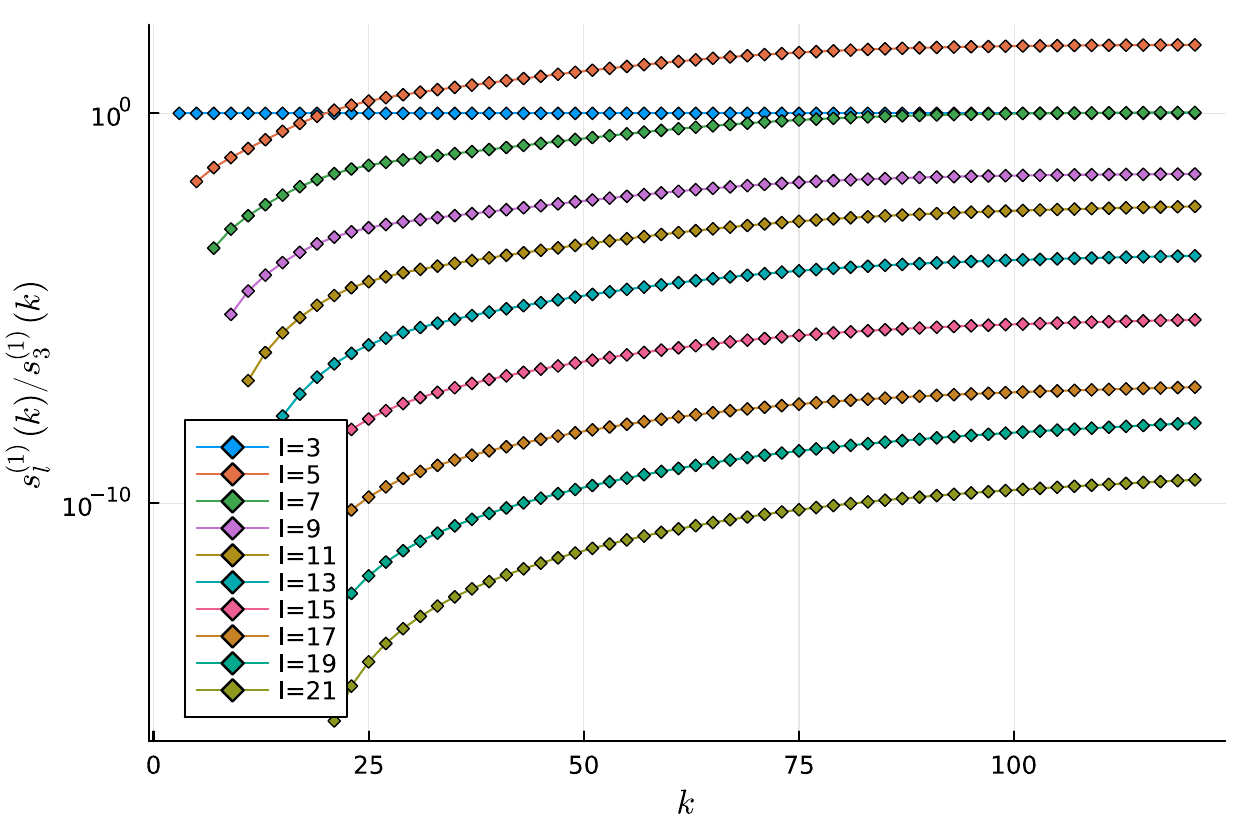}
    \includegraphics*[width=0.99\columnwidth]{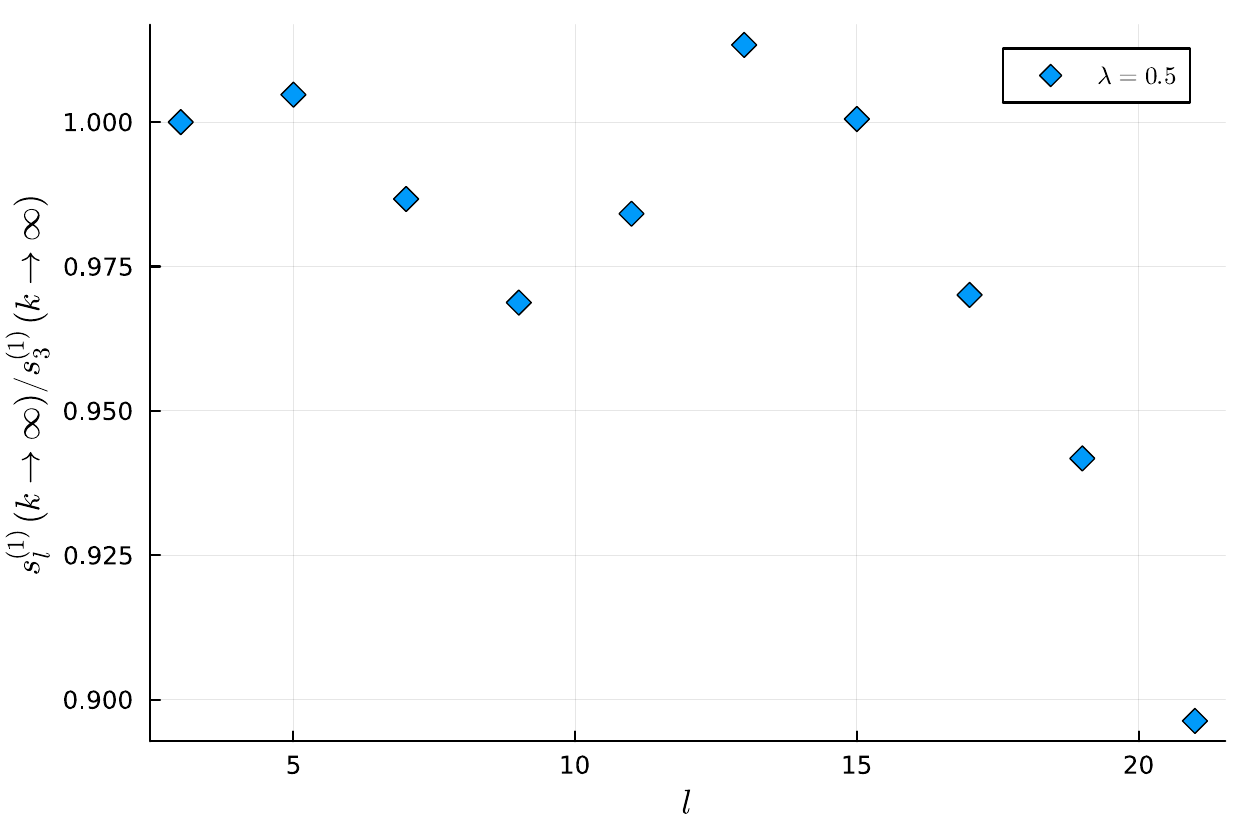}
    \includegraphics*[width=0.99\columnwidth]{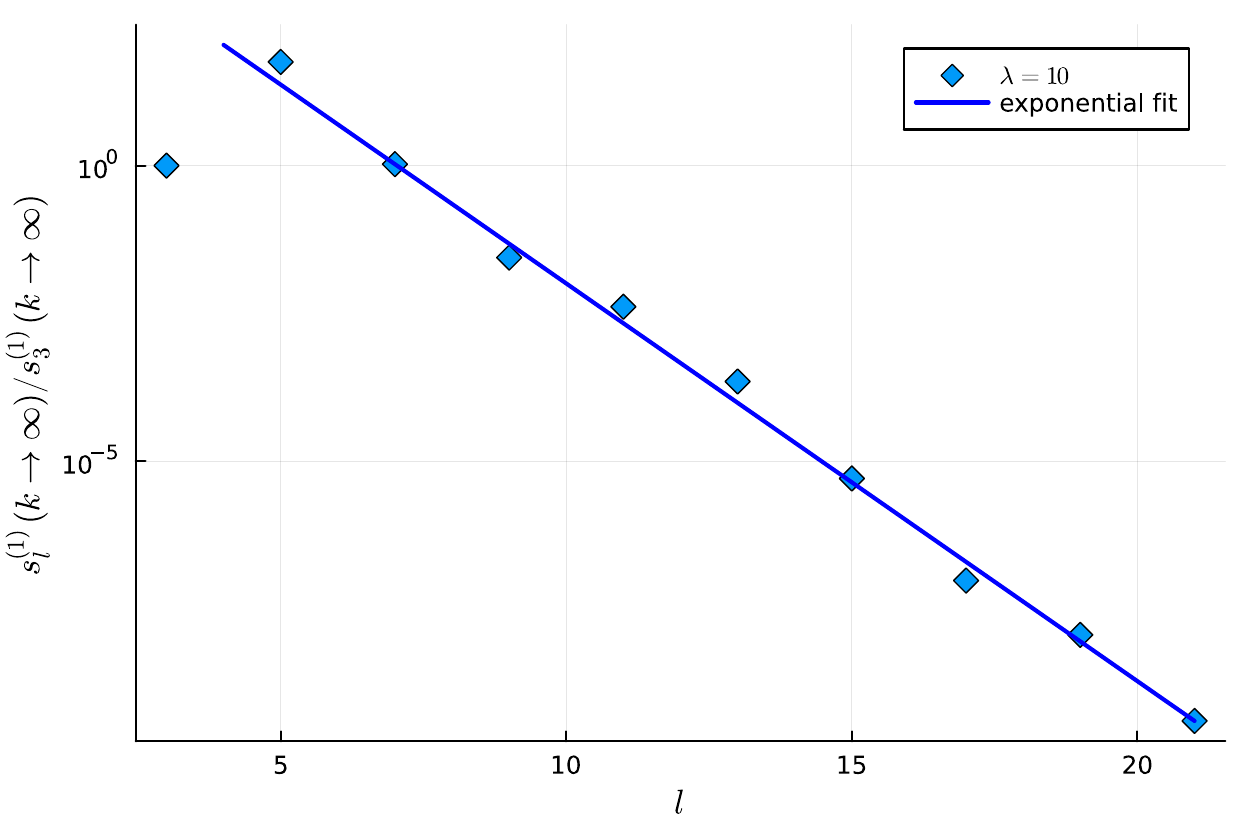}
\caption{Aubry-Andr\'e model for $\beta=(1+\sqrt{5})/2$, $\phi=0$ where $\lambda=0.5$ (left column, linear scale) and $\lambda=10$ (right column, logarithmic scale). For $\lambda=0.5$, $s^{(1)}_l(k)$ is extrapolated (blue lines) and the results are consistent with $s^{(1)}_l(k)\sim C \e^k$ with a constant $C$ independent of $l$. This leads to $s^{(1)}_l(k\to\infty)/s^{(1)}_3(k\to\infty)\approx 1$ (bottom left panel) clearly showing that the model is not in the localized phase. In contrast, for $\lambda=10$ the norm scales as $s^{(1)}_l(k)\sim \e^k \e^{-l}$, i.e., is exponentially suppressed with increasing $l$ for $k$ fixed, demonstrating that the model is now in the localized phase (bottom right panel).}
\label{FigAA1}
\end{figure*}

To further check that the exponential decay of $s_l(k)=||[H,\sigma^z_0]^{(k)}_l||$  with $l$ is indeed the signature of a localized system, we consider the Aubry-Andr\'e model next. This is a model which is typically formulated as a fermionic model \eqref{Ham2} where instead of quenched disorder we have a periodic potential
\begin{equation}
    \label{AA_pot}
    h_j = \frac{\lambda}{2}\cos(2\pi\beta j +\phi) \, .
\end{equation}
Then, for $\beta$ Diophantine and almost any $\phi$---excluding a set of measure zero of resonant phases---the eigenstates of the Aubry-Andr\'e model are localized for $\lambda>2$ \cite{AubryAndre,Jitomirskaya}. A common choice is to use the golden ratio $\beta=(1+\sqrt{5})/2$. The model thus provides an important test case where both a non-localized and a localized phase exist. Below the transition in the non-localized phase we expect that $s_l(k)$ is qualitatively described by Eq.~\eqref{spat_structure}, i.e., grows with $k$ independent of $l$ for $k\gg l$. In the localized phase, on the other hand, we expect that $s_l(k)$ is exponentially localized, see Eq.~\eqref{spat_structure_Anderson}. Note that the graph for the Aubry-Andr\'e model is the same as the one in the Anderson case shown in Fig.~\ref{Fig_nonint_nodes} with $D\to\lambda/2$ on the edges. With this replacement, also the bound \eqref{1norm_bound_disordered} holds. We note that the considerations which let us to Eq.~\eqref{spat_structure_Anderson} in the Anderson case were valid only in the case of strong disorder. In this limit, they remain valid for the Aubry-Andr\'e model. That the exponential decaying structure for the spatial support of the norm remains unchanged for all $D\neq 0$ in the Anderson case while there is a transition to the structure \eqref{spat_structure}, which is independent of $l$ for $k\gg l$,  for the Aubry-Andr\'e model is beyond the strong disorder arguments we used in the previous section. All these results and expectations outlined above are fully consistent with the symbolic computations shown in Fig.~\ref{FigAA1}. Note that the data in the left panels for the non-localized phase are shown on a linear scale while the results in the right panels for the localized phase are shown on a logarithmic scale.

While other methods might be better suited to determine the exact phase transition point, there is a clear {\it qualitative} difference in the scaling of $s_l(k)$ between the extended and the localized phase. The operator growth $s_l(k)$ therefore allows for an unambiguous distinction between these two phases.

\subsection{Interacting case}
\label{Sec_int}
After having checked the analytical results derived in Sec.~\ref{Sec_general} and Sec.~\ref{localization} for the non-interacting non-localized and localized cases, we can now look at the interacting isotropic Heisenberg (XXX) model, i.e., the Hamiltonian \eqref{Ham} with $\Delta=1$. A fundamental general difference in interacting as compared to non-interacting models is that one-body operators do not remain one-body operators when commuted with the Hamiltonian. This leads to a qualitatively different phenomenology in the operator growth with exponentially many more terms and paths.

We also note that the Heisenberg model without disorder is a so-called integrable model which has, what is commonly called, an infinite set of local conserved charges $O$ with $[H,O]=0$ in the thermodynamic limit. It is crucial to note though that `local' is meant here in an entirely different way. These charges are in no way localized but rather can be written as $O=\sum_j o_j$ where $o_j$ is a local density acting on a finite number of lattice sites. I.e., the charge densities are local but the entire conserved operator $O$ is not and rather acts on the entire lattice. The trivial examples are the Hamiltonian itself, $H=\sum_ I h_I$ and the magnetization $M=\sum_I \sigma^z_I$. What is special about integrable models is that an infinite set of operators of this type exist which allows to uniquely characterize all eigenstates by the corresponding quantum numbers. The eigenspectrum of the model can thus be calculated exactly by using the Bethe ansatz \cite{Bethe,HubbardBook}. For the question of {\it local operator growth}, integrability is not relevant and we do not expect to see any difference to a non-integrable model. Even if we break integrability, the magnetization $M$ is still conserved as long as $U(1)$ symmetry is preserved. I.e., we are in any case studying the operator growth of a local operator $\sigma^z_j$ which belongs to a globally conserved charge. Differences between integrable and non-integrable models are, of course, expected if we study the Euclidean time dynamics of operators of the type $O=\sum_j o_j$ \cite{ParkerCao}. However, for the Euclidean time dynamics of {\it local} operators in a generic system, the only thing which matters is whether or not this system has local conserved charges, i.e., whether it is localized or not. We also note that integrability is immediately destroyed if we introduce even an infinitesimal amount of disorder.   

We will first briefly consider the case without disorder before turning to the most interesting case, the XXX model with random magnetic fields $h_j\in [-D,D]$ which has been suggested to have a localized phase for sufficiently strong disorder.

\subsubsection{No disorder}
\begin{figure}
    \includegraphics*[width=0.99\columnwidth]{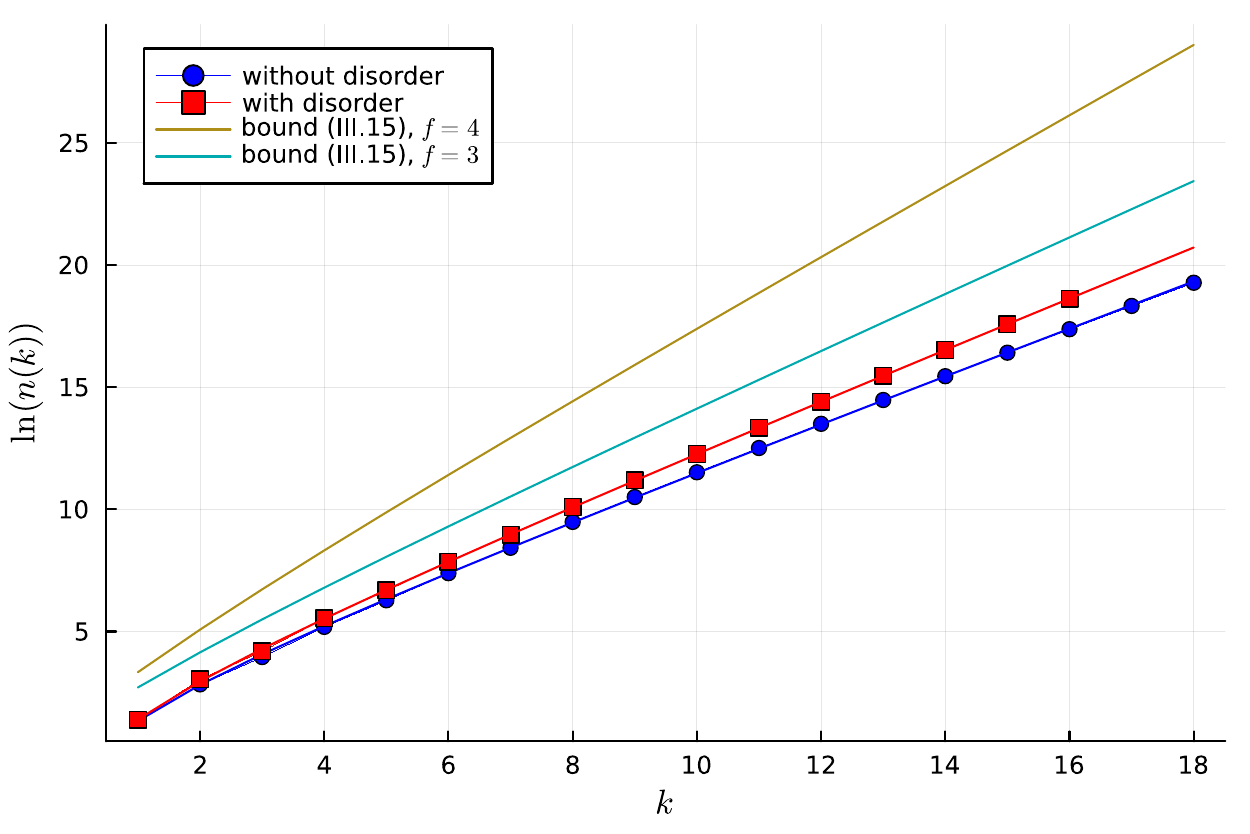}
\caption{In the interacting model with $\Delta=1$, the number of different terms $n(k)$ in the commutator $[H,\sigma_0^z]^{(k)}$ grows exponentially (symbols) and is very well fitted by $n(k)\sim 1.514\, k^{0.803}\exp(0.922k)$ in the non-disordered case and $n(k)\sim 1.492\, k^{0.815}\exp(0.998k)$ in the case with disorder (solid lines). The general bounds \eqref{nk_general} with $f=3$ and $f=4$ are shown as well.}
\label{Fig_XXX_number_of_terms}
\end{figure}

From the exact solution of the model, we know for sure that the eigenstates of the XXX model are not localized. We can calculate the number of distinct terms $n(k)$ in the commutator $[H,\sigma_0^z]^{(k)}$ at order $k$ by symbolic calculations up to order $k=18$ exactly. We have not been able to identify the series but asymptotically the number of terms $n(k)$ does grow exponentially with a multiplicative algebraic correction which is consistent with the general bound \eqref{nk_general} which we derived earlier. The data and a corresponding fit are shown in Fig.~\ref{Fig_XXX_number_of_terms}.
The fit works extremely well with a least square deviation of $\chi^2\sim 10^{-2}$. The local operator basis for the Heisenberg model has dimension $f=4$ and consists of the operators ${\mathbbm{1},\sigma^x,\sigma^y,\sigma^z}$. We note, however, that only a small subset of terms in $[H,\sigma^z_0]^{(k)}$ contains an identity in between Pauli matrices. Ignoring such terms, we can set $f=3$ in Eq.~\eqref{nk_general} which gives a much tighter bound, see Fig.~\ref{Fig_XXX_number_of_terms}. The exponential growth of $n(k)$, instead of the quadratic growth in the free-fermion case, indicates that in the corresponding commutator graph the number of nodes and edges per level are no longer constant.

\begin{figure}
    \includegraphics*[width=0.99\columnwidth]{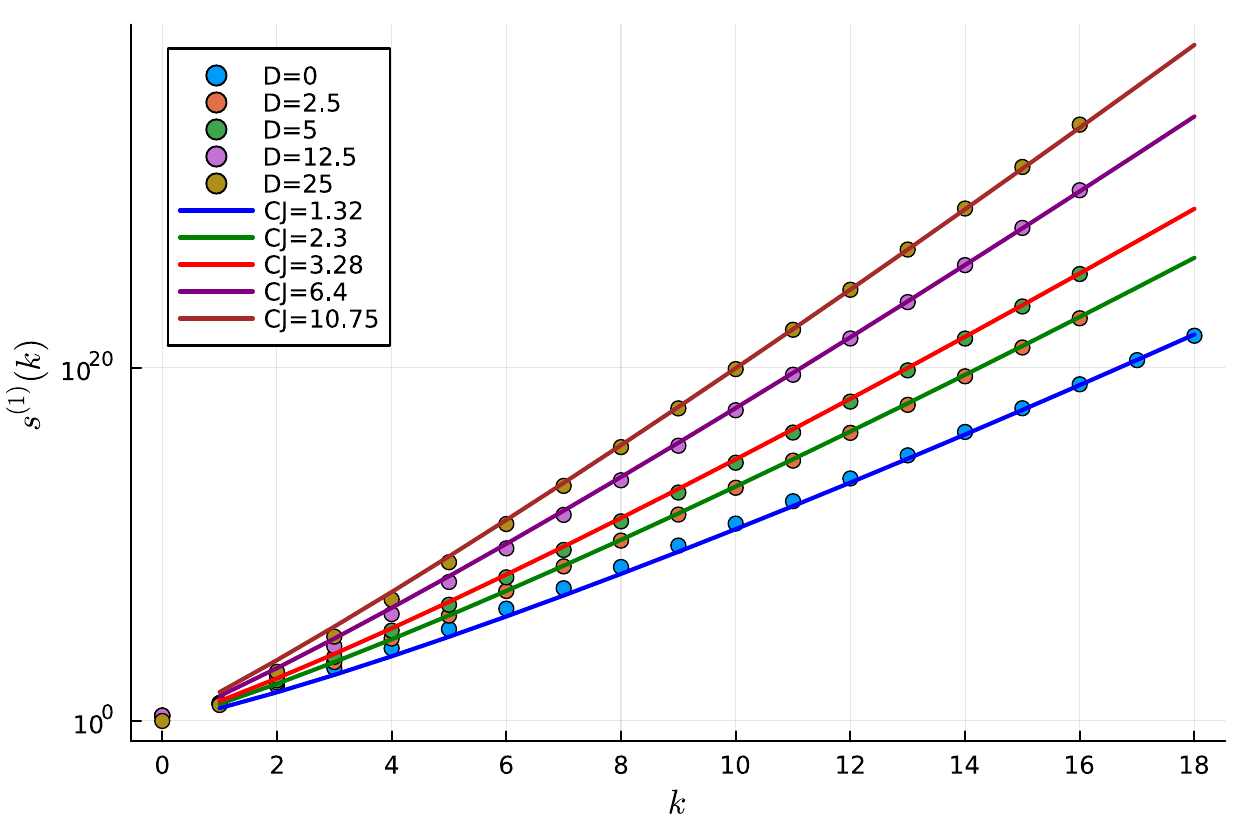}
\caption{The 1-norm in the interacting case with $\Delta=1$ grows faster than exponential for all disorder strengths shown, consistent with the rescaled general bound, Eq.~\eqref{generalB}. The data are averaged over $30$ disorder realizations for $D=2.5$, $D=5$ and $D=12.5$ and over 50 realizations for $D=25$.}
\label{Fig_XXX_1-norm}
\end{figure}

\begin{figure*}
    \includegraphics*[width=0.99\textwidth]{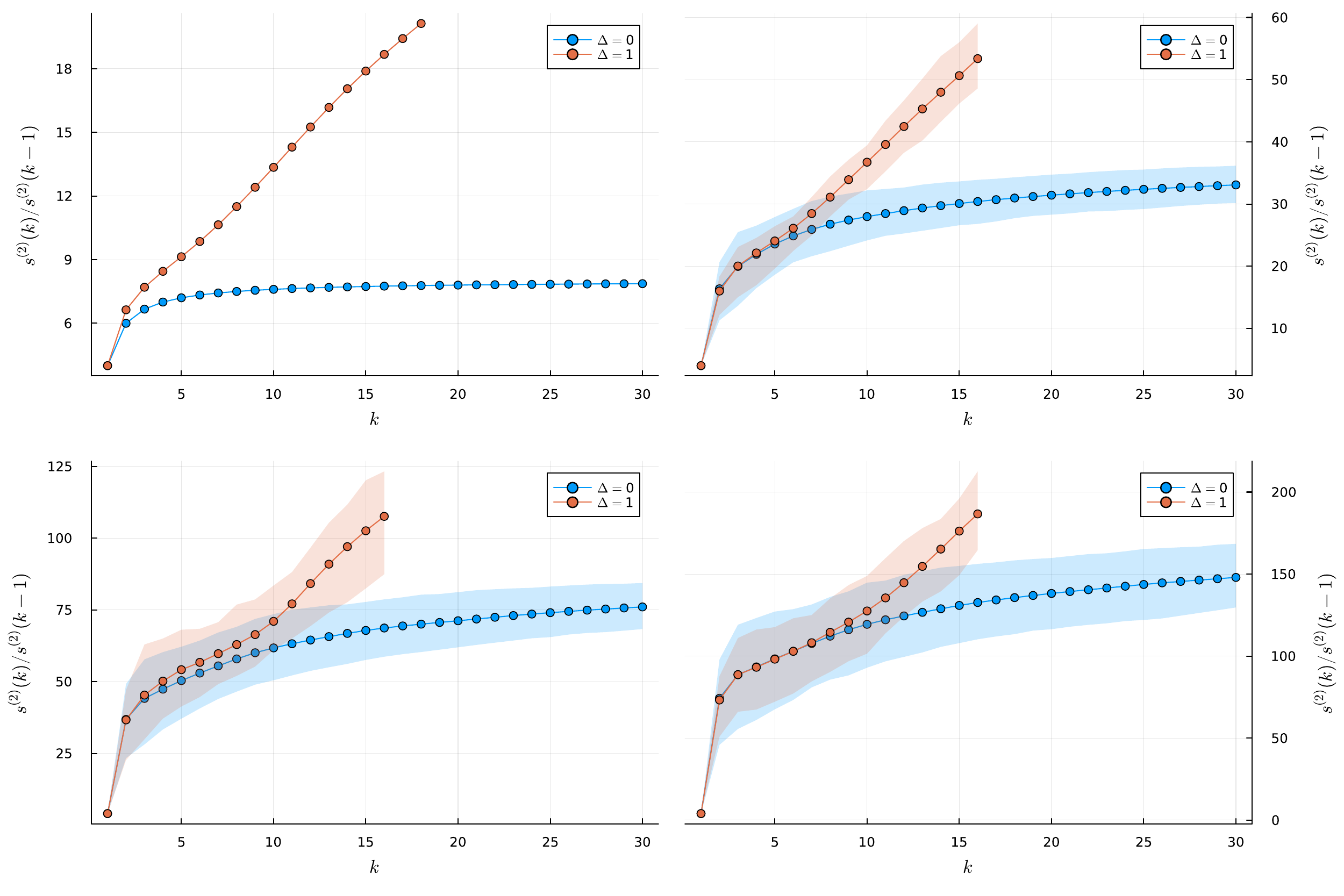}
\caption{Average of the $2$-norm ratio $s^{(2)}(k)/s^{(2)}(k-1)$ for the XXZ chain with $\Delta=0$ and $\Delta=1$  for disorder strengths $D=0,5,12.5,25$. The data for $D=5$ and $D=12.5$ are averaged over $50$, and the data for $D=25$ over $70$ realizations. The shaded regions indicate the intervals into which 50\% of all realizations fall. The norm ratio has to become a constant for large $k$ if the norm grows exponentially.}
\label{XXX_norm_ratios}
\end{figure*}

\begin{figure*}
    \includegraphics*[width=0.99\textwidth]{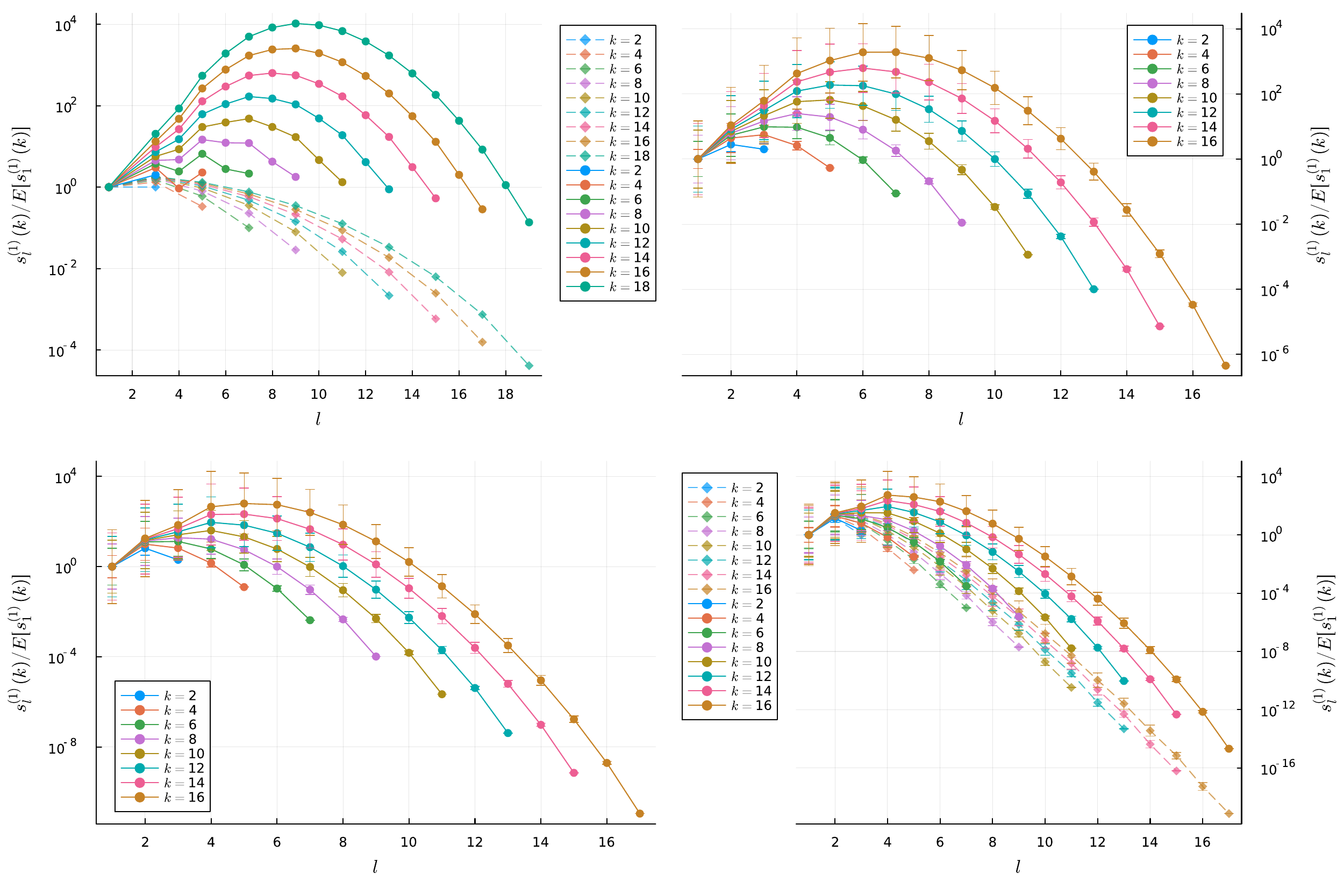}
\caption{The norm of operators $s^{(1)}_l(k)/E[s^{(1)}_1(k)]$ with support on $l$ sites for the Heisenberg model with disorder $D=0,5,12.5,25$. The data for $D=5$ and $D=12.5$ are averaged over $50$, the data for $D=25$ over $70$ disorder realizations. There is no indication for a convergence with increasing order $k$ to an exponential \ch{or almost exponential} decay with $l$ even for very strong disorder. For $D=0$ and $D=25$ the results for the non-interacting case are shown for comparison (symbols with dotted lines).}
\label{XXX_support}
\end{figure*}

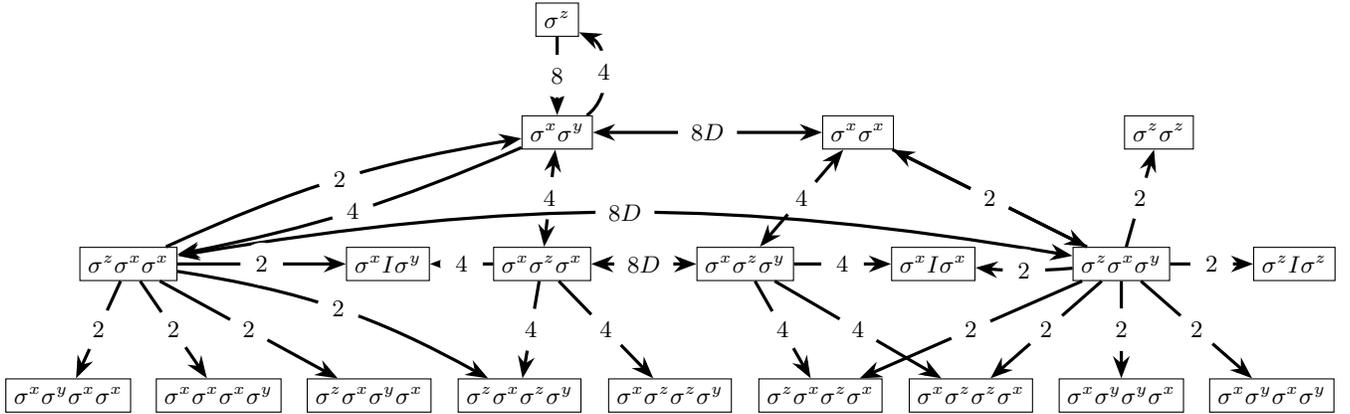
\begin{figure*}
\centering
\begin{tikzpicture}
\begin{scope}[every node/.style={rectangle,draw}]
    \node (A) at (0,0) {$\sigma^z$};
    \node (B) at (0,-1.5) {$\sigma^x \sigma^y$};
    \node (C) at (-0.2,-3.25) {$\sigma^x \sigma^z \sigma^x$};

    \node (E) at (4,-1.5) {$\sigma^x \sigma^x$};
    \node (F) at (2.5,-3.25) {$\sigma^x \sigma^z \sigma^y$};

    \node (H) at (-5.7, -3.25) {$\sigma^z \sigma^x \sigma^x$};
    \node (I) at (7.5, -3.25) {$\sigma^z \sigma^x \sigma^y$};

    \node (K) at (-6.5, -5) {$\sigma^x \sigma^y \sigma^x \sigma^x$};
    \node (L) at (-4.5, -5) {$\sigma^x \sigma^x \sigma^x \sigma^y$};
    \node (M) at (-2.5, -5) {$\sigma^z \sigma^x \sigma^y \sigma^x$};
    \node (N) at (-0.5, -5) {$\sigma^z \sigma^x \sigma^z \sigma^y$};
    \node (O) at (-2.25, -3.25) {$\sigma^x I \sigma^y$};
    \node (P) at (1.5, -5) {$\sigma^x \sigma^z \sigma^z \sigma^y$};

    \node (Q) at (3.5, -5) {$\sigma^z \sigma^x \sigma^z \sigma^x$};
    \node (R) at (5.5, -5) {$\sigma^x \sigma^z \sigma^z \sigma^x$};
    \node (S) at (7.5, -5) {$\sigma^x \sigma^y \sigma^y \sigma^x$};
    \node (T) at (9.5, -5) {$\sigma^x \sigma^y \sigma^x \sigma^y$};

    \node (U) at (5, -3.25) {$\sigma^x I \sigma^x$};
    \node (V) at (8, -1.5) {$\sigma^z \sigma^z$};
    \node (W) at (9.8, -3.25) {$\sigma^z I \sigma^z$};

\end{scope}

\begin{scope}[>={Stealth[black]},
              every node/.style={fill=white,circle},
              every edge/.style={draw=black,very thick}]
    \path [->] (A) edge node {$8$} (B);
    \path [<->] (B) edge node {$4$} (C);
    \path [->] (B) edge[bend right=60] node {$4$} (A); 
    \path [<->] (E) edge node {$4$} (F);
    \path [<->] (B) edge node {$8D$} (E);
    \path [<->] (C) edge node {$8D$} (F);
    
    \path [->] (B) edge[bend left=6] node {$4$} (H);
    \path [->] (H) edge[bend left=7.5] node {$2$} (B);
    \path [->] (E) edge node {$4$} (I);
    \path [<->] (H) edge[bend left=9.5] node {$8D$} (I);

    \path [->] (H) edge node {$2$} (K);
    \path [->] (H) edge node {$2$} (L);
    \path [->] (H) edge node {$2$} (M);
    \path [->] (H) edge[bend left=10] node {$2$} (N);
    \path [->] (H) edge node {$2$} (O);
    \path [->] (C) edge node {$4$} (O);
    \path [->] (C) edge node {$4$} (P);

    \path [->] (C) edge node {$4$} (N);
    \path [->] (F) edge node {$4$} (Q);
    \path [->] (F) edge node {$4$} (R);
    \path [->] (F) edge node {$4$} (U);
    \path [->] (I) edge node {$2$} (Q);
    \path [->] (I) edge node {$2$} (R);
    \path [->] (I) edge[bend left=5] node {$2$} (U);
    \path [->] (I) edge node {$2$} (S);
    \path [->] (I) edge node {$2$} (T);
    \path [->] (I) edge node {$2$} (V);
    \path [->] (I) edge node {$2$} (W);
    \path [->] (I) edge node {$2$} (E);

\end{scope}
\end{tikzpicture}
\caption{Graph for the $1$-norm $s^{(1)}(k)$ up to order $k=3$ for the Heisenberg model with random fields $h_j \in [-D,D]$. The number of nodes grows exponentially with the length of the support.}
\label{Graph_interacting}
\end{figure*}

We note that the number of distinct terms is different and always smaller than the number of connected clusters which we have calculated in Eq.~\eqref{Bell}. The reason is that different connected clusters can result in the same term. For the calculation of the norm this does not matter though. We can bound the contribution of each connected cluster as in Eq.~\eqref{generalB} even if some of the resulting terms are the same. In Fig.~\ref{Fig_XXX_1-norm} we show that this general bound indeed describes the norm growth well if we use the bound on the local Hamiltonian, $||h^a_I||\leq J$, as a fitting parameter.
In particular, the norm clearly grows faster than exponential for the orders of the commutator which we can handle symbolically. We also note that we obtain qualitatively similar results if we use the $2$-norm instead of the $1$-norm, see Fig.~\ref{XXX_norm_ratios}.  
This difference between the non-interacting and the interacting case becomes even more evident if we consider the norm $s^{(1)}_l(k)$ of terms supported on $l$ sites, see Fig.~\ref{XXX_support}.
Whereas in the non-interacting, non-disordered case the norm of the terms supported on $l$ sites grows asymptotically in the same way independent of $l$ if the order of the commutator $k$ is much larger than $l$, see Eq.~\eqref{spat_structure}, $s^{(1)}_l(k)/s^{(1)}_1(k)$ has a maximum which shifts to larger and larger $l$ with increasing $k$ in the interacting case. For large $k$, the total norm is thus completely dominated by terms which have support on a large number of lattice sites. This is to be expected, given that the number of distinct terms in the commutator is increasing exponentially which also leads to a rapidly growing number of possible paths between those terms, see also the graph in Fig.~\ref{Graph_interacting}. Note that the graph is no longer regular; the number of edges per node is not constant and is increasing with the level the node is situated on. Note that this increase of the number of edges is a prerequisite for a faster than exponential growth of the norm. 

\subsubsection{Many-body localization}
\label{Symbolic_localization}
\ch{Based on this graph representation, it is obvious that a strict exponential localization of operators in a many-body system is thus an extremely challenging requirement.} In a graph with an equal number of edges at each level as realized in a non-interacting model, see Fig.~\ref{Fig_nonint_nodes}, it is easy to see---at least in the limit of large disorder---how the norm of terms with support on only a few sites can dominate the total norm. The exponential growth of edges in the interacting case (see Fig.~\ref{Graph_interacting}), on the other hand, suggests that the total norm for large $k$ will be entirely dominated by terms with support on many lattice sites. I.e., a local operator is generically expected to delocalize. This aspect also makes it difficult to see how a perturbative construction of local integrals of motion can be arranged. This is a point which seems to have not been sufficiently appreciated in Refs.~\cite{Imbrie2016,ImbrieMBL} where such a construction was suggested. We will get back to this point in Sec.~\ref{Sec_SW}. The only way a local operator can avoid to spread through the entire system in the many-body case thus seems to be intricate interference effects between the exponentially many different paths in the graph, Fig.~\ref{Graph_interacting}. \js{While such fine-tuned destructive interference seems implausible as a generic mechanism for many-body localization we cannot, in general, show that such effects do not exist. We note, however, that for the transverse Ising chain with random fields---the model considered in the proof of MBL in Refs.~\cite{Imbrie2016,ImbrieMBL}---a gauge can be defined such that the commutator has a stoquastic-type property \cite{Cao}. I.e., all terms in the commutator in this gauge have a positive amplitude. For this model, interference effects as a mechanism for localization are thus strictly excluded \ch{and we have shown in Sec.~\ref{localization} that the almost factorial norm growth is indeed incompatible with exponential localization}. For the Heisenberg model considered here, such a gauge does not exist. We will instead consider data from symbolic computations up to order $k=16$ and show that there is no evidence for destructive interference effects and a corresponding localization. \ch{The data are consistent with a stronger than exponential norm growth.}} We note that at order $k$, we are considering operators which are spread over $2k+1$ sites in the lattice. I.e., at order $k=16$ we are dealing effectively with a chain of $33$ lattice sites which is well beyond what is possible in exact diagonalizations. We will also consider disorder strengths up to $D=25$ which is much larger than the critical values which have been suggested in the past \cite{PalHuse,Luitz1,Luitz2,MorningstarHuse}. 

Let us start by considering the number of distinct terms in the commutator at order $k$ which is shown in Fig.~\ref{Fig_XXX_number_of_terms}. We find that the number of terms grows faster in the disordered than in the case without disorder. This is easy to understand: additional terms appear in the commutator due to the random magnetic field terms while all the terms present before remain. The different classes of terms which appear up to order $k=3$ are shown in the graph in Fig.~\ref{Graph_interacting}. Next, we consider the growth of the total norm which should be at most exponential in a \ch{strictly exponentially localized} phase as we have proven in Eq.~\eqref{loc3}. The data, however, are overall much better described by the general bound \eqref{generalB} than by an exponential, with the norm bound of the local Hamiltonian, $||h^a_I||\leq J$, used as a fitting parameter. While this does not exclude the possibility that for larger $k$ the scaling eventually does become exponential, we can say with confidence that no indications for a strict exponential localization are observed in spin chains up to $33$ sites. These system sizes are much larger than those used in exact diagonalizations, the method which claims of a many-body localized phase have been mostly based on \cite{PalHuse,Luitz1,Luitz2,HuseNandkishore}. The stronger than exponential growth of the norm is further supported by considering the ratio of norms $s^{(2)}(k)/s^{(2)}(k-1)$ which is shown in Fig.~\ref{XXX_norm_ratios}. For the orders of the commutator which we can handle symbolically, there is no indication that this norm ratio saturates, even for extremely strong disorder, as it has to in the strictly localized case where $s(k)\sim\exp(k)$. As the figure shows, the behavior in the interacting case is in stark contrast to the localized non-interacting case.

Finally, let us also consider the norm ratio $s^{(1)}_l(k)/E[s^{(1)}_1(k)]$ of terms with support on $l$ sites which is shown in Fig.~\ref{XXX_support} for various disorder strengths. Here $E[s^{(1)}_1(k)]$ denotes the mean of terms with support on one site. Even for very strong disorder $D=25$ there is no indication of a convergence to an \ch{exponential decay with $l$ as is expected in the strictly exponentially localized case}. To the contrary, there is always a maximum which shifts to larger $l$ with increasing order $k$ just as in the case without disorder. 
\begin{figure}
    \includegraphics*[width=0.99\columnwidth]{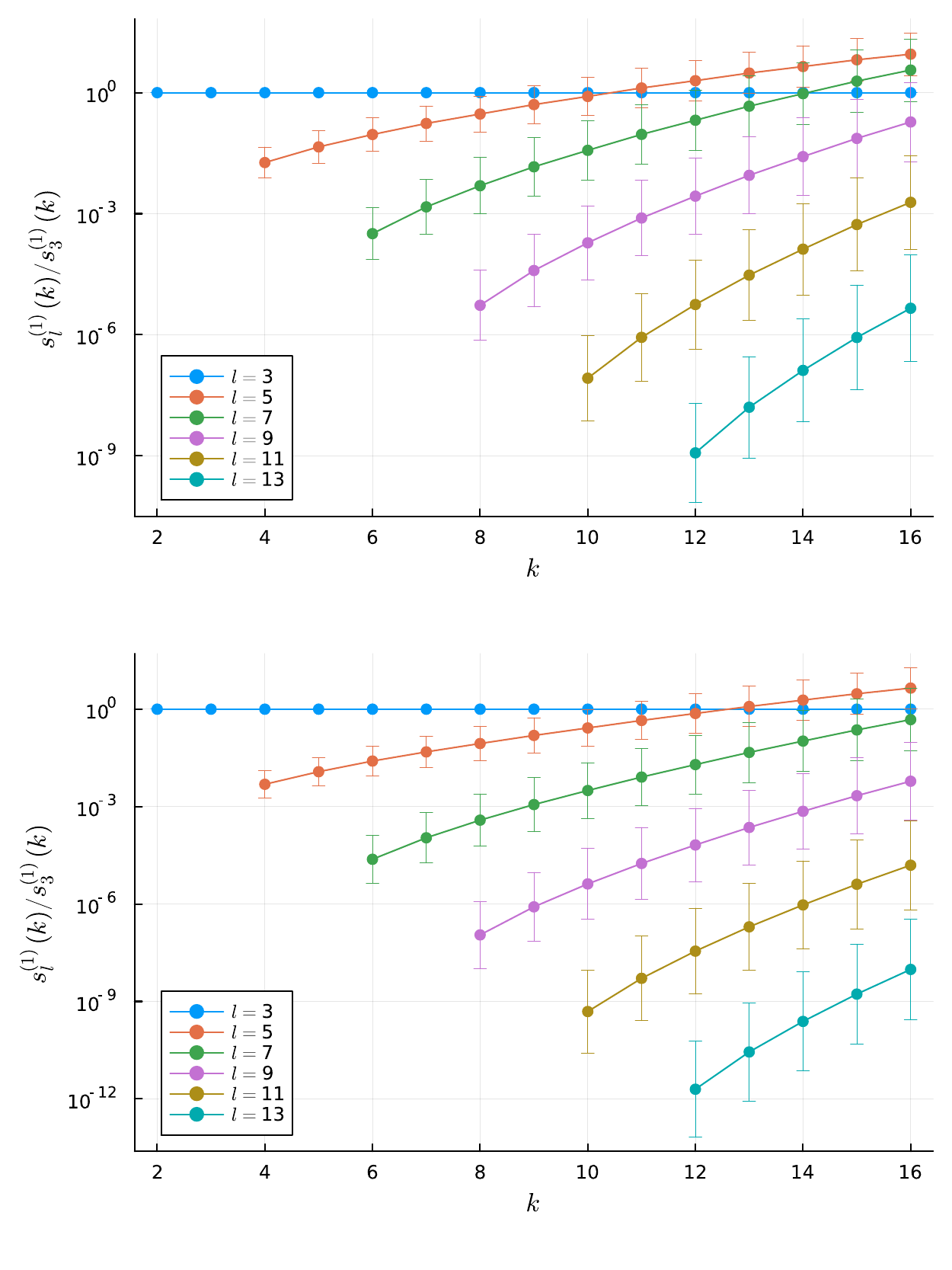}
    \caption{Norm ratio $s^{(1)}_l(k)/s^{(1)}_3(k)$ for a fixed $l$ as a function of $k$ for the Heisenberg model with $D=12.5$ (top) and $D=25$ (bottom). These results are to be compared with the non-interacting case shown in Fig.~\ref{Fig_XX_disorder_analysis}.}
    \label{Fig_XXX_disorder_analysis}
\end{figure}
To further highlight this point, we show in Fig.~\ref{Fig_XXX_disorder_analysis} the norm ratio $s^{(1)}_l(k)/s^{(1)}_3(k)$ as a function of the order $k$ for various fixed $l$. In contrast to the non-interacting case shown in Fig.~\ref{Fig_XX_disorder_analysis}, the curves do not saturate for large $k$ but rather are consistent with the weight in the total norm shifting to larger and larger $l$ with increasing $k$. While we cannot exclude that this trend changes for even larger $k$ than the ones we can handle in symbolic calculations, we want to stress once more that the terms in the commutator are spread over $33$ sites for $k=16$. This is already much larger than the system sizes typically considered in exact diagonalizations. At the very least, we can say that for chains of lengths up to $33$ sites, there is no indication that a phase with strict exponential localization is realized even at disorder strengths up to $D=25$.

\ch{This leaves open the second scenario where the Hamiltonian is unitarily equivalent to the effective Hamiltonian $\tilde H$ and where instead of a strict exponential localization, local operators spread across the lattice according to Eq.~\eqref{local_bound}. First, we note that the decay rate $\tilde\kappa_2=\tilde\kappa_2(D)$ is increasing with increasing $D$ so that for very strong disorder there should be an almost exponential decay with $l$. There is no indication for such a behavior in Fig.~\ref{XXX_support} even for $D=25$. However, it is of course possible that $D$ is still not large enough. Second, we note that Eq.~\eqref{local_growth} predicts a maximum at $l=k/\tilde\kappa_2$. This is not inconsistent with Fig.~\eqref{XXX_support} where we indeed observe an approximately linear shift of the maximum with increasing commutator order $k$. For the norm ratios shown in Fig.~\ref{Fig_XXX_disorder_analysis} the result \eqref{local_growth} predicts that $s_l(k)/s_3(k)\sim l^k$ which is also not inconsistent with the data from the symbolic calculations (note the logarithmic scale).} 

\ch{To summarize, the symbolic data are consistent with a maximal total norm growth in the disordered Heisenberg chain for all disorder strengths as in the random field Ising model. This and the data for the norms with support on $l$ sites seem to exclude the existence of a phase where operators remain strictly exponentially localized. However, the data are not inconsistent with the second scenario where $H$ is unitarily equivalent to the effective Hamiltonian $\tilde H$ describing a non-ergodic phase where local operators, however, eventually spread over the entire lattice and do not remain localized. We therefore investigate next if the unitary transform $U$ can be constructed perturbatively.}

\section{Schrieffer-Wolff transformations}
\label{Sec_SW}
\begin{figure*}
    \includegraphics*[width=0.99\textwidth]{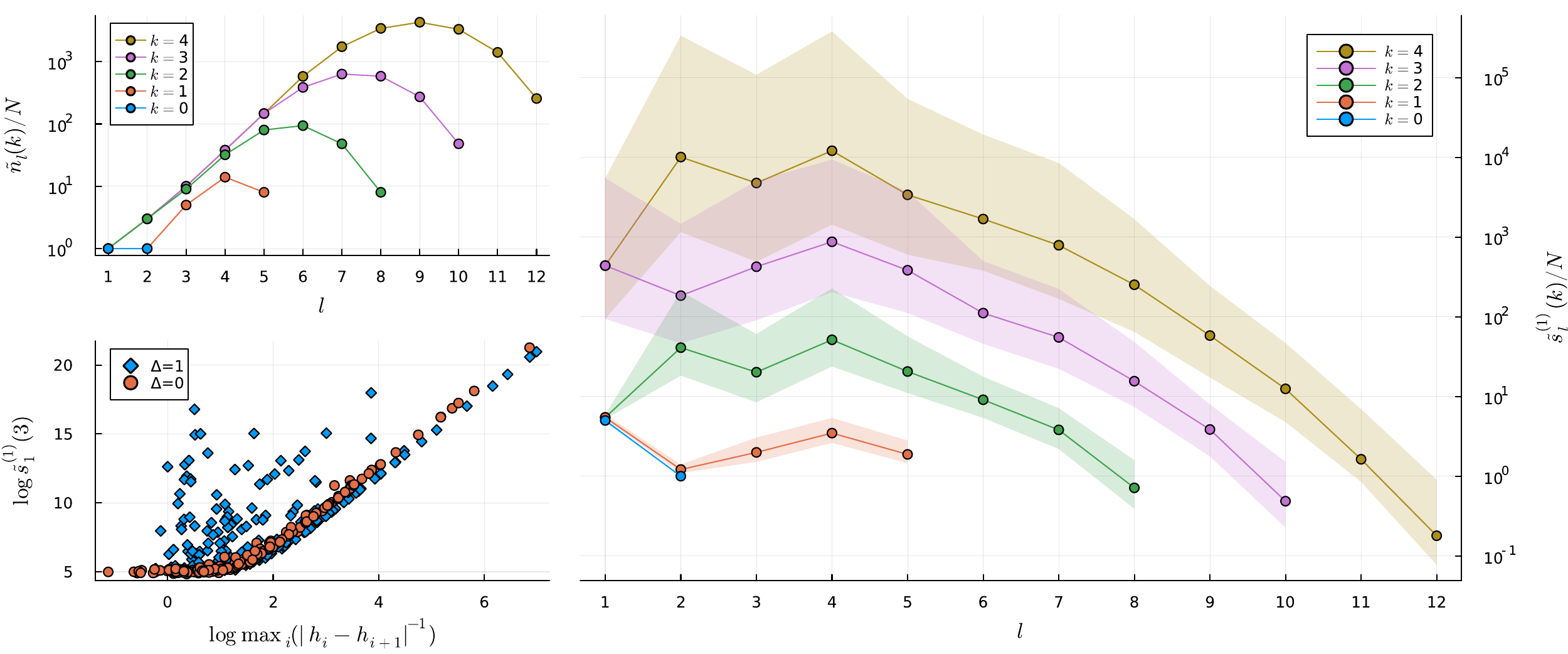}
\caption{Top left: Number of terms in the Hamiltonian with support on $l$ sites after performing the Schrieffer-Wolff transformation \eqref{SW_int1} up to order $k$ for the XXX chain with 
disorder and $N=100$ lattice sites. For large $k$, this number increases approximately exponentially with $l$. Note that $\tilde n_l(k)$ does not depend on the disorder realization; the results are exact. Bottom left: Total norm of terms with support on one site after $k=3$ transformations for $200$ samples versus $\text{max}_i|h_i-h_{i+1}|^{-1}$. Right: Total norm of terms with support on $l$ sites after $k$ transformations averaged over $200$ realizations for $N=100$ and $D=5$. While each term is expected to have an amplitude $\sim D^{-2l}$ if there are no resonances, the exponentially increasing number of terms means that the total coupling of distant sites is no longer controlled by a small parameter. For example, we have $\tilde s_1^{(1)}(k)\approx \tilde s_8^{(1)}(k)$ for $k=4$.}
\label{SW_fig}
\end{figure*} 
We have so far considered the commutator of a microscopic lattice Hamiltonian $H$ with a local operator. If this Hamiltonian is in a \ch{ strictly exponentially localized phase, then the operator will remain localized---up to exponential tails---under Euclidean time evolution. However, we have also seen that if we only demand that the Hamiltonian is unitarily equivalent to an effective Hamiltonian of (interacting) conserved charges, then this might open up the possibility for a non-ergodic phase where local operators, however, do not remain localized. Our symbolic calculations of the commutator were unable to completely rule out this scenario.} 

\ch{We therefore consider next if we can  explicitly construct the unitarily equivalent Hamiltonian $\tilde H$ written in terms of local conserved charges $\tau^z_j$ \cite{Imbrie2016,ImbrieMBL,ImbrieRosScardicchio,LuBertoni,ThomsonSchiro,SelsPolkovnikov2}}. It is important to note that these charges are not uniquely defined: if, for example, $\tau^z_j$ and $\tau^z_{j+1}$ are conserved charges localized near lattice site $j$ with $[H,\tau^z_j]=[H,\tau^z_{j+1}]=0$ then $\tau^z_j\tau^z_{j+1}$ is also local and conserved, $[H,\tau^z_j\tau^z_{j+1}]=0$. 

One way to obtain these charges is to try to perturbatively construct them in the limit of strong disorder \cite{Imbrie2016,ImbrieMBL,ImbrieRosScardicchio}. In this limit, one can think of these charges as the local operators $\sigma^z_i$ smeared out by the hopping and dressed by the interactions. As in our previous considerations, there is then again an important difference between the non-interacting and the interacting case: While in the former the one-particle operators $\sigma^z_i=c^\dagger_i c_i - 1$ remain one-particle operators, they become many-particle operators in the latter case. As we will see, the perturbative transformation of the Hamiltonian will lead us again to multiple commutators which are closely related to the ones studied in the previous sections.

The Hamiltonian of a quantum system can be diagonalized by a unitary transformation
\begin{equation}
    \label{SW}
H \to \exp(T) H \exp(-T) =\sum_{k=0}^\infty \frac{[T,H]^{(k)}}{k!} 
\end{equation}
with $T^\dagger = -T$ being an anti-unitary operator. For a Hamiltonian $H=H_0+V$ where $H_0$ is diagonal in the operator basis chosen, we can instead of a full diagonalization demand that the new Hamiltonian $H^{(1)}$ is diagonal only to first order in the perturbation $V$. This is the Schrieffer-Wolff transformation \cite{SchriefferWolff,BravyiDiVincenzo}. Going one step further, we can consider successive Schrieffer-Wolff transformations where in each step the perturbation is eliminated to leading order. This is closely related to the approach in Ref.~\cite{Imbrie2016,ImbrieMBL} which tries to generalize a proof of Anderson localization to the many-body case. In these works, the author is mostly concerned with the impact of resonances and splits the off-diagonal part of the Hamiltonian into a resonant and a non-resonant part. We will discuss the impact of resonances as well but will point out an even more fundamental issue which might call into question the entire perturbative construction of local conserved charges in the interacting case even for disorder configurations where no resonances are present.

To be specific, we consider again the XXZ Hamiltonian \eqref{Ham}. We note that in Refs.~\cite{Imbrie2016,ImbrieMBL} the transverse field Ising model with random couplings and random fields was considered instead. However, the arguments made in the following are quite general and will apply to the latter model as well \cite{Cao}. We want to investigate the case of large disorder where the magnetic field and the interaction terms constitute $H_0$ with $[H_0,\sigma^z_i]=0$ and the hopping terms are the perturbation. The condition the operator $T$ has to fulfill to lowest order is then given by
\begin{equation}
    \label{SW2}
 V + [T,H_0] = 0 
\end{equation}
and the new, transformed Hamiltonian is obtained as
\begin{equation}
    \label{SW3}
H^{(1)} = H_0 +\frac{1}{2}[T,V]+\frac{1}{3}[T,[T,V]]+\cdots \, .    
\end{equation}
In the Schrieffer-Wolff transformation, one typically stops at the first order correction to $H_0$. To justify this perturbative approach, we can rescale the Hamiltonian \eqref{Ham} by $1/D$ which changes the energy scales but leaves the physical properties unchanged. Then, the hopping terms have an amplitude $\bar{J}^{(0)}_{j,j+1}=1/D$ and the interaction terms an amplitude $\bar{\Delta}=\Delta/D$ which are both small parameters for large $D$. After this rescaling, the random magnetic fields are always given by $h_j\in[-1,1]$. 

\subsection{Non-interacting case}
First, we consider again the non-interacting case. An ansatz for the operator $T$ can be obtained by commuting $H_0$ and $V$. This leads to $T = \sum_j A_j(\sigma^+_j\sigma^-_{j+1} - \sigma^+_{j+1}\sigma^-_j)$. The amplitudes $A_j$ can then be determined from Eq.~\eqref{SW2} leading to
\begin{equation}
    \label{SW_non1}
T = \frac{1}{2}\sum_j \frac{\bar{J}^{(0)}_{j,j+1}}{h_{j}-h_{j+1}}(\sigma^+_j\sigma^-_{j+1} - \sigma^+_{j+1}\sigma^-_j) \, .     
\end{equation}
We note that the Schrieffer-Wolff operator $T$ is thus proportional to the spin current operator $\mathcal{J}=\sum_i j_i$ which can be obtained from the lattice continuity equation \cite{SirkerLectureNotes}
\begin{equation}
    \label{current}
    \partial_t \sigma^z_i =-i [\sigma^z_i,H] = - (j_i -j_{i-1}) \, .
\end{equation}
We note, furthermore, that $\overline{|h_{j+1}-h_j|}=2/3$, i.e., the factor $\bar{J}^{(0)}_{j,j+1}/|h_{j}-h_{j+1}|$ is small on average if the disorder strength $D$ is large. However, if two neighboring sites have roughly the same potential, then this factor can become large or might even diverge. Rare resonances thus might destroy the perturbative construction if they proliferate. An example how such resonances can affect the transformed Hamiltonian is shown in Fig.~\ref{SW_fig}, bottom left panel. We will come back to this point below.

After the Schrieffer-Wolff transformation, the new Hamiltonian is given by 
\begin{eqnarray}
    \label{SW_non2}
    H^{(1)} &=& H^{(0)}_0 +\frac{1}{2}[T^{(0)},V^{(0)}] +\frac{1}{3} [T^{(0)},[T^{(0)},V^{(0)}]]\nonumber \\
    &=& \sum_j \left\{\bar{J}^{(1)}_{j,j+2}(\sigma^+_j\sigma^z_{j+1}\sigma^-_{j+2} + h.c.) + 2h_j^{(1)} \sigma^z_j \right\} \nonumber \\
    &+& \sum_j \bar{J}^{(1)}_{j,j+1} (\sigma^+_j\sigma^-_{j+1}+h.c.)
\end{eqnarray}
with $H^{(0)}_0=H_0$, $V^{(0)}=V$, and $T^{(0)}=T$. The coupling constants and renormalized fields are given by 
\begin{eqnarray}
    \label{SW_non3}
    \bar{J}^{(1)}_{j,j+2} &=& \frac{(\bar{J}^{(0)}_{j,j+1})^2}{2(h_{j+1}-h_{j})}+\frac{(\bar{J}^{(0)}_{j,j+1})^2}{2(h_{j+1}-h_{j+2})} \\
    h_j^{(1)}&=& h_j^{(0)}+\frac{(\bar{J}^{(0)}_{j,j+1})^2}{4(h_j-h_{j-k})}+\frac{(\bar{J}^{(0)}_{j,j+1})^2}{4(h_{j}-h_{j+1})} \nonumber
\end{eqnarray}
and $J^{(1)}_{j,j+1}\sim 1/D^3$ (the exact expression is lengthy and not needed in the following). The first two terms correspond to those generated by $[T^{(0)},V^{(0)}]$ while the renormalized nearest-neighbor hopping $J^{(1)}_{j,j+1}$ stems from the next higher commutator $[T^{(0)},[T^{(0)},V^{(0)}]]$.  I.e., the transformation to lowest order eliminates the nearest-neighbor hopping processes but generates a next-nearest neighbor hopping $\bar{J}^{(1)}_{j,j+2}\sim 1/D^2$ (note that $\sigma^+_j\sigma^z_{j+1}\sigma^-_{j+2} + h.c. = c^\dagger_j c_{j+2} + h.c.$ in the fermionic language) and renormalizes the local fields $h^{(0)}_j\to h^{(1)}_j$. 

If we stop at the leading order of the transformation, then we obtain a Hamiltonian $H^{(1)}$ which looks like the original Hamiltonian $H^{(0)}$ but with the nearest-neighbor hopping replaced by a weaker next-nearest neighbor hopping. We can then again define a Schrieffer-Wolff transformation to eliminate these next-nearest hopping terms which has the form
\begin{equation}
    \label{SW_non4}
T^{(k)} = \frac{1}{2}\sum_j \frac{\bar{J}^{(k)}_{j,j+k+1}}{h^{(k)}_{j+k+1}-h^{(k)}_j}(\sigma^+_j\sigma^z_{j+1}\dots\sigma^z_{j+k}\sigma^-_{j+k+1} - h.c.)     
\end{equation}
with $k=1$. This will again, to leading order, renormalize the local fields and generate a third-nearest neighbor hopping which again can be eliminated to leading order by a the Schrieffer-Wolff transformation \eqref{SW_non4} with $k=2$. I.e., we can define a series of successive Schrieffer-Wolff transformations which lead to  longer and longer range interactions which are, however, suppressed as $D^{-2k}$. At the same time, they define an exponentially converging series for the renormalized random fields $h^{(k)}_j$. If there are no resonances, then we can drop the hopping term when all the amplitudes $J^{(k)}$ have fallen below some threshold, giving us an effective Hamiltonian with renormalized magnetic fields. In this approximation, which can be justified at very strong disorder and which corresponds to summing up exactly certain classes of diagrams, the conserved local charges are just the $\sigma^z_j$ themselves but with renormalized amplitudes. This can be seen as a consistency check that at very large disorder such an effective description exists because the hopping is an irrelevant perturbation. 

In the non-interacting case, it is also well understood that distance resonances have a vanishingly small probability to proliferate \cite{Anderson58,AbrahamsAnderson,AndersonLocalization}. In our construction, this scaling argument arises as follows: Two lattice sites $j$ and $j+\ell$ a distance $\ell$ apart are coupled by an effective long-range hopping $\bar{J}^{(\ell-1)}_{j,j+\ell}(c_j^\dagger c_{j+\ell} + h.c.)$ with an effective amplitude obtained by $\ell-1$ successive Schrieffer-Wolff transformations. The amplitude of this long-range effective hopping term is given by $\bar{J}^{\ell-1}\sim D^{-2\ell+2}\times f[\{(h^{\ell-1}_{j}-h^{\ell-1}_k)^{-1}\}]$ where $f$ is some function of the inverse differences of the random fields within the cluster of $\ell$ sites. Now the smallest pairwise difference on this cluster on average scales as $\ell^{-2}$. Thus the effective long-range hopping amplitude scales as $\bar{J}^{(\ell-1)}_{j,j+\ell}\sim D^{-2\ell}\ell^2$ and is thus typically exponentially suppressed. Therefore resonances have a vanishingly small probability to proliferate and to destroy the perturbative construction above. This is also supported by our symbolic calculations on finite chains of length $N$. First, we note that the size of the norm of terms with support on one site is entirely determined by how close to a nearest-neighbor resonance a given sample is. Distant resonances play no role; all the data fall onto a 'hockey stick' curve, see Fig.~\ref{SW_fig}, bottom left panel.
Second, we consider successive lowest order Schrieffer-Wolff transformations and stop the transformations if the total 1-norm of the off-diagonal part is below some threshold value $\varepsilon$. We have $\sim N$ possible off-diagonal terms of a given range and if the amplitudes of all those terms, on average, are suppressed as $D^{-2k}$ at order $k$ of the transformation then our condition to stop is $N D^{-2k}<\varepsilon$. We thus expect that we need $k\sim\ln(N/\varepsilon)$ many transformation steps. This is consistent with our observations. 

Quite generally we may ask why we expect a perturbative construction of the local conserved charges to work in the non-interacting case? In the lowest-order construction discussed above, the Hamiltonian $H^{(k)}$ always has $3N$ terms for a chain of length $N$ for every $k$. I.e., the number of terms does not increase during successive Schrieffer-Wolff transformations. If we take higher order commutators in the transformation into account, then the number of terms will increase, however, the Hamiltonian $H^{(k)}$ and the transformation $T^{(k)}$ will always consist of a sum of one-body operators and, for a chain of length $N$, there are only $N^2$ distinct one-body operators. Therefore the problem of finding local conserved charges is at most quadratic, and, in any finite order approximation, one in fact always deals with a constant, non-increasing number of terms in the transformed Hamiltonians.  

\subsection{Interacting case}
The interacting case, on the other hand, is completely different and none of the above arguments hold. We are, however, able to generalize Eq.~\eqref{SW_non1} and to obtain the exact Schrieffer-Wolff transformation for the XXZ chain and arbitrary interaction strength $\Delta$. We find
\begin{eqnarray}
    \label{SW_int1}
    T&=&\sum_j\frac{\bar{J}^{(0)}_{j,j+1}(\sigma^+_j\sigma^-_{j+1}-h.c.)}{2(h_j-h_{j+1})(2\bar{\Delta}^2-2(h_j-h_{j+1})^2)} \nonumber \\
    &\times& [\bar{\Delta}^2 - 2(h_j-h_{j+1})^2+ \bar{\Delta}^2 \sigma^z_{j-1}\sigma^z_{j+2}\nonumber \\
    && +\bar{\Delta}(h_j-h_{j+1})(\sigma^z_{j-1}-\sigma^z_{j+2})] 
\end{eqnarray}
with $\bar{J}^{(0)}_{j,j+1}=1/D$ and $\bar{\Delta}=\Delta/D$. We note that for $\Delta\neq 0$, the transformation $T$ consists of $8N$ distinct operators for a chain of length $N$, including operators acting on $2,3$ and $4$ neighboring sites.

If we stop the transformation \eqref{SW3} at first order, the transformed Hamiltonian $H^{(1)}$ already has $29N$ distinct operators if the chain length $N$ is sufficiently large. This includes terms acting on $5$ neighboring sites but also terms such as $\sigma^z_j\sigma^z_{j+1}$ and $\sigma^z_j\sigma^z_{j+2}\sigma^z_{j+3}$. Even for very strong disorder, the transformation thus does not just lead to a renormalization of the random fields but necessarily leads to a broadening of the diagonal basis. Symbolically, we can also construct $T^{(1)}$ to eliminate the off-diagonal terms in $H^{(1)}$. We find that $T^{(1)}$ already consists of $786 N$ distinct operators. The transformed Hamiltonian $H^{(2)}$ then has $30618 N$ many terms. I.e., even if we stop the Schrieffer-Wolff transformation at lowest order, the number of terms in $T^{(k)}$ and $H^{(k)}$ is growing exponentially. Furthermore, there is no distinct set of diagrams which we can simply sum up to all orders. Exponentially many new diagrams are created at each order.

The related and most fundamental problem, however, is that lattice sites $j$ and $j+\ell$ a distance $\ell$ apart will be coupled by $\sim 3^2 4^{\ell-1}$ {\it distinct} operators because on each site we can put in principle one of the operators $\{\sigma^x,\sigma^y,\sigma^z,\mathbb{I}\}$ but the string of operators of length $\ell+1$ has to start and end with one of the three Pauli matrices. While not all of these combinations of operators will be realized in a microscopic model such as the XXX chain, there will be exponentially many distinct operators which couple these two sites, see Fig.~\ref{SW_fig}.  Thus there is, in general, no reason to expect this perturbative construction to work because the overall effective amplitude of terms coupling these two sites will be $\bar{J}_{j,j+\ell}\sim D^{-\ell}4^\ell\sim\mathcal{O}(1)$. I.e., there is not necessarily a small parameter anymore. This reasoning is supported by the results of symbolic calculations shown in Fig.~\ref{SW_fig}. Here we have calculated the sum $\tilde s_\ell^{(1)}(k)$ of the amplitudes of all terms obtained in the Schrieffer-Wolff transformation \eqref{SW_int1} to order $k$ which have support on $\ell$ sites, i.e., which connect sites a distance $\ell-1$ apart. For a periodic chain with even length $N$, there are $\sim N$ distinct pairs of sites which are a distance $1,\cdots,N/2$ apart. I.e., for a specific pair the overall coupling is on average $\sim \tilde s_\ell^{(1)}(k)/N$ and, as can be seen in Fig.~\ref{SW_fig}, is in general not exponentially decreasing with $\ell$. That the number of terms generated in such a transformation with support on $\ell$ sites is exponentially increasing with $\ell$ makes it very questionable that a construction of local conserved charges in this perturbative manner is a controlled approximation. This is a major issue even before addressing the issue of resonances and thermal inclusions (regions for a given disorder configuration which, by chance, are only weakly disordered). It seems to us that the exponential increase of the number of terms in such transformations is a point which has not been properly addressed in the proof of MBL in Refs.~\cite{Imbrie2016,ImbrieMBL} where the author seems to be only concerned about the effect of resonances. While it is not impossible that for disorder configurations without resonances and thermal inclusions the perturbative construction of local charges is working in the thermodynamic limit for very strong disorder, any proof needs to provide rigorous arguments that the exponentially many terms do not overwhelm the exponential small parameter in the expansion. Such arguments seem to be missing so far. We note that for a finite system the construction will always eventually converge because the finite length provides a cutoff and $\tilde s_\ell^{(1)}(k)$ will eventually drop off exponentially with $\ell$.

The destabilizing effects of resonances and thermal inclusions have received considerable attention in the literature already \cite{MorningstarColmenarez,HaMorningstar,Sels2022}. From the perspective of perturbative Schrieffer-Wolff transformations, they lead to the following additional problems: The renormalized couplings for an operator acting on $\ell$ lattice sites will in its denominator typically contain a linear combination of {\it all} the random fields on these $\ell$ sites. Given that there are exponentially many such operators, the simple scaling argument for the non-proliferation of resonances in the Anderson case is clearly no longer applicable in the interacting case. This is confirmed by the symbolic calculations shown in Fig.~\ref{SW_fig}, bottom left panel, where the sum of the amplitudes of terms in the Hamiltonian with support on one site, $\tilde s_1(k)$, in the interacting case is large even for samples which do not have nearest-neighbor resonances. Thermal inclusions---regions of the lattice which have similar potentials---lead, furthermore to an additional destabilization of the perturbative transformation. This can be seen, for example, in symbolic calculations of the unitary transformation \eqref{SW} with the operator \eqref{SW_int1} to higher orders. In configurations where such regions exist, the transformation generates long-range terms with large amplitudes even if the disorder $D$ is extremely strong. If, on the other hand, we exclude configurations with such regions and with any pairwise resonances by hand, then the amplitudes of any individual operator typically decay with the length of its support.  Another manifestation of this issue can be seen when performing $k$ successive lowest order Schrieffer-Wolff transformations for a finite chain of length $N$. We want to stop the transformations again once the 1-norm of the off-diagonal part is below some threshold $\varepsilon$. In contrast to the non-interacting case, we now have exponentially many terms connecting sites a given distance apart. If we assume that individual amplitudes in the off-diagonal part are suppressed as $D^{-2k}$, then the condition for stopping the transformations becomes $\e^N D^{-2k}<\varepsilon$ and we would thus expect that we need $k\sim N$ steps for $N$ sufficiently large. This is, however, not what we observe for typical disorder configurations. The results of our symbolic calculations point instead to an increase of the number of transformations $k\sim\e^N$. This implies that some amplitudes in the off-diagonal part are $\sim\mathcal{O}(1)$ and that we have to eliminate almost all of the off-diagonal terms to achieve convergence. 

In summary, the iterative transformation of the Hamiltonian cannot simply be transferred from the non-interacting to the interacting case. There are two main issues: Even if one considers only a lowest order transformation, one is not dealing with a constant number of operators but rather with an exponentially increasing number of operators, both in the Schrieffer-Wolff transformation $T^{(k)}$ as well as in the transformed Hamiltonian $H^{(k)}$. Even if one tries to only perform a single unitary transformation \eqref{SW} with the operator $T$, the number of distinct operators in the transformed Hamiltonian is increasing exponentially with the support of the operator which thus can overwhelm the exponential decay of the amplitude of individual operators with the length of their support. We note that different but related arguments for the failure of a perturbative construction of conserved charges in the interacting case have recently been made in Ref.~\cite{SelsPolkolnikovPRX}. Furthermore, regions with similar lattice potentials result in individual operators with large support which, nevertheless, have large amplitudes even for very strong disorder. The latter issue is related to the well-known avalanche instability \cite{MorningstarColmenarez,HaMorningstar,PeacockSels,Sels2022}.

\section{Conclusions}
We have considered operator growth in one-dimensional lattice models and its connection to localization from two different perspectives. First, we studied the multiple commutator $[H,A]^{(k)}$, which appears \ch{both in real-time and Euclidean time evolution}, for a local operator $A$ and a Hamiltonian $H$ which is a sum of nearest-neighbor terms. For a generic, ergodic system, we have shown that the number of distinct terms generated is exponentially increasing with the order $k$ of the commutator. This leads to a strict bound for the commutator norm $s(k)$ in nearest-neighbor models which grows \ch{almost factorially} with $k$, a result which was first demonstrated in Ref.~\cite{AvdoshkinDymarsky}. Since this result is based on a simple counting of connected clusters generated by the commutator, it is expected that this bound is asymptotically tight for an ergodic system. This {\it operator growth hypothesis} is challenged by recent results for systems such as the random field Ising chain which are supposed to have a localized phase but, nevertheless, always show almost factorial operator norm growth \ch{independent of the strength of the disorder}. Here we have systematically addressed this issue. We started by \ch{analytically obtaining strict upper bounds for the operator norm for free fermion and for localized systems. For a free fermion system}, all the terms generated by the commutator of the Hamiltonian $H$ with a one-body operator $A$ will also always be one-body operators. For a short-range Hamiltonian, the spatial range of operators generated at order $k$ will be proportional to $k$. The number of distinct terms in the commutator can therefore only grow $\sim k^2$ as compared to the exponential growth in the general, interacting case. This leads to a rigorous exponential upper bound for the commutator norm in this case. Second, we considered the case of \ch{strict exponential localization where we demanded that a local operator $A$ remains exponentially localized when commuted with $H$ up to exponential tails. We proved that in this case a strict upper exponential norm bound does exist as well. The almost factorial norm growth proven for the Ising model with random fields for all disorder strengths and strongly suggested for the random field Heisenberg model by symbolic calculations in this work is thus {\it inconsistent with a strict exponential localization of operators.} We then considered the alternative definition of localization based on the unitary transformation $U$ of a microscopic Hamiltonian $H$ to an effective Hamiltonian $\tilde H$ which has an infinite set of local conserved charges. Importantly, local operators in $H$ are mapped by $U$ onto quasi-local operators. For the non-interacting case we proved that this definition is equivalent to the strict exponential localization of local operators. The contribution of terms in the commutator with support on $l$ sites to the norm does scale as $s_l(k)\sim \exp(k)\exp(-l)$. For the interacting case, however, we were only able to prove a weaker bound $s_l(k)\sim l^k\exp(-l)$ which is consistent with an almost factorial growth of the total norm. Furthermore, this bound means that for very large $k$, $s_l(k)$ grows in exactly the same way as in a generic ergodic system. The question then becomes whether the transformation $U$ exists and models such as the random field Ising and Heisenberg models have a non-ergodic phase in which, however, local operators eventually completely delocalize or if the transformation $U$ simply does not exist. We have also shown that even if the quasi-local unitary $U$ does exist, the spreading of local operators through the entire lattice is generically expected to lead to transport of globally conserved charges so that such phases, if they do exist, should not be called many-body localized.}

\ch{To further study the existence of the transformation $U$}, we considered unitary Schrieffer-Wolff transformations of the XXZ chain to perturbatively eliminate the hopping terms and construct $U$. We showed that such an approach will succeed in the Anderson case because the Hamiltonian in this case consists of one-body operators and this property is not changed by unitary transformations. Consequently, there are only two terms possible in the effective Hamiltonian which can couple two sites $j$ and $k$: $c^\dagger_j c_k$ and $c_j c_k^\dagger$. This allows to define consecutive lowest order Schrieffer-Wolff transformations and to sum up the entire series. In this approximation, valid at very strong disorder, the local conserved charges are simply the $\sigma^z_j$ operators with renormalized fields. The longer-range hoppings generated by the transformations have amplitudes which decay exponentially with the range of the hopping process. The minimum gap between random fields on a cluster of length $\ell$ on the other hand---which enters the effective hopping in the denominator---scales as $\ell^{-2}$ and thus cannot overcome the general suppression of the hopping process which scales as $D^{-2\ell}$. This means, in particular, that resonances cannot proliferate.

All of this changes drastically once interactions are included. In particular, there are then $\sim 4^{\ell}$ different terms which can connect sites a distance $\ell$ apart. We confirm this exponential growth with the length of the support by symbolic calculations. A priori, there is thus no reason to believe that a perturbative construction of the local conserved charges by unitary transformations is still possible in the interacting case. Even if we assume that there are no resonances or thermal inclusions (regions with random fields of similar strengths), there is potentially a fundamental problem with this construction: while the amplitude of an individual term connecting sites a distance $\ell$ apart will still be $\sim D^{-2\ell}$, there are exponentially many such terms so that the total effective amplitude is $\sim 4^\ell D^{-2\ell}\sim \mathcal{O}(1)$. The small parameter in the perturbative construction is in general lost even before considering the issue of resonances. \ch{While this simple scaling argument does not preclude that the perturbative construction in the absence of resonances can succeed for very strong disorder such that $4^\ell D^{-2\ell}\ll 1$} our symbolic calculations show that the total effective amplitude is not decaying with the support of the operator in contrast to the non-interacting case. As far as we can tell, this issue is not addressed in the proof of MBL by Imbrie \cite{Imbrie2016,ImbrieMBL} which uses the same kind of unitary transformations but seems to be only concerned about resonances. The latter is also a much more severe problem in the interacting case. Even if we assume that the perturbative construction is convergent without resonances and thermal inclusions, many-body resonances will destroy the approach when present. A long-range term typically involves operators on all sites between the left most and the right most site. Consequently, the renormalization will involve linear combinations of all the random fields on this cluster and there are exponentially many of such terms. Clearly, the argument for the non-proliferation of resonances made in the non-interacting case is then no longer valid. At the very least, we believe that we have shown that \ch{there are certain aspects of the perturbative proof in Refs.~\cite{Imbrie2016,ImbrieMBL} which would need to be addressed to make it more convincing}. \ch{In particular, the discussion of the construction seems to ignore that, in general, the number of terms connecting distant sites grows exponential with the distance between the sites. It is not obvious why the construction remains controlled by a small parameter despite the exponential number of terms and why resonances do not proliferate.} Our symbolic calculations for finite chains and finite orders of the transformation suggest that such an argument, in fact, cannot be made and that a perturbative construction of local conserved charges in this way is impossible except for $D=\infty$ where the model is trivially localized. We therefore conclude that we have found strong evidence that the unitary $U$ does not exist and that the almost factorial growth of the operator norm for the random field Ising and Heisenberg models indicates that these models are ergodic in the thermodynamic limit for all disorder strengths, consistent with the operator growth hypothesis. \ch{In a generic scenario, this implies that these systems in the thermodynamic limit and at very long times will ultimately exhibit finite transport coefficients.}

\ch{As an outlook, we note that under the assumption that the unitary transformation $U$ exists, we have only been able to prove the bound $s_l(k)\sim l^k\exp(-l)$ which is consistent with an almost factorial growth of the total norm, $s(k)\sim(k/\ln k)^k$. However, we have not used any specific properties of a microscopic model such as the random Ising or Heisenberg model so it might be possible that a stricter bound for $s_l(k)$ can be derived for special cases which leads to a parametrically slower growth than $s(k)\sim(k/\ln k)^k$ and is thus in direct contradiction with the rigorous result for the norm growth from Ref.~\cite{Cao}. This seems to be a promising direction for future research.}

\acknowledgments
J.S. acknowledges discussions with D. Sels and M. Kiefer-Emmanouilidis and funding by the NSERC Discovery program and by the DFG via the research unit FOR 2316.

\newpage
\clearpage
\begin{widetext}


\section*{Supplementary material}

and in fact, $S(6,4) = 65$.

We now denote by $\tilde{S}(k,j)$ our 'adjusted' Stirling number which inserts the requirement that contributions from sites of increasing distance from our initial site are exponentially decaying. Let $k_n$ denote the number of bonds at a distance $n$ from our initial site, where a distance of zero corresponds to a bond involving our initial site. We want exponential decay of the ratio of the contributions from bonds at $k_n$ to bonds at $k_0$. To insert this condition, we multiply by a factor of $e^{-nk_n/\xi}$ for each $k_n$ and for any constant $\xi > 0$ which plays the role of a localization length. We note that this is normalized so that at $k_0$ this factor is 1, meaning the initial site is not reduced. This separate treatment of the $k_n$'s, which represent the number of times each bond appears in the connected cluster, is why the form (A.1) of $S(k,j)$ is desirable. Applying the decaying factors, we get

$$
\begin{aligned}
&\tilde{S}(k,j) \\
&= \sum_{k_0+\ldots+k_{j-1}=k} \frac{1}{j!} \prod_i \frac{1}{a_i!} \binom{k}{k_0,\ldots,k_{j-1}} \prod_{n=0}^{j-1} e^{-nk_n/\xi} \\
&\leq \sum_{k_0+\ldots+k_{j-1}=k} \binom{k}{k_0,\ldots,k_{j-1}} \prod_{n=0}^{j-1} \frac{e^{-nk_n/\xi}}{n+1} \\
&= e^{k/\xi} \left( \sum_{n=1}^j \frac{e^{-n/\xi}}{n} \right)^k .
\end{aligned}
\tag{A.4}
$$

We also loosened the condition on the $k_i$'s in the second line so that they are now greater than or equal to 0 instead of 1 so that we may apply the multinomial theorem in the third line. Since this may only increase the sum due to allowing more solutions to $k_0 + \ldots + k_{j-1} = k$, the upper bound is still valid. For $k \gg j \gg 1$ we can extend the sum up to infinity and obtain

$$
\tilde{S}(k,j) \leq \left( -e^{1/\xi} \ln\left(1 - e^{-1/\xi}\right) \right)^k .
\tag{A.5}
$$

Then the adjusted number of connected clusters, with the inserted requirement of the exponentially decreasing contribution with cluster length, is bounded from above by

$$
\begin{aligned}
\sideset{}{'}\sum_{\{I_1,\cdots,I_k\}} &= \sum_{j=1}^k 2^j \tilde{S}(k,j) \\
&\leq 2(2^k - 1) \left( -e^{1/\xi} \ln\left(1 - e^{-1/\xi}\right) \right)^k \\
&\approx 2 \left( -2e^{1/\xi} \ln\left(1 - e^{-1/\xi}\right) \right)^k
\end{aligned}
\tag{A.6}
$$

where in the last line we took the large $k$ limit. Defining the constant $\alpha = -2e^{1/\xi} \ln\left(1 - e^{-1/\xi}\right)$, we therefore obtain the bound

$$
||[H,A]^{(k)}|| \leq 2||A||(2\alpha\tilde{J})^k .
\tag{A.7}
$$

We note that this more careful calculation is qualitatively fully consistent with Eq. (II.21) in the main text where we made the simplifying assumption that clusters longer than the localization length do not contribute at all to the norm. We thus conclude that the norm growth of the commutator in any localized model is at most exponential.

## Appendix B: Lieb-Robinson bounds

Another physically relevant property which can be obtained from our results for the $k$-fold commutator are Lieb-Robinson bounds for the spreading of information in Euclidean time.

### 1. General bound

We start by summarizing the result for the generic, quantum chaotic case which was first derived in Ref. [30]. We consider the commutator between two local operators a distance $\ell$ apart

$$
\begin{aligned}
||[A_0(\tau), B_\ell(0)]|| &= ||[\sum_k [H, A_0]^{(k)} \frac{\tau^k}{k!}, B_\ell]|| \\
&\leq 2||A||\,||B|| \sum_k \frac{(2CJ\tau)^k}{k!} \sum_{j=\ell}^k 2^j S(k,j)
\end{aligned}
\tag{B.1}
$$

where we have used the fact that the lattice animal has to consist of at least $j = \ell$ bonds for the commutator to be non-zero. We can now use the representation of the Stirling numbers of the second kind

$$
S(k,j) = \sum_{s=1}^j \frac{(-1)^{j-s} s^{k-1}}{(j-s)!(s-1)!}
\tag{B.2}
$$

to rewrite the bound of the commutator as

$$
2||A||\,||B|| \sum_{j=\ell}^\infty 2^j \sum_{s=1}^j \frac{(-1)^{j-s}}{s(j-s)!(s-1)!} \underbrace{\sum_{k=1}^\infty \frac{(2CJ\tau s)^k}{k!}}_{e^{2CJ\tau s}-1}
$$

$$
= 2||A||\,||B|| \sum_{j=\ell}^\infty \frac{(2q)^j}{j!}, \quad q = e^{2CJ\tau} - 1
\tag{B.3}
$$

$$
= 2||A||\,||B|| \frac{e^{2q}}{(l-1)!} \int_0^{2q} x^{\ell-1} e^{-x} dx .
$$

For $\ell \gg q$ we obtain the scaling

$$
\begin{aligned}
||[A_0(\tau), B_\ell(0)]|| &\sim 2||A||\,||B|| \frac{(2q)^l}{l!} \\
&\sim 2||A||\,||B||(2q)^\ell \left(\frac{e}{\ell}\right)^\ell .
\end{aligned}
\tag{B.4}
$$

This means that the commutator is $\mathcal{O}(1)$ if $q \sim \ell$ which is equivalent to $\exp(2CJ\tau) \sim \ell$. I.e., in a generic, chaotic

one-dimensional quantum system with short-range interactions, information is spreading exponentially fast in Euclidean time. From the derivation we see that this exponentially fast spreading is directly related to the almost factorial growth of the commutator $[H, A]^{(k)}$.

### 2. Non-interacting bound

Let us first consider the case without disorder. In this case we have an explicit result for the contribution of lattice animals with $j$ bonds, see Eq. (III.11). Generalizing this to the case of an operator $A$, we find

$$||[A_0(\tau), B_\ell(0)]|| \leq 2||A||\,||B|| \sum_{k=\ell-1}^{\infty} \frac{(4\tau)^k}{k!} \sum_{j=\ell}^{k+1} \binom{k}{\frac{k+1-j}{2}}$$

$$= 2||A||\,||B|| \sum_{k=\ell-1}^{\infty} \frac{(4\tau)^k}{k!} \sum_{j=0}^{(k+1-\ell)/2)} \binom{k}{j} \qquad (B.5)$$

for the case where $k$ is odd, and $\ell$ is even. This becomes non-zero if $k \gtrsim \ell$ and is of order $\mathcal{O}(1)$ if $\tau \sim \ell$. I.e., for a non-interacting system the spreading is ballistic in Euclidean time. This is directly related to the exponential instead of almost factorial growth of $[H, A]^{(k)}$ which is due to the fact that most lattice animals are zero.

Importantly, this scaling remains intact in the non-interacting disordered case. I.e., the spreading in Euclidean time is still ballistic but the time scale is renormalized. This can be seen by considering

$$||[A_0(\tau), B_\ell(0)]|| \leq 2||A||\,||B|| \sum_{k=\ell-1}^{\infty} \frac{(\tau)^k}{k!} \sum_{j=\ell}^{k+1} \underbrace{||[H, A_0]^{(k)}||}_{2^{2j-1}\eta^{k-j+1}}$$

$$(B.6)$$

where we have used that $s_j^{(1)}(k) = ||[H, A_0]_j^{(k)}|| \sim \theta(k - (j-1))2^{2j-1}\eta^{k-j+1}$. I.e., independent of the disorder we always have that $s_j^{(1)}(j-1) = 2^{2j-1}$, see Eq. (III.13). Only the further increase of $s_j^{(1)}(k)$ for $k \geq j-1$ is affected by disorder, $\eta = \eta(D)$. However, this does not affect the scaling when the commutator becomes of order $\mathcal{O}(1)$ which is still given by $\tau \sim \ell$.

### 3. Many-body localization

In the interacting, disordered case all the paths which exist in the non-disordered case remain present, see Fig. 13. This suggests that information will still spread exponentially fast, $e^\tau \sim \ell$, in Euclidean time albeit with a renormalized time scale. I.e., based on the view of the commutator $[H, A_0]^{(k)}$ as a graph we expect that while the spreading of information does slow down with increasing disorder, this is merely a rescaling of time but the functional form remains unaffected.

The only way how the spreading of information can change qualitatively is if disorder introduces destructive

interferences between different paths in the graph 13. We note again that this is not possible in the transverse Ising model with random fields because the model is stoquastic and all the weights along the edges in the graph can be taken to be positive.

The XXZ chain does not have the stoquastic property so that, in principle, destructive interferences are possible. If we assume this to be the case, then a minimal model is the one discussed in Sec. A where the Stirling numbers of the second kind are replaced by modified Stirling numbers, see Eq. (A.4). I.e., our Lieb-Robinson bound will then be given by Eq. (B.1) with $S(k, j) \to \tilde{S}(k, j)$. We can then obtain a bound for information spreading by using

$$\tilde{S}(k, j) \leq (-e^{1/\xi} \ln\left(1 - e^{-1/\xi}\right))^k \equiv \alpha^k. \qquad (B.7)$$

Now the two sums can be performed and we find

$$||[A_0(\tau), B_\ell(0)]|| \leq \frac{4||A||\,||B||}{(\ell-1)!} e^{4CJ\alpha\tau} \qquad (B.8)$$

$$\times \left(\gamma(\ell, 4CJ\alpha\tau) - 2^{\ell-1}e^{-2CJ\alpha\tau}\gamma(\ell, 2CJ\alpha\tau)\right)$$

where $\gamma(\ell, x)$ is the lower incomplete Gamma function. For large $\ell$ we can approximate the Gamma function and find

$$||[A_0(\tau), B_\ell(0)]|| \lesssim \frac{2||A||\,||B||}{\ell!}(4CJ\alpha\tau)^\ell \qquad (B.9)$$

Using the Stirling approximation this means that the commutator is of order $\mathcal{O}(1)$ if $\tau \sim \ell/(4CJ\alpha)$. This means that the spreading is ballistic and since $\alpha = \alpha(\xi)$ increases with increasing localization length $\xi$, the velocity decreases with increasing disorder as expected. We stress once more though that we have not found any indications for such destructive interference—which would result in the modified Stirling numbers $\tilde{S}(k, j)$—to be present in the Heisenberg chain.

To summarize, an almost factorial growth of the commutator $[H, A_0]^{(k)}$, as expected in a generic quantum chaotic system, leads to exponential spreading while an exponential growth leads to ballistic spreading of information in Euclidean time.

## Appendix C: Off-diagonal matrix elements

We can also use our bounds on the growth of the commutator $[H, A]^{(k)}$ for a local operator $A$ to derive results for the asymptotic scaling of off-diagonal matrix elements of $A$ in the energy eigenbasis for a finite system.

We have

$$\langle E_i|A(\tau)|E_j\rangle = A_{ij}e^{\omega\tau} \qquad (C.1)$$

with $\omega = E_i - E_j$ and $A_{ij} = \langle E_i|A|E_j\rangle$ where $|E_j\rangle$ are energy eigenstates. We note that $|A_{ij}| \leq \sqrt{\text{tr}\,\mathbb{1}_{M \times M}}||A||_p \equiv ||A||$ where $||A||_p$ with $p = 1, 2$ are the norms defined in Eq. (II.4). This implies that

$$|A_{ij}| \leq ||A(\tau)|| e^{-\omega\tau} \qquad (C.2)$$

which is valid for any time $\tau$. We can thus minimize the r.h.s. with respect to $\tau$ to obtain an optimal bound.

### 1. General case

The general, quantum chaotic case was already analyzed in Ref. [30]. Here we provide a short summary. From Eq. (B.3) we see that

$$||A(\tau)|| \leq ||A|| e^{2q}, \quad q = e^{2CJ\tau} - 1. \qquad (C.3)$$

Minimizing the function $e^{2q}e^{-\omega\tau}$ w.r.t. $\tau$ one finds

$$|A_{ij}| \leq ||A|| \begin{cases} \exp(2\tilde{\omega}(1 - \ln \tilde{\omega}) - 2) & \tilde{\omega} > 1 \\ 1 & \tilde{\omega} \leq 1 \end{cases} \qquad (C.4)$$

with $\tilde{\omega} = \omega/(4CJ)$. This means that matrix elements are in general non-zero even for extensive energy differences $\omega$. Asymptotically, these matrix elements decay faster than exponential.

### 2. Non-interacting case

In the non-interacting case, energy differences $\omega$ cannot be extensive. A more precise bound can be obtained by using again Eq. (C.2) together with the bound

$$||A(\tau)|| \leq \sum_k \frac{\tau^k}{k!} \underbrace{||[H, A]^{(k)}||}_{\leq ||A||(8CJ)^k} \qquad (C.5)$$

where we have used Eq. (II.17). This leads to the upper bound

$$|A_{ij}| \leq ||A|| \exp[(8cJ - \omega)\tau]. \qquad (C.6)$$

Minimizing the r.h.s. we thus find $|A_{ij}| = 0$ if $\omega > 8CJ$ and the generic bound $|A_{ij}| \leq ||A||$ if $\omega \leq 8CJ$.

In the disordered, non-interacting case a sharp cut-off will remain. To be concrete, we can for example consider the $\sigma^z$ operator in the Anderson case. Using Eq. (C.5) together with the bound (III.15) we find

$$|\sigma_{ij}^z| = 0, \qquad \omega > 8(D+1) \qquad (C.7)$$
$$|\sigma_{ij}^z| \leq \frac{1}{D+1}, \qquad \omega \leq 8(D+1)$$

In addition to the sharp cut-off in energy, we also find a non-trivial bound for the magnitude of the matrix elements below that cutoff in this case which is consistent with the fact that the eigenstates of the Anderson model are eigenstates of $\sigma^z$ for $D \to \infty$. Off-diagonal matrix elements therefore vanish entirely in this limit.

### 3. Many-body localization

As for the Lieb-Robinson bounds, the general result (C.4) will still apply qualitatively as long as $[H, A]^{(k)}$ does grow almost factorially. We know this to be the case for the transverse Ising model with random fields and found strong evidence that this is also true in the disordered Heisenberg model.

We can, however, check again what would happen if there are destructive interferences between different paths in the commutator graph, leading to our model with the modified Stirling numbers $\tilde{S}(k, j)$, see Eq. (A.6). Using the bound (C.5) with $||[H, A]^{(k)}|| \leq (2CJ)^k ||A|| \sum_{I_1, \dots, I_k}' \leq 2(4CJ\alpha)^k ||A||$ with $\alpha = -e^{1/\xi} \ln(1 - e^{-1/\xi})$ we find

$$||A(\tau)|| \leq 2||A|| \exp(4CJ\alpha\tau) \qquad (C.8)$$

This then implies

$$|A_{ij}| = 0, \qquad \omega > 4CJ\alpha \qquad (C.9)$$
$$|A_{ij}| \leq ||A||, \qquad \omega \leq 4CJ\alpha.$$

It is instructive to consider the limits of weak and strong disorder. For weak disorder, the norm of the local Hamiltonian scales as $||h|| \sim CJ \sim 1$ if we set the exchange couplings to be of order $\mathcal{O}(1)$. For the parameter $\alpha$ we find that $\alpha \sim \ln \xi$ for $\xi \gg 1$. This means that there is no cut-off for $\xi \to \infty$; extensive energy differences are allowed, qualitatively consistent with the general bound (C.4). For very strong disorder, on the other hand, $||h|| \sim CJ \sim D$ where $D$ is the disorder strength. The parameter $\alpha$ for $\xi \ll 1$ becomes $\alpha \sim 1$. For very strong disorder in a finite system the prediction of this model is therefore that a sharp cut-off of order $4D$ exists. This is consistent with the infinite disorder limit where the eigenstates are Ising states and local operators can at most couple states which differ by a local spin flip.

To summarize, an almost factorial growth of the commutator $[H, A]^{(k)}$ implies that matrix elements are non-zero even for extensive energy differences but that they decay faster than exponential with the magnitude of the energy difference. For an exponential growth, on the other hand, there is a sharp cut-off and matrix elements for states with an energy difference larger than this cut-off are zero.

### Appendix D: Graphs

The growth of the 1-norm of the commutator $[H, \sigma_0^z]^{(k)}$ can also be understood using graphs both without and with disorder. Here we provide some more details how such graphs can be constructed and how they have to be read.

### 1. No disorder

We use a rooted, directed, weighted, infinite graph to model the commutator growth. In the graph, each node represents classes of terms formed by the commutator and each directed edge represents connections between classes of terms formed via commutation with the Hamiltonian. For example, if commuting a term of class $A$ with $H$ produces a term of class $B$ then an arrow is drawn on the graph from $A$ to $B$. Further, the 1-norm of terms of class $B$ in that commutation is assigned to the weight of that edge. The classes are organized such that each term of that class produces the same result when commuting with $H$.

In the non-interacting, non-disordered case, whose exact commutator structure is given by Eq. (III.4) and Eq. (III.5), we find that the growth is modelled by the left column of the graph of Fig. 4. By convention, we label the classes by the string of Pauli operators beginning with $\sigma^x$, and we omit index labels with the understanding that a string of operators acts on consecutive sites. Note that terms are invariant with respect to commutation with $H$ under the transformation $\sigma^x \leftrightarrow \sigma^y$, so we group terms of the form $\sigma^x \cdots \sigma^y$ and $\sigma^y \cdots \sigma^x$ into the same class, and $\sigma^x \cdots \sigma^x$ and $\sigma^y \cdots \sigma^y$, but not $\sigma^x \cdots \sigma^y$ and $\sigma^x \cdots \sigma^x$.

Enumerative calculations on the graph Fig. 4 now allow us to calculate the 1-norm growth and spatial structure of the commutator $[H, \sigma_0^z]^{(k)}$. We always begin with a single $\sigma^z$ operator which we label node 1. We then commute with the Hamiltonian $k$ times which corresponds to a walk of length $k$ on the graph starting at node 1. To get the total 1-norm, we must sum over all weighted walks of length $k$ which begin at node 1 with no restrictions on the end node. By a weighted walk we mean a walk through the edges of the graph in which the running count is multiplied by the weight of the traversed edge.

It is simple to count the total number of weighted walks on the graph Fig. 4 for $D = 0$ (left column) if we make the following observation. If on the $i$-th step we are at node 1, then we have a single option for the next node, and this traversal has weight 8. Thus, the weighted number of options is $8 \cdot 1 = 8$. If we are at any other node, there are two options for the next node, and each traversal has weight 4, so the weighted number of options is $4 \cdot 2 = 8$. Since the number of weighted options is thus equal at each step and there are $k$ total steps, the 1-norm must be equal to $8^k$, reproducing our result of Eq. (III.8).

We label node $l$ to be the node corresponding to the class of terms which are a string of $l$ Pauli matrices. The support on $l$ sites then corresponds to the weighted number of walks which begin at node 1 and end at node $l$. We first consider the support on one site. We note that any walk that begins and ends at node 1 on the graph Fig. 4 must have even $k$ and traces out a Dyck path. Thus, the total unweighted number of walks is equal to $C\left(\frac{k}{2}\right)$, where $C(n)$ in this case is the Catalan number. However, due to the edge leaving the first node having a different

weight, to find the total weighted number of Dyck paths one must use the following formula

$$s_1^{(1)}(k) = \sum_m w(m,k) p(m,k) \tag{D.1}$$

where $s_1^{(1)}(k)$ means the 1-norm of terms with support on one site as a function of $k$, $w(m,k)$ is the weight of a path as a function of $m$ and $k$ where $m$ is the number of traversals of the single edge of weight 8, and $p(m,k)$ is the number of paths as a function of $m$ and $k$. Note that we use the term paths instead of walks here to refer to the fact that these are Dyck paths. It is clear that the bounds on $m$ are $1 \leq m \leq \frac{k}{2}$ and that $w(m,k) = 4^{k-m} 8^m = 4^k 2^m$. We calculate $p(m,k)$ by partitioning the Dyck path into $m$ parts according to intersections with the $x$-axis. We then have

$$
\begin{aligned}
p(m,k) &= \sum_{i_1 + \ldots + i_m = \frac{k}{2}} \prod_{j=1}^m C(i_j - 1) \\
&= \frac{m}{k-m} \binom{k-m}{\frac{k}{2}}
\end{aligned}
\tag{D.2}
$$

where we used Catalan's k-fold convolution formula. This leads to

$$s_1^{(1)}(k) = 4^k \sum_{m=1}^{\frac{k}{2}} 2^m \frac{m}{k-m} \binom{k-m}{\frac{k}{2}} = 4^k \binom{k}{\frac{k}{2}}. \tag{D.3}$$

The support on 2 sites is then easily found using a recursive strategy as

$$s_2^{(1)}(k) = \frac{1}{4} s_1(k+1) = 2^{2k+1} \binom{k}{\frac{k-1}{2}}. \tag{D.4}$$

We see that this reproduces our previous result Eq. (III.11) with $l = 2$. Another recursive equation gives the support on 3 sites

$$
\begin{aligned}
s_3^{(1)}(k) &= \frac{1}{4} s_2(k+1) - 2 s_1(k) \\
&= 2^{2k+1} \left[ \binom{k+1}{\frac{k}{2}} - \binom{k}{\frac{k}{2}} \right] = 2^{2k+1} \binom{k}{\frac{k-2}{2}}.
\end{aligned}
\tag{D.5}
$$

We see that this also agrees with our previous result Eq. (III.11) for $l = 3$. Finally, we get the following recursive equation for $l \geq 4$:

$$s_l^{(1)}(k) = \frac{1}{4} s_{l-1}(k+1) - s_{l-2}(k). \tag{D.6}$$

Simple inductive arguments show that the solution to Eq. (D.6) with base cases (D.4) and (D.5) is indeed our previous result (III.11).

### 2. Disordered case

In the disordered case, we find that we have new classes of terms in the commutator with edges given by the magnetic field values $h_j$. For the Anderson case, we know that

$h_j \in [-D, D]$, so the maximum absolute value of the disorder at each site is $D$. For the Aubry-André model, the same additional terms in the commutator appear but in this case the maximum value is $\lambda/2$. This allows us to create a modified graph which contains an additional column of terms, see Fig. 4, and which allows us to derive again an upper bound of the 1-norm. We discuss the Anderson case here and note that the Aubry-André case follows immediately with the replacement $D \to \lambda/2$.

Since $D$ is an upper bound on the disorder at each site, we can only compute an upper bound on the 1-norm growth of the commutator. Following the same strategy as for the non-disordered case, we see that at a given step, if we are at node 1, then we have a weighted number of options of 8. If we are at the $\sigma^x \sigma^x$ node, we have a weighted number of options of $4 + 8D$. If we are at any other node, we have a weighted number of options of $8 + 8D$. The maximum of these is $8 + 8D$, so we get the following upper bound on the 1-norm growth of the commutator, keeping in mind that the first step must have a contribution of 8:

$$||[H, \sigma_0^z]||_1 \leq 8^k (D+1)^{k-1}. \tag{D.7}$$

This is the result Eq. (III.15). The bound is exponential as expected, with an increasing disorder strength corresponding to an increased growth rate. We also see that setting $D = 0$ gives us $8^k$, the exact result in the case without disorder.

Now we consider the spatial structure of the commutator. We can get a qualitative sense of why localization occurs in this case. Recall that the support on $l$ sites in the graphical language is the weighted number of walks of length $k$ that start at node 1 and ends at node $l$. First note that in this case there are two nodes for each $l$. Additionally, with the new terms due to disorder, there will be more possible walks of length $k$ as one may also traverse the edges with weight $8D$. However, traversing these edges does not result in a node with an increased $l$. Therefore, in contrast to the non-disordered case, there will be a greater number of walks that end at nodes with smaller $l$, and the relative number of walks should decrease as $l$ increases. Furthermore, for $D \gg 1$ traversing the edges with weight $8D$, which does not increase $l$, leads to a much larger contribution to the norm than traversing those edges which connect nodes with different $l$'s. This means that the 1-norm of term with support on $l$ sites decays with $l$, indicating localization.

For a more quantitative analysis, let us consider the upper bound on the support at $l$ sites $s_l^{(1)}(k)$. Numerical calculations using weighted adjacency matrices led to the following result for the support on one site

$$s_1^{(1)}(k) \leq 2^{3k-1} D^{k-2} \sum_{i=0}^{\frac{k-2}{2}} C(i) \binom{k-1}{2i} (2D)^{-2i} \tag{D.8}$$

$$= 2^{3k-1} D^{k-2} {}_2F_1\left(\frac{1-k}{2}, \frac{2-k}{2}; 2; \frac{1}{D^2}\right)$$

where ${}_2F_1$ is the hypergeometric function. Note that this result is derived for even $k$. For $k$ odd the support will be zero; one can infer this from Fig. 4 since there is no possible walk on the graph beginning and ending at the top node with an odd number of steps. The upper bound in Eq. (D.8) only arises from the fact that $D$ is an upper bound on the local magnetic fields; the enumerative calculations are exact. Similar numerical calculations show that the support on $l > 1$ sites is bounded from above by a hypergeometric function as well but with a different leading term

$$s_l^{(1)}(k) \leq 2^{3k+2-l} \binom{k-1}{l-2} D^{k-l+1} \tag{D.9}$$
$$\times \ {}_2F_1\left(\frac{1-k}{2}, \frac{2-k}{2}; 2; \frac{1}{D^2}\right).$$

The upper bound is looser in this case because the enumerative calculations are no longer exact. We are then interested in the ratio $s_l^{(1)}(k)/s_1^{(1)}(k)$ in the limit $k \gg l \gg 1$. Since the upper bound given for $s_l^{(1)}(k)$ is in fact looser than that given for $s_1^{(1)}(k)$, we may bound the ratio by the ratio of the two given upper bounds, leading to the result

$$\frac{s_l^{(1)}(k)}{s_1^{(1)}(k)} \leq \binom{k-1}{l-2} (2D)^{3-l}$$
$$\approx \frac{1}{\sqrt{2\pi l}} \left(\frac{ke}{l}\right)^l e^{-\frac{l^2}{2k}} (2D)^{3-l} \tag{D.10}$$

This coincides with Eq. (III.18). As mentioned in the main text, there is a maximum in this ratio which shifts to larger $l$ with increasing $k$, which is not expected for a localized model. This is likely explained by the looseness of the bound Eq. (D.9); while it is correct for the leading term in $D$, the actual prefactors for terms of decreasing order in $D$ become increasingly smaller than those in the given bound. We expect that a formula replacing Eq. (D.9) in which the enumerative calculations are exact and the upper bound only comes from $D$ would not possess this unexpected behaviour, although we have been unable to derive such a formula. We nevertheless see the expected result that with $k$ held constant we have an exponential decay of the ratio of support for increasing $l$, and that increasing $D$ corresponds to an increasing rate of decay.


\end{widetext}

\end{document}